\documentclass[preprint]{JHEP3} 

\JHEPspecialurl{http://jhep.sissa.it/JOURNAL/JHEP3.tar.gz}

\usepackage{epsfig,multicol,bbm}
\usepackage{longtable}
\usepackage{amsthm,latexsym,multibox,amssymb,amsfonts,array}
\usepackage[centertags,intlimits]{amsmath}
\usepackage{eufrak}
\usepackage[mathscr]{eucal}
\usepackage{graphicx}
\usepackage{epstopdf}
\newcommand\fverb{\setbox\pippobox=\hbox\bgroup\verb}
\newcommand\fverbdo{\egroup\medskip\noindent%
            \fbox{\unhbox\pippobox}\ }
\newcommand\fverbit{\egroup\item[\fbox{\unhbox\pippobox}]}
\newbox\pippobox

\newcommand{\beq}{\begin{equation}}
\newcommand{\eeq}{\end{equation}}
\newcommand{\eqnlab}[1]{\label{eqn:#1}}
\newcommand{\Eqnref}[1]{Eq.~(\ref{eqn:#1})}
\newcommand{\beqa}{\begin{eqnarray}}
\newcommand{\eqa}{\end{eqnarray}}

\newcommand{\al}{\alpha}
\newcommand{\be}{\beta}

\newcommand{\de}{\delta}

\newcommand{\f}{\frac}

\newcommand{\nn}{\nonumber}

\newcommand{\g}{\mathrm{g}}

\newcommand{\mf}{\mathfrak}
\newcommand{\mg}{\mathfrak{g}}
\newcommand{\mk}{\mathfrak{k}}
\newcommand{\mh}{\mathfrak{h}}
\newcommand{\mgb}{\bar{\mathfrak{g}}}

\title{Geometric Configurations, Regular Subalgebras of $E_{10}$ and M-Theory Cosmology}

\author{Marc Henneaux$^{1,3}$, Mauricio Leston$^{2,4}$, Daniel Persson$^{1}$ and Philippe Spindel$^{5}$\\
     $^{1}$
Physique Th\'{e}orique et Math\'{e}matique, \\ Universit\'{e} Libre
de Bruxelles and
 International Solvay Institutes, \\
 ULB-Campus Plaine C.P.231, B-1050 Bruxelles,
Belgium. \\
 $^2$ Theoretische Naturkunde, Vrije Universiteit Brussel and   \\
  International Solvay Institutes, Pleinlaan 2, B-1050 Brussels, Belgium.\\
   $^{3}$  Centro de Estudios Cient\'{\i}ficos
(CECS), Casilla 1469, Valdivia, Chile.\\
 $^{4}$ Instituto de Astronomica y Fisica del Espacio (IAFE), \\
  Casilla de Correo 67, Sucursal 28, 1428 Buenos Aires, Argentina.\\
  $^{5}$ M\'ecanique et Gravitation,
 Universit\'e de Mons-Hainaut, \\
Acad\'emie Wallonie-Bruxelles, Place du Parc 20, 7000 Mons,
Belgium.\\
    E-mail: \email{henneaux@ulb.ac.be}, \email{mauricio@iafe.uba.ar}, \email{dpersson@ulb.ac.be}, \email{Philippe.Spindel@umh.ac.be}}
\received{????, 2006}       
\revised{????, 2006}
\accepted{???? 2006}        

\preprint{\hepth{0606123}}  

\abstract{We re-examine previously found cosmological solutions to
eleven-dimensional supergravity in the light of the
$E_{10}$-approach to M-theory. We focus on the solutions with non
zero electric field determined by geometric configurations $(n_m,
g_3)$, $n\leq 10$. We show that these solutions are associated with
rank $g$ regular subalgebras of $E_{10}$, the Dynkin diagrams of
which are the (line) incidence diagrams of the geometric
configurations. Our analysis provides as a byproduct an interesting
class of rank-10 Coxeter subgroups of the Weyl group of $E_{10}$.}


\begin{document}


\section{Introduction}
\setcounter{equation}{0} Even in the cosmological context of
homogeneous fields $G_{\alpha \beta}(t)$, $F_{\alpha \beta \gamma
\delta}(t)$ that depend only on time (``Bianchi I cosmological
models" \cite{Demaret:1985js}), the equations of motion of
eleven-dimensional supergravity remain notoriously complicated.
This is because the dynamical behavior of the system is, for
generic initial conditions, a never ending succession of ``free"
Kasner regimes interrupted by ``collisions" against ``symmetry" or
``electric" walls \cite{Damour:2000th,Damour:2000hv}.  During a
given Kasner regime, the energy-momentum of the $3$-form potential
can be neglected and the scale factors of the spatial metric have
the typical Kasner power law behavior $\sim t^{2 p_i}$ with
$\sum_i p_i = 1 = \sum_i p_i^2$ in terms of the proper time $t$.
Any of these free flight motions ultimately ends with a collision,
leading to a transition to a new Kasner regime characterized by
new Kasner exponents.  In the collision against an electric wall,
the energy density of the electric field becomes comparable to the
Ricci tensor for the short time of duration of the collision. The
localization in time of the collision and hence of the
corresponding electric energy density gets sharper and sharper as
one goes to the cosmological singularity\footnote{We are
simplifying the discussion by assuming the collisions to be
clearly separated in time.  In general, ``multiple collisions"
might take place, without changing the qualitative picture
\cite{Damour:2000th}. Also, when the magnetic field is non zero,
there can be collisions against magnetic walls.}.  The model is a
simple example exhibiting the intricate BKL-type phenomenon
\cite{BKL}.

The dynamical behavior of the system can be represented as a
billiard motion
\cite{Misner,Kirillov,Kirillov:1994fc,Ivashchuk:1999rm,Damour:2000hv,Damour:2002et}
in the fundamental Weyl chamber of the Lorentzian Kac-Moody
algebra $E_{10}$ \cite{Damour:2000hv}. The hyperbolic character of
$E_{10}$ accounts for the chaotic properties of the dynamics
\cite{Damour:2001sa}.

In order to get a more tractable dynamical system, one may impose
further conditions on the metric and the $4$-form. This must be
done in a manner compatible with the equations of motion: if the
additional conditions are imposed initially, they should be
preserved by the time evolution.  One such set of conditions is
that the spatial metric be diagonal, \beq ds^2 = - N^2(x^0)
(dx^0)^2 + \sum_{i = 1}^{10} a_i^2(x^0) (dx^i)^2 \, .\eeq
Invariance under the ten distinct spatial reflections $\{x^j
\rightarrow -x^j, x^{i\neq j}\rightarrow x^{i\neq j}\}$  of the
metric is compatible with the Einstein equations only if the
energy-momentum tensor of the $4$-form is also diagonal. Although
one cannot impose on the 4-form itself to be reflection invariant
without forcing it to vanish, one can ensure that the
energy-momentum tensor is reflection invariant.

A large class of electric solutions to the question of finding
$F_{\alpha \beta \gamma \delta}$ such that $T_{\mu \nu}$ is
invariant under spatial reflections was found in
\cite{Demaret:1985js}.  These can be elegantly expressed in terms
of geometric configurations $(n_m, g_3)$ of $n$ points and $g$
lines (with $n \leq 10$). That is, for each geometric
configuration $(n_m, g_3)$ (whose definition is recalled below),
one can associate diagonal solutions with some non-zero electric
field components $F_{0ijk}$ determined by the configuration.  The
purpose of this paper is to re-examine this result in the light of
the attempt to reformulate M-theory as an $E_{10}$ non linear
$\sigma$-model in one dimension.

It has recently been shown in \cite{Damour:2002cu} that the
dynamical equations of eleven-dimensional supergravity can be
reformulated as the equations of motion of the one-dimensional non
linear $\sigma$-model
$\mathcal{E}_{10}/\mathcal{K}(\mathcal{E}_{10})$, where
$\mathcal{K}(\mathcal{E}_{10})$ is the subgroup of
$\mathcal{E}_{10} \equiv \exp{E_{10}}$ obtained by exponentiating
the subalgebra $K(E_{10})$ of $E_{10}$ invariant under the
standard Chevalley involution\footnote{In the infinite-dimensional
case of $E_{10}$, the connection between the Lie algebra and the
corresponding group is somewhat subtle.  We shall proceed formally
here, as in the finite-dimensional case. This is possible because,
as a rule, the quantities of direct interest for our analysis will
be elements of the algebra.}. Although the matching works only at
low levels (with the dictionary between the two theories derived
so far), it provides further intriguing evidence that infinite
dimensional algebras of $E$-type might underlie the dynamics of
M-theory \cite{Julia:1982gx,Damour:2000hv,West:2001as}.

We prove here that the conditions on the electric field embodied
in the geometric configurations $(n_m, g_3)$ have a direct
Lie-algebraic interpretation.  They simply amount to consistently
truncating the $E_{10}$ non linear $\sigma$-model to a $\mgb$ non
linear $\sigma$-model, where $\mgb$ is a rank-$g$ Kac-Moody
subalgebra of $E_{10}$ (or a quotient of such a Kac-Moody
subalgebra by an appropriate ideal when the relevant Cartan matrix
has vanishing determinant), which has three properties: (i) it is
regularly embedded in $E_{10}$, (ii) it is generated by electric
roots only, and (iii) every node $P$ in its Dynkin diagram
$\mathbb{D}_{\mgb}$ is linked to a number $k$ of nodes that is
independent of $P$ (but depend on the algebra). The Dynkin diagram
$\mathbb{D}_{\mgb}$ of $\mgb$ is actually the line incidence
diagram of the geometric configuration $(n_m, g_3)$ in the sense
that (i) each line of $(n_m, g_3)$ defines a node of
$\mathbb{D}_{\mgb}$, and (ii) two nodes of $\mathbb{D}_{\mgb}$ are
connected by a single bond iff the corresponding lines of $(n_m,
g_3)$ have no point in common. None of the algebras $\mgb$
relevant to the truncated models turn out to be hyperbolic: they
can be finite, affine, or Lorentzian with infinite-volume Weyl
chamber. Because of this, the solutions are non chaotic. After a
finite number of collisions, they settle asymptotically into a
definite Kasner regime (both in the future and in the past).
Disappearance of chaos for diagonal models was also recently
observed in \cite{Ivashchuk:2005kn}.

In the most interesting cases, $\mgb$ is a rank-10, Lorentzian
(but not hyperbolic) Kac-Moody subalgebra of $E_{10}$. We do, in
fact, get six rank-10 Lorentzian Kac-Moody subalgebras of
$E_{10}$, which, to our knowledge, have not been previously
discussed. We also get one rank-10 Kac-Moody algebra with a Cartan
matrix that is degenerate but not positive semi-definite, and
hence is not of affine type. Its embedding in $E_{10}$ involves
the quotient by its center.  We believe that the display of these
subalgebras might be in itself of some mathematical interest in
understanding better the structure of $E_{10}$.  At the level of
the corresponding reflection groups, our method exhibits seven
rank-10 Coxeter subgroups of the Weyl group of $E_{10}$ which have
the property that their Coxeter exponents are either $2$ or $3$ -
but never $\infty$, and which are furthermore such that there are
exactly three edges that meet at each node of their Coxeter
diagrams.

Our paper is organized as follows.  In the next section, we recall
the equations of motion of eleven-dimensional supergravity for
time dependent fields (Bianchi type I models) and the consistent
truncations associated with geometric configurations.  We then
recall in Section \ref{symmetric} the $\sigma$-model formulation
of the bosonic sector of eleven-dimensional supergravity. In
Section \ref{Regular} we consider regular subalgebras and
consistent subgroup truncations of $\sigma$-models. This is used
in Section \ref{geometric} to relate the consistent truncations
of eleven-dimensional supergravity associated with geometric
configurations to the consistent truncations of the $E_{10}$-sigma
model based on regular subalgebras with definite properties that
are also spelled out.  The method is illustrated in the case of
configurations associated with subalgebras of $E_8$.  We then turn
in Section \ref{Affine} to the geometric configurations leading to
affine subalgebras, naturally embedded in $E_9$. Section
\ref{Rank10} is devoted to the rank 10 case.  Finally, we close
our paper with conclusions and directions for future developments.

\section{Bianchi I Models and Eleven-Dimensional Supergravity}
\setcounter{equation}{0} \subsection{Equations of motion} For
time-dependent fields, \beqa
&& ds^2 = - N^2(x^0) (dx^0)^2 + G_{ab}(x^0)dx^a dx^b \\
&&F_{\lambda \rho \sigma \tau} = F_{\lambda \rho \sigma
\tau}(x^0), \eqa the equations of motion of eleven-dimensional
supergravity read \beqa \frac{d\left(K^a_{\; \; b}
\sqrt{G}\right)}{dx^0} &=& -\frac{N}{2} \sqrt{G} F^{a \rho \sigma
\tau} F_{b \rho \sigma \tau} + \frac{N}{144} \sqrt{G} F^{\lambda
\rho \sigma \tau} F_{\lambda \rho \sigma \tau}
\delta^a_b \label{Einstein0}\\
\frac{d\left(F^{0abc} N \sqrt{G} \right)}{dx^0} &=& \frac{1}{144}
\varepsilon^{0 a b c d_1 d_2 d_3 e_1 e_2 e_3 e_4} F_{0 d_1 d_2
d_3} F_{e_1 e_2 e_3 e_4} \\
\frac{d F_{a_1 a_2 a_3 a_4}}{dx^0} &=& 0 \eqa (dynamical
equations) and \beqa && K^a_{\; \;b} K^b_{\; \; a} - K^2
+\frac{1}{12} F_{\perp abc}F_\perp^{\; \; abc} + \frac{1}{48}
F_{abcd}F^{abcd} = 0 \label{HamiltonianC}\\ && \frac{1}{6} N
F^{0bcd}F_{abcd} = 0 \label{momentum}
\\ &&\varepsilon^{0abc_1 c_2 c_3 c_4 d_1 d_2 d_3 d_4} F_{c_1 c_2 c_3
c_4} F_{d_1 d_2 d_3 d_4} = 0\eqa (Hamiltonian constraint, momentum
constraint and Gauss law). Here, we have set $K_{ab} = (-1/2N)
\dot{G}_{ab}$ and $F_{\perp abc} = (1/N) F_{0 abc}$.

\subsection{Diagonal Metrics and Geometric Configurations}
If the metric is diagonal, the extrinsic curvature $K_{ab}$ is
also diagonal.  This is consistent with Eq. (\ref{Einstein0}) only
if $F^{a \rho \sigma \tau} F_{b \rho \sigma \tau}$ is diagonal,
i.e., taking also into account Eq. (\ref{momentum}), if the
energy-momentum tensor $T^\alpha_{\; \; \beta}$ of the $4$-form
$F_{\lambda \rho \sigma \tau}$ is diagonal. Assuming zero magnetic
field (this restriction will be lifted below), one way to achieve
this condition is to assume that the non-vanishing components of
the electric field $F^{\perp a b c}$ are determined by ``geometric
configurations" $(n_m,g_3)$ with $n \leq 10$
\cite{Demaret:1985js}.

A geometric configuration $(n_m,g_3)$ is a set of $n$ points and
$g$ lines with the following incidence rules
\cite{Kantor,Hilbert,Page}:
\begin{enumerate}
\item \label{rule1} Each line contains three points. \item \label{rule2} Each point is on $m$
lines. \item \label{rule3} Two points determine at most one line.
\end{enumerate} It follows that two lines have at most one point
in common.  It is an easy exercise to verify that $m n = 3 g $. An
interesting question is whether the lines can actually be realized
as straight lines in the (real) plane, but, for our purposes, it
is not necessary that it should be so; the lines can be bent.

We shall need the configurations with $n \leq 10$ points. These
are all known and are reproduced in the appendix B of
\cite{Demaret:1985js} and listed in Sections \ref{geometric},
\ref{Affine} and \ref{Rank10} below. There are:
\begin{itemize}
\item one configuration $(3_1,1_3)$ with 3 points; \item two
configurations with 6 points, namely $(6_1,2_3)$ and $(6_2, 4_3)$;
\item one configuration $(7_3,7_3)$, which is related to the
octonions and which cannot be realized by straight lines; \item
one configuration $(8_3, 8_3)$, which cannot be realized by
straight lines; \item one configuration $(9_1, 3_3)$, two
configurations $(9_2,6_3)$, three configurations $(9_3,9_3)$, and
finally one configuration $(9_4,12_3)$ that cannot be drawn with
straight lines; \item ten configurations $(10_3,10_3)$, with one
of them, denoted $(10_3,10_3)_1$, not being realizable in terms of
straight lines.
\end{itemize}
Some of these configurations are related to theorems of projective
geometry and are given a name - e.g. the Desargues configuration
$(10_3,10_3)_3$ explicitly discussed below; but most of them,
however, bear no name.

Let $(n_m,g_3)$ be a geometric configuration with $n \leq 10$
points.  We number the points of the configuration $1, \cdots, n$.
We associate to this geometric configuration a pattern of
electric field components $F^{\perp a b c}$ with the following
property: $F^{\perp a b c}$ can be non-zero only if the triplet
$(a,b,c)$ is a line of the geometric configuration.  If it is
not, we take $F^{\perp a b c} = 0$.  It is clear that this
property is preserved in time by the equations of motion (in the
absence of magnetic field).  Furthermore, because of Rule {\bf
\ref{rule3}} above, the products $F^{\perp a b c} F^{\perp a' b'
c'} g_{bb'} g_{cc'}$ vanish when $a \not= a'$ so that the
energy-momentum tensor is diagonal.

We shall now show that these configurations have an
algebraic interpretation in terms of subalgebras of $E_{10}$. This
will also enable us to relax the condition that the magnetic field
should be zero while preserving diagonality.  To that end, we need
first to recall the $\sigma$-model reformulation of
eleven-dimensional supergravity.

\section{Geodesics on the Symmetric Space
$\mathcal{E}_{10}/\mathcal{K}(\mathcal{E}_{10})$: An Overview}
\label{symmetric} \setcounter{equation}{0}
\subsection{Borel Gauge}
Let $\mg$ be the split real form of a rank $n$ Kac-Moody algebra
with generators $h_i,e_i,f_i$ ($i = 1, 2 \cdots , n$) and Cartan
matrix $A_{ij}$ (see \cite{Kac} for more information). This
algebra may be finite or infinite dimensional.  We assume in this
latter case that the Cartan matrix is symmetrizable and that the
symmetrization of $A_{ij}$ is invertible.  In fact, only
Lorentzian Kac-Moody algebras, for which the symmetrization of
$A_{ij}$ has signature $(-, +, +, \cdots, +)$, will be of
immediate concern in this paper. We write the triangular
decomposition of $\mg$ as $\mg = \mf{n}_- \oplus \mf{h} \oplus \mf{n}_+$, where $\mf{h}$
is the Cartan subalgebra of $\mg$ and $\mf{n}_+$ (respectively, $\mf{n}_-$) its
upper (respectively, lower) triangular subalgebra containing the
``raising" (respectively, ``lowering") operators $e_i, [e_i, e_j],
[e_i, [e_j, e_k]]$ etc (respectively, $f_i, [f_i, f_j], [f_i, [f_j,
f_k]]$ etc) that do not vanish on account of the Serre relations
or the Jacobi identity.  The raising and lowering operators are
collectively called ``step operators".

The split real form of $\mg$ is the real algebra containing all the
real linear combinations of the generators $h_i,e_i,f_i$ and their
multiple commutators. Although non split real forms of Kac-Moody
algebras are relevant to some supergravity models \cite{MHBJ}, we
shall for definiteness not consider them explicitly here as the
case under central consideration in this paper is $E_{10,10}$, the
split real form of $E_{10}$.  In the non split case, it is the
real roots and the real Weyl groups that play the role of the
roots and Weyl groups introduced below.

Let $\mf{k}$ be the maximal ``compact subalgebra" of $\mg$, i.e. the
subalgebra pointwise invariant under the Chevalley involution
$\tau$ defined by
\begin{equation} \tau(h_i) = - h_i , \; \; \; \; \tau(e_i) = -
f_i, \; \; \; \; \tau(f_i) = -e_i.
\end{equation}
\noindent Consider the
symmetric space $\mathcal{G}/\mathcal{K}$, where $\mathcal{G}$ is
the group obtained by exponentiation of $\mg$ and $\mathcal{K}$ is
its maximal compact subgroup, obtained by exponentiation of $\mf{k}$
(as already mentioned previously, in the case where the Cartan
matrix is of indefinite type, this is somewhat formal).

The construction of the Lagrangian for the geodesic motion on
$\mathcal{G}/\mathcal{K}$ follows a standard pattern
\cite{CWZ,CCWZ}. The motion is formulated in terms of a
one-parameter dependent group element $g(x^0) \in \mathcal{G}$,
with the identification of $g$ with $kg$, where $k \in
\mathcal{K}$. The Lie algebra element $\nu(x^0)= \dot{g} g^{-1}
\in \mg$ is invariant under multiplication of $g(x^0)$ to the right
by an arbitrary constant group element $h$, $g(x^0) \rightarrow
g(x^0)h$. Decompose $\nu(x^0)$ into a part along $\mf{k}$ and a part
perpendicular to $\mf{k}$, \beq \nu(x^0) = Q(x^0) + P(x^0), \; \; \;
Q(x^0) = \frac{1}{2}\left(\nu + \tau(\nu)\right) \in \mk, \; \; \;
P(x^0) = \frac{1}{2}\left(\nu - \tau(\nu)\right). \eeq Under left
multiplication by an element $k(x^0)\in \mathcal{K}$, $Q(x^0)$
behaves as a $\mk$-connection, while $P(x^0)$ transforms in a
definite $\mk$-representation (which depends on the coset space at
hand). The Lagrangian from which the geodesic equations of motion
derive is an invariant built out of $P$ which reads explicitly
\beq \mathcal{L}= n(x^0)^{-1} \left<P(x^0) \vert P(x^0) \right>
\eqnlab{Lagrangian}\eeq where $\left<\; \vert\; \right>$ is the
invariant bilinear form on $\mg$. In the case of relevance to
gravity, diffeomorphism invariance with respect to time is
enforced by introducing the (rescaled) lapse variable $n(x^0)$, as
done above, whose variation implies that the geodesic on
$\mathcal{G}/\mathcal{K}$ -- whose metric has Lorentzian signature
-- is lightlike.  The connection between $n(x^0)$ and the
standard lapse $N(x^0)$ will be given below.

The Lagrangian is invariant under the gauge transformation $g(x^0)
\rightarrow k(x^0) g(x^0)$ with $k(x^0) \in \mathcal{K}$.  One can
use this gauge freedom to go to the Borel gauge.  The Iwasawa
decomposition states that $g$ can be uniquely written as $g = k a
n$, where $k \in \mathcal{K}$, $a \in \mathcal{H}$ and $n \in
\mathcal{N}_+$.  Here, $\mathcal{H}$ is the abelian group obtained
by exponentiating the Cartan subalgebra $\mh$ while $\mathcal{N}_+$
is the group obtained by exponentiating $\mf{n}_+$.  The Borel gauge is
defined by $k = e$ so that $g = an$ contains only the Cartan
fields (``scale factors") and the off-diagonal fields associated
with the raising operators $e_\alpha = \{e_i, [e_i, e_j], \cdots
\}$, namely, $a = \exp{\beta^i(x^0) h_i}$ and $n = \exp{a^\alpha
(x^0) e_\alpha}$.

We shall write more explicitly the Lagrangian in the Borel gauge
in the $E_{10}$ case.  To that end, we recall first the structure
of $E_{10}$ at low levels.

\subsection{$E_{10}$ at Low Levels}

We describe the algebra $E_{10}$ using the level decomposition of
\cite{Damour:2002cu}.  The level zero elements are all the
elements of the $A_{9,9} \equiv \mf{sl}(10,\mathbb{R})$ subalgebra
corresponding to the Dynkin subdiagram with nodes 1 to 9, together
with the tenth Cartan generator corresponding to the exceptional
root labeled ``$10$'' in Figure \ref{figure:E10} and given
explicitly in \Eqnref{exceptionalCartan} below. Thus, the level
zero subalgebra is enlarged from $\mf{sl}(10,\mathbb{R})$ to
$\mf{gl}(10,\mathbb{R}) = \mf{sl}(10,\mathbb{R}) + \mh_{E_{10}} $ (the sum is not
direct) and contains all Cartan generators. The
$\mf{sl}(10,\mathbb{R})$ Chevalley generators are given by \beq
e_{i}={K^{i+1}}_{i} \qquad f_{i}={K^{i}}_{i+1} \qquad
h_{i}={K^{i+1}}_{i+1}-{K^{i}}_{i} \qquad (i=1,\dots , 9),
\eqnlab{A9generators} \eeq while the commutation relations of the
level $0$ algebra $\mf{gl}(10, \mathbb{R})$ read \beq
[{K^{a}}_{b},{K^{c}}_{d}]=
\delta^{c}_{b}{K^{a}}_{d}-\delta^{a}_{d}{K^{c}}_{b}. \eqnlab{GL10}
\eeq

\begin{figure}[ht]
\begin{center}
\includegraphics[width=120mm]{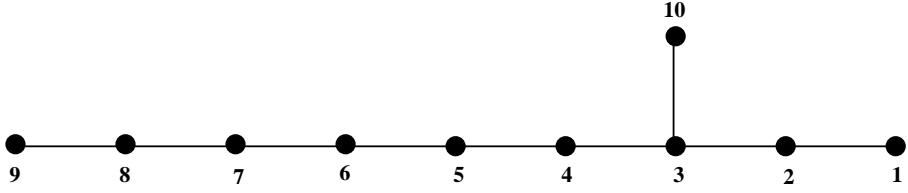}
\caption{The Dynkin diagram of
$E_{10}$. Labels $i=1,\dots, 9$ enumerate the nodes corresponding
to simple roots, $\al_{i}$, of the $A_{9}$ subalgebra and the
exceptional node, labeled ``$10$'', is associated to the root
$\al_{10}$ that defines the level decomposition.}
\label{figure:E10}
\end{center}
\end{figure}

\noindent The level $1$ generators $E^{abc}$ and their
``transposes"\footnote{As in \cite{Damour:2002et}, we formally
define the transpose of a Lie algebra element $u$ through $u^T = -
\tau(u)$ and extend to formal products - including group elements
- through $(uv)^T = v^T u^T$.  Elements $k$ of
$\mathcal{K}(\mathcal{E}_{10})$ are such that $k^T = k^{-1}$. One
has also $Q^T = - Q$ and $P^T = P$.} $F_{abc}= - \tau(E^{abc})$ at
level $-1$ transform contravariantly and covariantly with respect to
$\mf{gl}(10,\mathbb{R})$, \beq
[{K^{a}}_{b},E^{cde}]=3\delta^{[c}_{b}E^{de]a}\qquad
[{K^{a}}_{b},F_{cde}]=-3{\delta^{a}}_{[c}F_{de]b}. \eqnlab{level1}
\eeq We further have \beq
[E^{abc},F_{def}]=18\delta^{[ab}_{[de}{K^{c]}}_{f]}-
2\delta^{abc}_{def}\sum_{a=1}^{10}{K^{a}}_{a},
\eqnlab{level1relations} \eeq where we defined \beq
\delta^{ab}_{cd}=\f{1}{2}(\de^{a}_{c}\de^{b}_{d}-\de^{b}_{c}\de^{a}_{d}),
\eqnlab{delta} \eeq and similarly for $\delta^{abc}_{def}$. The
exceptional generators associated with the roots $\alpha_{10}$ and
$-\alpha_{10}$ are \beq e_{10}=E^{123} \qquad f_{10}=F_{123}.
\eqnlab{exceptionalgenerators} \eeq From the Chevalley relation
$[e_{10},f_{10}]=h_{10}$, one identifies, upon examination of
\Eqnref{level1relations},  the remaining Cartan generator as \beq
h_{10}=-\f{1}{3}\sum_{i\neq 1,2,3}
{K^{a}}_{a}+\f{2}{3}({K^{1}}_{1}+{K^{2}}_{2}+{K^{3}}_{3}).
\eqnlab{exceptionalCartan} \eeq It is a straightforward exercice
to check from the above commutation relations that the generators
$e_{i}, e_{10}, f_{i}, f_{10}, h_{i}$ and $h_{10}$ in the
$\mf{gl}(10,\mathbb{R})$-form given above satisfies indeed the standard
Chevalley-Serre relations associated with the Cartan matrix of
$E_{10}$.

The bilinear form of $E_{10}$ is given up to level $\pm 1$ by
\beqa {}æ\left<{K^{a}}_{b}|{K^{c}}_{d}\right> &=&
\de^{a}_{d}\de^{c}_{b}-\de^{a}_{b}\de^{c}_{d}\\
{}æ\left<E^{abc}|F_{def}\right> &=& 3! \de^{abc}_{def},
\eqnlab{bilinearform} \eqa where the second relation is normalized
such that $æ\left<E^{123}|F_{123}\right> =1$.

We will explicitly need the generators of $E_{10}$ up to level $3$.
These are constructed from multiple commutators of the level
$1$-generators, i.e. at level $2$ we find a $6$-form \beq
[E^{a_{1}a_{2}a_{3}},E^{a_{4}a_{5}a_{6}}]\equiv E^{a_{1}\dots a_{6}}
\eqnlab{level2} \eeq (and similarly for the transposes at level
$-2$). The commutators at this level are \beq [E^{a_{1}\dots
a_{6}},F_{b_{1}\dots b_{6}}]=6\cdot 6! \delta^{[a_{1}\dots
a_{5}}_{[b_{1}\dots b_{5}}{K^{a_{6}]}}_{b_{6}]}-\f{2}{3}\cdot 6!
\delta^{a_{1}\dots a_{6}}_{b_{1}\dots
b_{6}}\sum_{a=1}^{10}{K^{a}}_{a}. \eqnlab{level2commutator} \eeq At
level $3$ the Jacobi-identity leaves as sole representation
occurring in $E_{10}$ the mixed representation \beq
[E^{a_{1}a_{2}a_{3}},E^{a_{4}\dots
a_{9}}]=E^{[a_{1}|a_{2}a_{3}]\dots a_{9}}, \eqnlab{level3} \eeq
where the level $3$-generator $E^{a_{1}|a_{2}\dots a_{9}}$ is
antisymmetric in the indices $a_{2}\dots a_{9}$ and such that
antisymmetrizing over all indices gives identically zero, \beq
E^{[a_{1}|a_{2}\dots a_{9}]}=0. \eqnlab{level3identity} \eeq These
are all the relations that we will need in this paper. For more
details on the decomposition of $E_{10}$ into representations of
$\mf{gl}(10,\mathbb{R})$ see
\cite{KMW,NicolaiFischbacher,Damour:2004zy}.

\subsection{Lagrangian and Conserved Currents}

The Lagrangian \Eqnref{Lagrangian} has been explicitly written
down in the Borel gauge in \cite{Damour:2002cu}. Parametrizing the
group element $g(x^0)$ as \beq g(x^0) = \exp{X_h(x^0)}
\exp{X_A(x^0)} \eeq where $$X_h(x^0) = h^a_{\; \; b}(x^0) K^b_{\;
\; a}$$ ($a \geq b$) contains the level zero fields and
\beq
X_A(x^0) = \frac{1}{3!} A_{a_1 a_2 a_3}(x^0) E^{a_1 a_2 a_3} +
\frac{1}{6!} A_{a_1 \cdots a_6}(x^0) E^{a_1 \cdots a_6} +
\frac{1}{9!} A_{a_1\vert a_2 \cdots a_9}(x^0) E^{a_1 \vert a_2
\cdots a_9} + \cdots
\eqnlab{levelzerofields}
\eeq
\noindent contains all fields at positive levels, one finds
\beqa n \,
\mathcal{L} &=& \frac{1}{4}\left(\g^{ac} \g^{bd} - \g^{ad} \g^{bc}
\right) \dot{\g}_{ab} \dot{\g}_{cd} + \frac{1}{2} \frac{1}{3!}
DA_{a_1 a_2 a_3} DA^{a_1 a_2 a_3}
 \nn \\
 && + \frac{1}{2}
\frac{1}{6!} DA_{a_1 \cdots a_6} DA^{a_1 \cdots a_6} + \frac{1}{2}
\frac{1}{9!} DA_{a_1\vert a_2 \cdots a_9} DA^{a_1\vert a_2 \cdots
a_9} + \cdots
\nn \\
 \eqa
Here, the metric $\g_{ab}$ and its inverse are
constructed from the level zero vielbein, while the ``covariant
time
derivatives" $DA$ are defined by \beqa DA_{a_1 a_2 a_3}
&=& \dot{A}_{a_1 a_2 a_3} \nn \\
DA_{a_1 \cdots a_6} &=& \dot{A}_{a_1 \cdots a_6} + 10 A_{[a_1 a_2
a_3} \dot{A}_{a_4 a_5 a_6]} \nn \\DA_{a_1\vert a_2 \cdots a_9} &=&
\dot{A}_{a_1\vert a_2 \cdots a_9} + 42 A_{<a_1 a_2 a_3}
\dot{A}_{a_4 \cdots a_9>} - 42 \dot{A}_{<a_1 a_2 a_3} A_{a_4
\cdots a_9>} \nn \\ && \hspace{1cm} +
280 A_{<a_1 a_2 a_3} A_{a_4 a_5 a_6}\dot{A}_{a_7 a_8 a_9>} \nn \\
\cdots && \nn \eqa where $<>$ denotes projection on the level 3
representation.   At level $k$, each term in $DA_{a_1 \cdots
a_{3k}}$ contains one time derivative and is such that the levels
match (a typical term in $DA^{(k)}$ has thus the form
$\dot{A}^{(i_1)} A^{(i_2)} \cdots A^{(i_f)}$ with $i_1 + i_2 +
\cdots i_f = k$).

The Lagrangian is not only gauge invariant under left
multiplication by an arbitrary time-dependent element of
$\mathcal{K}(\mathcal{E}_{10})$, it is also invariant under right
multiplication by an arbitrary constant element of
$\mathcal{E}_{10}$. Invariance under this rigid symmetry leads to
an infinite set of $E_{10}$-valued conserved currents
\cite{Damour:2002et,Damour:2002cu}, which are, in the gauge $n=1$,
\beq J = g^{-1} P g = \frac{1}{2}\mathcal{M}^{-1}
\dot{\mathcal{M}} \label{ConCurr}\eeq where the gauge invariant
infinite ``symmetric" matrix $\mathcal{M}$ is defined by \beq
\mathcal{M} = g^T g.\eeq  The current fulfills \beq J^T
\mathcal{M} = \mathcal{M} J. \eeq Eq. (\ref{ConCurr}) can formally
be integrated to yield \beq \mathcal{M}(x^0) = \mathcal{M}(0) e^{2
x^0 J} = e^{ x^0 J^T}\mathcal{M}(0) e^{ x^0 J}.\eeq  From this, one
can read off the group element, \beq g(x^0) = k(x^0) g(0) e^{x^0J}
\label{solutiong}\eeq where the compensating
$\mathcal{K}(\mathcal{E}_{10})$-transformation $k(x^0)$ is such
that $g(x^0)$ remains in the Borel gauge.  The explicit
determination of $k(x^0)$ may be quite a hard task.

\subsection{Consistent Truncations}

The $\sigma$-model can be truncated in various consistent ways. By
``consistent truncation", we mean a truncation to a sub-model
whose solutions are also solutions of the full model.

\subsubsection{Level Truncation}
The first useful truncation was discussed in
\cite{Damour:2002et,Damour:2002cu} and consists in setting all
covariant derivatives of the fields above a given level equal to
zero.  This is equivalent to equating to zero the momenta
conjugate to the $\sigma$-model variables above that given level.

Imposing in particular $DA^{(k)} = 0$ for $k \geq 3$ leads to
equations of motion which are not only consistent from the
$\sigma$-model point of view, but which are also equivalent to the
dynamical equations of motion of eleven-dimensional supergravity
restricted to homogeneous fields $G_{ab}(t)$ and $F_{\alpha \beta
\gamma \delta}(t)$ (and no fermions).  The dictionary that makes
the equivalence between the $\sigma$-model and supergravity is
given by
\cite{Damour:2002cu} \beqa && \g_{ab} = G_{ab}\\
&& DA_{a_1 a_2 a_3} = F_{0 a_1 a_2 a_3}\\ && DA^{a_1 a_2 a_3 a_4
a_5 a_6} = - \frac{n}{4!} \varepsilon^{a_1 a_2 a_3 a_4 a_5 a_6 b_1
b_2 b_3 b_4} F_{b_1 b_2 b_3 b_4} \eqa together with $n =
N/\sqrt{G}$.  Furthermore, the $\sigma$-model constraint obtained
by varying $n(x^0)$, which enforces reparametrization invariance,
is just the supergravity Hamiltonian constraint
(\ref{HamiltonianC}) in the homogeneous setting.

Full equivalence of the level 3 truncated $\sigma$-model with
spatially homogeneous supergravity requires that one imposes also
the momentum constraint as well as the Gauss law.

\subsubsection{Subgroup Truncation}

Another way to consistently truncate the $\sigma$-model equations
of motion is to restrict the dynamics to an appropriately chosen
subgroup.

We shall consider here only subgroups obtained by exponentiating
regular subalgebras of $\mg$, a concept to which we now turn.

\section{Regular Subalgebras}\label{Regular}
\setcounter{equation}{0}
\subsection{Definitions}
\label{definitions} Let $\mgb$ be a Kac-Moody subalgebra of $\mg$,
with triangular decomposition $\mgb= \bar{\mf{n}}_- \oplus
\bar{\mh} \oplus \bar{\mf{n}}_+ $. We assume that $\mgb$ is
canonically embedded in $\mg$, i.e., that the Cartan subalgebra
$\bar{\mh}$ of $\mgb$ is a subalgebra of the Cartan subalgebra
$\mh$ of $\mg$, $\bar{\mh} \subset \mh$, so that $\bar{\mh}= \mgb
\cap \mh$. We shall say that $\mgb$ is regularly embedded in $\mg$
(and call it a ``regular subalgebra") iff two conditions are
fulfilled~: (i) the step operators of $\mgb$ are step operators of
$\mg$; and (ii) the simple roots of $\mgb$ are real roots of
$\mg$. It follows that the Weyl group of $\mgb$ is a subgroup of
the Weyl group of $\mg$ and that the root lattice of $\mgb$ is a
sublattice of the root lattice of $\mg$.

The second condition is automatic in the finite-dimensional case
where there are only real roots.  It must be separately imposed in
the general case. Consider for instance the rank 2 Kac-Moody
algebra $\mathcal{A}$ with Cartan matrix $$ \left(
\begin{array}{cc}
2 & -3\\
-3 & 2\\
\end{array}\right).$$ Let \beqa x &=& \frac{1}{\sqrt{3}}[e_1,e_2]
\\ y &=& \frac{1}{\sqrt{3}}[f_1,f_2]  \\ z &=& -(h_1 + h_2).
\eqa  It is easy to verify that $x,y,z$ define an $A_1$ subalgebra
of $\mathcal{A}$ since $[z, x] = 2 x$, $[z,y] = -2 y$ and $[x,y] =
z$.  Moreover, the Cartan subalgebra of $A_1$ is a subalgebra of
the Cartan subalgebra of $\mathcal{A}$, and the step operators of
$A_1$ are step operators of $\mathcal{A}$.  However, the simple
root $\alpha = \alpha_1 + \alpha_2$ of $A_1$ (which is an
$A_1$-real root since $A_1$ is finite-dimensional), is an
imaginary root of $\mathcal{A}$: $\alpha_1 + \alpha_2$ has norm
squared equal to $-2$.  Even though the root lattice of $A_1$
(namely, $\{\pm \alpha\}$) is a sublattice of the root lattice of
$\mathcal{A}$, the reflection in $\alpha$ is not a Weyl reflection
of $\mathcal{A}$. According to our definition, this embedding of
$A_1$ in $\mathcal{A}$ is not a regular embedding.

\subsection{Examples}

We shall be interested in regular subalgebras of $E_{10}$.

\subsubsection{$A_9 \subset \mathcal{B} \subset E_{10}$}
\begin{figure}[ht]
\begin{center}
\includegraphics[width=120mm]{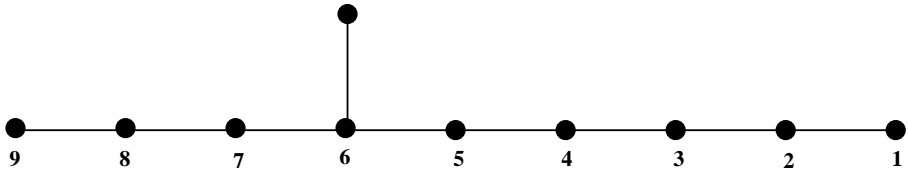}
\caption{The Dynkin diagram of $\mathcal{B} \equiv E^{+++}_7$. The
root without number is the root denoted $\bar{\alpha}_{10}$ in the
text.} \label{figure:E7ppp}
\end{center}
\end{figure}

A first, simple, example of a regular embedding is the embedding of
$A_9$ in $E_{10}$ used to define the level. This is not a maximal
embedding since one can find a proper subalgebra $\mathcal{B}$ of
$E_{10}$ that contains $A_9$.  One may take for $\mathcal{B}$ the
Kac-Moody subalgebra of $E_{10}$ generated by the operators at
levels $0$ and $\pm 2$, which is a subalgebra of the algebra
containing all operators of even level\footnote{We thank Axel
Kleinschmidt for an informative comment on this point.}. It is
regularly embedded in $E_{10}$. Its Dynkin diagram is shown on
Figure \ref{figure:E7ppp}.

\noindent In terms of the simple roots of $E_{10}$, the simple roots of
$\mathcal{B}$ are $\alpha_1$ through $\alpha_9$ and
$\bar{\alpha}_{10} = 2 \alpha_{10} + \alpha_1 + 2 \alpha_2 + 3
\alpha_3 + 2 \alpha_4 + \alpha_5$.  The algebra $\mathcal{B}$ is
Lorentzian but not hyperbolic.  It can be identified with the
``very extended" algebra $E^{+++}_7$ \cite{Gaberdiel:2002db}.

\subsubsection{$DE_{10} \subset E_{10}$}

In \cite{Dynkin}, Dynkin has given a method for finding all
maximal regular subalgebras of finite-dimensional simple Lie
algebras. The method is based on using the highest root and is not
generalizable as such to general Kac-Moody algebras for which
there is no highest root.  Nevertherless, it is useful for
constructing regular embeddings of overextensions of finite
dimensional simple Lie algebras.  We illustrate this point in the
case of $E_8$ and its overextension $E_{10}\equiv E_{8}^{++}$. In
the notation of Figure \ref{figure:E10}, the simple roots of
$E_{8}$ (which is regularly embedded in $E_{10}$) are $\alpha_1,
\cdots , \alpha_7$ and $\alpha_{10}$.

Applying Dynkin's procedure to $E_8$, one easily finds that $D_8$
can be regularly embedded in $E_8$.  The simple roots of $D_8
\subset E_8$ are $\alpha_2, \alpha_3, \alpha_4, \alpha_5,
\alpha_6, \alpha_7$, $\alpha_{10}$ and $\beta \equiv - \theta$,
where $\theta = 3 \alpha_{10} + 6 \alpha_3 + 4 \alpha_2 + 2
\alpha_1 + 5 \alpha_4 + 4 \alpha_5 + 3 \alpha_6 + 2 \alpha_7 $ is
the highest root of $E_8$ (which, incidentally, has height 29).
One can replace this embedding, in which a simple root of $D_8$,
namely $\beta$, is a negative root of $E_8$ (and the corresponding
raising operator of $D_8$ is a lowering operator for $E_8$), by an
equivalent one in which all simple roots of $D_8$ are positive
roots of $E_8$.

\begin{figure}[ht]
\begin{center}
\includegraphics[width=110mm]{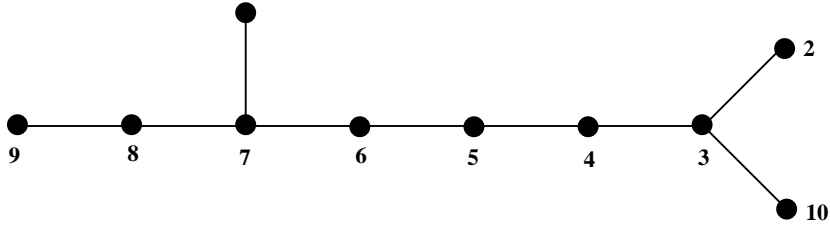}
\caption{$DE_{10}\equiv D_{8}^{++}$ regularly embedded in
$E_{10}$. Labels $2,\dots, 10$ represent the simple roots
$\al_2,\dots, \al_{10}$ of $E_{10}$ and the unlabeled node
corresponds to the positive root $\bar{\be}=2\al_1+
3\al_2+4\al_3+3\al_4+2\al_5+\al_6+2\al_{10}$.} \label{figure:DE10}
\end{center}
\end{figure}

This is done as follows.  It is reasonable to guess that the
searched-for Weyl element that maps the ``old" $D_8$ on the ``new"
$D_8$ is some product of the Weyl reflections in the four
$E_8$-roots orthogonal to the simple roots $\alpha_3$, $\alpha_4$,
$\alpha_5$, $\alpha_6$ and $\alpha_7$, expected to be shared (as
simple roots) by $E_8$, the old $D_8$ and the new $D_8$ - and
therefore to be invariant under the searched-for Weyl element.
This guess turns out to be correct: under the action of the
product of the commuting $E_8$-Weyl reflections in the $E_8$-roots
$\mu_1 = 2 \alpha_1 + 3 \alpha_2 + 5 \alpha_3 + 4 \alpha_4 + 3
\alpha_5 + 2 \alpha_6 + \alpha_7+ 3 \alpha_{10}$ and $\mu_2= 2
\alpha_1 + 4 \alpha_2 + 5 \alpha_3 + 4 \alpha_4 + 3 \alpha_5 + 2
\alpha_6 + \alpha_7+ 2 \alpha_{10}$, the set of $D_8$-roots
$\{\alpha_2,\alpha_3,\alpha_4,\alpha_5,\alpha_6, \alpha_7,
\alpha_{10}, \beta \}$ is mapped on the equivalent set of positive
roots $\{\alpha_{10},\alpha_3,\alpha_4,\alpha_5,\alpha_6,
\alpha_7, \alpha_{2}, \bar{\beta}  \}$ where $\bar{\beta} =2
\alpha_1 + 3 \alpha_2 + 4 \alpha_3 + 3 \alpha_4 + 2 \alpha_5 +
\alpha_6 + 2 \alpha_{10}$. In this equivalent embedding, all
raising operators of $D_8$ are also raising operators of $E_8$.
What is more, the highest root of $D_8$, $\theta_{D_8} =
\alpha_{10} + 2 \alpha_3 + 2\alpha_4 + 2 \alpha_5 + 2 \alpha_6+ 2
\alpha_7 + \alpha_{2}+ \bar{\beta}$ is equal to the highest root
of $E_8$. Because of this, the affine root $\alpha_8$ of the
untwisted affine extension $E_8^+$ can be identified with the
affine root of $D_8^+$, and the overextended root $\alpha_9$ can
also be taken to be the same. Hence, $DE_{10}$ can be regularly
embedded in $E_{10}$ (see Figure \ref{figure:DE10}).

The embedding just described is in fact relevant to string theory
and has been discussed from various points of view in previous
papers \cite{AKHN,JBrown}. By dimensional reduction of the bosonic
sector of eleven-dimensional supergravity on a circle, one gets,
after dropping the Kaluza-Klein vector and the 3-form, the bosonic
sector of pure N=1 ten-dimensional supergravity. The simple roots of
$DE_{10}$ are the symmetry walls and the electric and magnetic walls
of the 2-form and coincide with the positive roots given above
\cite{Damour:2000hv}.

A similar construction shows that $A_8^{++}$ can be regularly
embedded in $E_{10}$, and that $DE_{10}$ can be regularly embedded
in $BE_{10}\equiv B_8^{++}$.

\subsection{Further Properties}
As we have just seen, the raising operators of $\mgb$ might be
raising or lowering operators of $\mg$. We shall consider here only
the case when the positive (respectively, negative) step operators
of $\mgb$ are also positive (respectively, negative) step operators
of $\mg$, so that $\bar{\mf{n}}_- = \mf{n}_- \cap \mgb$ and $\bar{\mf{n}}_+ = \mf{n}_+ \cap \mgb$
(``positive regular embeddings").  This will always be assumed
from now on.

In the finite dimensional case, there is a useful criterion to
determine regular algebras from subsets of roots.  This criterion
has been generalized to Kac-Moody algebras in
\cite{Feingold:2003es}.  It goes as follows.

\vspace{.3cm} \noindent {\bf Theorem:} Let $\Phi^+_{real}$ be the
set of positive real roots of a Kac-Moody algebra $\mathcal{A}$.
Let $\beta_1, \cdots, \beta_n \in \Phi^+_{real}$ be chosen such
that none of the differences $\beta_i - \beta_j$ is a root of
$\mathcal{A}$.  Assume furthermore that the $\beta_i$'s are such
that the matrix $C =[C_{ij}] = [2 \left< \beta_i \vert \beta_j
\right> /\left<\beta_i \vert \beta_i \right>]$ has non-vanishing
determinant. For each $1 \leq i \leq n$, choose non-zero root
vectors $E_i$ and $F_i$ in the one-dimensional root spaces
corresponding to the positive real roots $\beta_i$ and the
negative real roots $-\beta_i$, respectively, and let $H_i = [E_i,
F_i]$ be the corresponding element in the Cartan subalgebra of
$\mathcal{A}$. Then, the (regular) subalgebra of $\mathcal{A}$
generated by $\{E_i, F_i, H_i\}$, $i= 1, \cdots, n$, is a
Kac-Moody algebra with Cartan matrix $[C_{ij}]$.

\vspace{.3cm}

\noindent {\bf Proof:} The proof of this theorem is given in
\cite{Feingold:2003es}.  Note that the Cartan integers $2 \f{\left<
\beta_i \vert \beta_j \right>}{ \left<\beta_i \vert \beta_i
\right>}$ are indeed integers (because the $\beta_i$'s are positive
real roots), which are non positive (because $\beta_i - \beta_j$
is not a root), so that $[C_{ij}]$ is a Cartan matrix.

\vspace{.3cm}

\noindent {\bf Comments:} \begin{enumerate}  \item When the Cartan
matrix is degenerate, the corresponding Kac-Moody algebra has non
trivial ideals \cite{Kac}. Verifying that the Chevalley-Serre
relations are fulfilled is not sufficient to guarantee that one
gets the Kac-Moody algebra corresponding to the Cartan matrix
$[C_{ij}]$ since there might be non trivial quotients. We will in
fact precisely encounter below situations in which the algebra
generated by the set $\{E_i, F_i, H_i\}$ is the quotient of the
Kac-Moody algebra with Cartan matrix $[C_{ij}]$ by a non trivial
ideal. \item If the matrix $[C_{ij}]$ is decomposable, say $C=
D\oplus E$ with $D$ and $E$ indecomposable, then the Kac-Moody
algebra $\mathbb{KM}(C)$ generated by $C$ is the direct sum of the
Kac-Moody algebra $\mathbb{KM}(D)$ generated by $D$ and the
Kac-Moody algebra $\mathbb{KM}(E)$ generated by $E$. The
subalgebras $\mathbb{KM}(D)$ and $\mathbb{KM}(E)$ are ideals.  If
$C$ has non-vanishing determinant, then both $D$ and $E$ have
non-vanishing determinant.  Accordingly, $\mathbb{KM}(D)$ and
$\mathbb{KM}(E)$ are simple \cite{Kac} and hence, either occur
faithfully or trivially. Because the generators $E_i$ are linearly
independent, both $\mathbb{KM}(D)$ and $\mathbb{KM}(E)$ occur
faithfully. Therefore, in the above theorem the only case that requires special treatment is when the Cartan matrix $C$ has vanishing
determinant.
\end{enumerate}

\vspace{.3cm}

\noindent It is convenient to universally normalize the
Killing form of Kac-Moody algebras in such a way that the long
real roots have always the same squared length, conveniently taken
equal to two. It is then easily seen that the Killing form of any
regular Kac-Moody subalgebra of $E_{10}$ coincides with the
invariant form induced from the Killing form of $E_{10}$ through
the embedding (``Dynkin index equal to one") since $E_{10}$ is
``simply laced". This property does not hold for non regular
embeddings as the example given in subsection \ref{definitions},
which has Dynkin index $-1$, shows.

\subsection{Reductive Subalgebras}

We shall also consider embeddings in a Kac-Moody algebra
$\mathcal{A}$ of algebras $\mathcal{B} = \mathcal{D} \oplus
\mathbb{R}^k$ which are the direct sums of a Kac-Moody algebra
$\mathcal{D}$ plus an Abelian algebra $\mathbb{R}^k$. One says
that the embedding is regular if $\mathcal{D}$ is regularly
embedded in the above sense and if $\mathbb{R}^k$ is a subalgebra
of the Cartan subalgebra $H_{\mathcal{A}}$ of $\mathcal{A}$.  The
abelian algebra $H_{\mathcal{D}}\oplus \mathbb{R}^k$ is called the
Cartan subalgebra of $\mathcal{B}$.  We take for invariant
bilinear form on $\mathbb{R}^k$ the invariant form induced from
the Killing form of $\mathcal{A}$ through the embedding.

\subsection{Back to Subgroup Truncations}

We now come back to consistent subgroup truncations of the
non-linear sigma model $\mathcal{G}/\mathcal{K}$.

Let $\bar{\mathcal{G}}$ be the subgroup of $\mathcal{G}$ obtained
by exponentiating a regular subalgebra $\bar{\mf{g}}$ of $\mf{g}$.
Assume that the initial conditions $g(0)$ and $\dot{g}(0)$ are
such that (i) the group element $g(0)$ is in $\bar{\mathcal{G}}$;
and (ii) the conserved current $J$ is in $\bar{\mf{g}}$.  Then,
$g(0) \exp{(x^0 J)}$ belongs to $\bar{\mathcal{G}}$ for all $x^0$.
Furthermore, there exists an element $k(x^0) \in
\bar{\mathcal{K}}$ (the maximal compact subgroup of
$\bar{\mathcal{G}}$) such that $k(x^0)g(0) \exp{(x^0 J)}$ fulfills
the Borel gauge from the point of view of $\bar{\mathcal{G}}$.
Because the embedding is regular, $k(x^0)g(0) \exp{(x^0 J)}$
fulfills also the Borel gauge from the point of view of
$\mathcal{G}$; $k(x^0)$ belongs in fact also to $\mathcal{K}$. But
$k(x^0)g(0) \exp{(x^0 J)}$ is precisely the solution of the
equations of motion (see Eq. (\ref{solutiong})). This shows that
one can consistently truncate the dynamics of the non-linear sigma
model $\mathcal{G}/\mathcal{K}$ to
$\bar{\mathcal{G}}/\bar{\mathcal{K}}$ since initial conditions in
$\bar{\mathcal{G}}$ remain in $\bar{\mathcal{G}}$.

Finally, because the Killing forms coincide, the constraints
resulting from time reparam-etrization invariance also agree.

The consistent truncation to a regular subgroup was used in
\cite{Englert} to investigate the compatibility of the non linear
$\sigma$-model $\mathcal{E}_{10}/\mathcal{K}(\mathcal{E}_{10})$
with the non linear $\sigma$-model
$\mathcal{E}_{11}/\mathcal{K}(\mathcal{E}_{11})$. Another
interesting consistent truncation is the truncation to the regular
subalgebra $E_8 \oplus A_1 \oplus \mathbb{R}l$.  Here, $E_8 \oplus
A_1$ is obtained by deleting the node numbered 8 in Figure
\ref{figure:E10}, and $\mathbb{R}l$ is the one-dimensional
subalgebra of the Cartan subalgebra $\mf{h}_{E_{10}}$ of $E_{10}$
containing the multiples of $l$, where $l$ is orthogonal to both
the Cartan subalgebra $\mf{h}_{E_8}$ of $E_8$ and the Cartan
subalgebra $\mf{h}_{A_1}$ of $A_1$. The subalgebra $\mathbb{R}l$
is timelike. Explicitly, $l = {K^9}_9 + {K^{10}}_{10}$. The
restriction to $E_8 \oplus A_1 \oplus \mathbb{R}l$ is not only
consistent with the sigma-model equations, but also with the
supergravity equations of motion because there is no root in $E_8
\oplus A_1$ of height $> 29$ and hence the argument of
\cite{Damour:2002cu} applies.  One way to realize the truncation
is to dimensionally reduce eleven-dimensional supergravity on the
3-torus, dualize all fields (except the three-dimensional metric)
to scalars, and then impose that the fields (metric and scalar
fields) depend only on time. Although non chaotic, this truncation
is interesting because it involves some level 3 fields,
corresponding to ``curvature walls" (the Kaluza-Klein vector
components of the eleven-dimensional metric do depend on space if
their duals depend only on time, so that there is some spatial
curvature).  One can further truncate to the subalgebra $E_8
\oplus \mathbb{R} l$ by assuming that the metric is diagonal with
equal diagonal components.  This case was thoroughly investigated
in \cite{Fre1,Fre2}.

\section{Geometric Configurations and Regular Subalgebras of
$E_{10}$}\label{geometric} \setcounter{equation}{0}

We will now apply the machinery from the previous sections to
reveal a ``duality'' between the geometric configurations and a
class of regular subalgebras of $E_{10}$.

\subsection{General Considerations}
In order to match diagonal Bianchi I cosmologies with the
$\sigma$-model, one must truncate the
$\mathcal{E}_{10}/\mathcal{K}(\mathcal{E}_{10})$ Lagrangian in
such a way that the metric $\g_{ab}$ is diagonal. This will be the
case if the subalgebra $S$ to which one truncates has no generator
$K^i_{\; \;j}$ with $i \not=j$. Indeed, the off-diagonal
components of the metric are precisely the exponentials of the associated
$\sigma$-model fields.  The set of simple roots of $S$ should
therefore not contain any level zero root.

Consider ``electric" regular subalgebras of $E_{10}$, for which
the simple roots are all at level one, where the 3-form electric
field variables live.  These roots can be parametrized by $3$
indices corresponding to the indices of the electric field, with
$i_1<i_2<i_3$.  We denote them $\alpha_{i_1i_2i_3}$.  For
instance, $\alpha_{123} \equiv \alpha_{10}$. In terms of the
$\beta$-parametrization of \cite{Damour:2000hv,Damour:2002et}, one
has $\alpha_{i_1i_2i_3} = \beta^{i_1} + \beta^{i_2} +
\beta^{i_3}$.

Now, for $S$ to be a regular subalgebra, it must fulfill, as we
have seen, the condition that the difference between any two of
its simple roots is not a root of $E_{10}$: $\alpha_{i_1i_2i_3} -
\alpha_{i'_1i'_2i'_3} \notin \Phi_{E_{10}}$ for any pair
$\alpha_{i_1i_2i_3}$ and $\alpha_{i'_1i'_2i'_3}$ of simple roots
of $S$.  But one sees by inspection of the commutator of
$E^{i_1i_2i_3}$ with $F_{i'_1i'_2i'_3}$ in
\Eqnref{level1relations} that $\alpha_{i_1i_2i_3} -
\alpha_{i'_1i'_2i'_3}$ is a root of $E_{10}$ if and only if the
sets $\{i_1, i_2, i_3\}$ and $\{i'_1, i'_2, i'_3\}$ have exactly
two points in common.  For instance, if $i_1= i'_1$,  $i_2 = i'_2$
and $i_3 \not= i'_3$, the commutator of $E^{i_1i_2i_3}$ with
$F_{i'_1i'_2i'_3}$ produces the off-diagonal generator
$K^{i_3}_{\; \; \; i'_3}$ corresponding to a level zero root of
$E_{10}$.  In order to fulfill the required condition, one must
avoid this case, i.e., one must choose the set of simple roots of
the electric regular subalgebra $S$ in such a way that given a
pair of indices $(i_1, i_2)$, there is at most one $i_3$ such that
the root $\alpha_{i j k}$ is a simple root of $S$, with $(i,j,k)$
the re-ordering of $(i_1, i_2, i_3 )$ such that $i<j<k$.

To each of the simple roots $\alpha_{i_1i_2i_3}$ of $S$, one can
associate the line $(i_1,i_2,i_3)$ connecting the three points
$i_1$, $i_2$ and $i_3$.  If one does this, one sees that the above
condition is equivalent to the following statement: {\em the set of
points and lines associated with the simple roots of $S$ must
fulfill the third Rule defining a geometric configuration,
namely, that two points determine at most one line}. Thus, this
geometric condition has a nice algebraic interpretation in terms
of regular subalgebras of $E_{10}$.

The first rule, which states that each line contains 3 points, is
a consequence of the fact that the $E_{10}$-generators at level
one are the components of a 3-index antisymmetric tensor.  The
second rule, that each point is on $m$ lines, is less fundamental
from the algebraic point of view since it is not required to hold
for $S$ to be a regular subalgebra. It was imposed in
\cite{Demaret:1985js} in order to allow for solutions isotropic in
the directions that support the electric field. We keep it here as
it yields interesting structure (see next subsection).  We briefly
discuss in the conclusions what happens when this condition is
lifted.

\subsection{Incidence Diagrams and Dynkin Diagrams}
We have just shown that each geometric configuration $(n_m,g_3)$
with $n \leq 10$ defines a regular subalgebra $S$ of $E_{10}$.  In
order to determine what this subalgebra $S$ is, one needs,
according to the theorem recalled in Section \ref{Regular}, to
compute the Cartan matrix \beq C = [C_{i_1 i_2 i_3, i'_1 i'_2
i'_3}] = \left[\left<\alpha_{i_1 i_2 i_3} \vert \alpha_{i'_1 i'_2
i'_3}\right>\right] \eeq (the real roots of $E_{10}$ have squared
length equal to 2).  According to that same theorem, the algebra
$S$ is then just the rank $g$ Kac-Moody algebra with Cartan matrix
$C$, unless $C$ has zero determinant, in which case $S$ might be
the quotient of that algebra by a non trivial ideal.

Using for instance the root parametrization of
\cite{Damour:2000hv,Damour:2002et} and the expression of the
scalar product in terms of this parametrization, one
easily verifies that the scalar product \phantom{hej} $\left<\alpha_{i_1 i_2
i_3} \vert \alpha_{i'_1 i'_2 i'_3}\right>$ is equal to: \beqa
\left<\alpha_{i_1 i_2 i_3} \vert \alpha_{i'_1 i'_2 i'_3}\right>
&=& 2 \hspace{2cm} \hbox{if all three indices coincide}, \\
&=& 1 \hspace{2cm} \hbox{if two and only two indices coincide}, \\
&=& 0 \hspace{2cm} \hbox{if one and only one index coincides}, \\
&=& -1 \hspace{1.7cm} \hbox{if no indices coincide}. \eqa  The second
possibility does not arise in our case since we deal with
geometric configurations.  For completeness, we also list the
scalar products of the electric roots $\alpha_{ijk}$ ($i<j<k$) with
the symmetry roots $\alpha_{\ell m}$ ($\ell < m$) associated with
the raising operators $K^m _{\; \; \ell}$: \beqa
\left<\alpha_{ijk} \vert \alpha_{\ell m}\right> &=& -1
\hspace{1.4cm} \hbox{if $\ell\in \{i,j,k\}$ and $m \notin
\{i,j,k\}$} \\ &=& 0 \hspace{1.7cm} \hbox{if $\{\ell, m\} \subset
\{i,j,k\}$ or $\{\ell, m\} \cap \{i,j, k\} = \emptyset$}, \\ &=& 1
\hspace{1.7cm} \hbox{if $\ell\notin \{i,j,k\}$ and $m \in
\{i,j,k\}$}\eqa as well as the scalar products of the symmetry
roots among themselves, \beqa \left<\alpha_{ij} \vert \alpha_{\ell
m}\right> &=& - 1 \hspace{1.7cm} \hbox{if $j = \ell$ or $i = m$},
\\ &=&  0
\hspace{2cm} \hbox{if $\{\ell, m\} \cap \{i,j\} = \emptyset$},
\\ &=& 1 \hspace{2cm} \hbox{if $i = \ell$ or $j \not= m$}, \\&=&  2
\hspace{2cm} \hbox{if $\{\ell, m\} = \{i,j\} $}.\eqa

\noindent Given a geometric configuration $(n_m,g_3)$, one can
associate with it a ``line incidence diagram" that encodes the
incidence relations between its lines.  To each line of
$(n_m,g_3)$ corresponds a node in the incidence diagram. Two nodes
are connected by a single bond if and only if they correspond to
lines with no common point (``parallel lines"). Otherwise, they
are not connected\footnote{One may also consider a point incidence
diagram defined as follows: the nodes of the point incidence
diagram are the points of the geometric configuration.  Two nodes
are joined by a single bond if and only if there is no straight
line connecting the corresponding points. The point incidence
diagrams of the configurations $(9_3,9_3)$ are given in
\cite{Hilbert}.  For these configurations, projective duality
between lines and points lead to identical line and point
incidence diagrams.  Unless otherwise stated, the expression
``incidence diagram" will mean ``line incidence diagram".}. By
inspection of the above scalar products, we come to the important
conclusion that {\em the Dynkin diagram of the regular, rank $g$,
Kac-Moody subalgebra $S$ associated with the geometric
configuration $(n_m,g_3)$ is just its line incidence diagram.}  We
shall call the Kac-Moody algebra $S$ the algebra ``dual" to the
geometric configuration $(n_m,g_3)$.

Because the geometric configurations have the property that the
number of lines through any point is equal to a constant $m$, the
number of lines parallel to any given line is equal to a number
$k$ that depends only on the configuration and not on the line.
This is in fact true in general and not only for $n\leq 10$ as can
be seen from the following argument. For a configuration with $n$
points, $g$ lines and $m$ lines through each point, any given line
$\Delta$ admits $3(m-1)$ true secants, namely, $(m-1)$ through
each of its points\footnote{A true secant is here defined as a
line, say $\Delta^{\prime}$, distinct from $\Delta$ and with a
non-empty intersection with $\Delta$.}. By definition, these
secants are all distinct since none of the lines that $\Delta$
intersects at one of its points, sayÊ $P$, can coincide with a
line that it intersects at another of its points, say
$P^{\prime}$, since the only line joining $P$ to $P^{\prime}$ is
$\Delta$ itself. It follows that the total number of lines that
$\Delta$ intersects is the number of true secants plus $\Delta$
itself, i.e. $3(m-1)+1$. As a consequence, each line in the
configuration admits $k=g-[3(m-1)+1]$ parallel lines, which is
then reflected in the fact that each node in the associated Dynkin
diagram has the same number, $k$, of adjacent nodes.

\subsection{Geometric Configuration $(3_{1},1_{3})$}

To illustrate the discussion, we begin by constructing the algebra
associated to the simplest configuration $(3_{1},1_{3})$.
\begin{figure}[ht]
\begin{center}
\includegraphics[width=50mm]{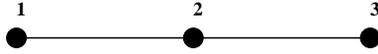}
\caption{$(3_{1},1_{3})$: The only allowed configuration for $n=3$.}
\label{figure:G31}
\end{center}
\end{figure}
\noindent This example also exhibits some subtleties associated with the
Hamiltonian constraint and the ensuing need to extend $S$ when the
algebra dual to the geometric configuration is
finite-dimensional.

In light of our discussion, considering the geometric
configuration $(3_{1},1_{3})$ is equivalent to turning on only the
component $A_{123}(x^0)$ of the $3$-form that multiplies the
generator $E^{123}$ in the group element $g$ and the diagonal
metric components corresponding to the Cartan generator $h =
[E^{123}, F_{123}]$. The algebra has thus basis $\{e,f,h\}$ with
\beq e\equiv E^{123} \qquad f\equiv F_{123} \qquad h= [e,f] =
-\f{1}{3}\sum_{a\neq 1,2,3}
{K^{a}}_{a}+\f{2}{3}({K^{1}}_{1}+{K^{2}}_{2}+{K^{3}}_{3}).
\eqnlab{A1generators} \eeq The Cartan matrix is just $(2)$ and is
not degenerate.  It defines an $A_1$ regular subalgebra. The
Chevalley-Serre relations, which are guaranteed to hold according to the
general argument, are easily verified. The configuration
$(3_{1},1_{3})$ is thus dual to $A_{1}$.

This $A_1$ algebra is simply the $\mf{sl}(2)$-algebra associated with
the simple root $\alpha_{10}$.  Because the Killing form on $A_1$
is positive definite, one cannot find a solution of the
Hamiltonian constraint if one turns on only $A_1$.  One needs to
enlarge $A_1$ (at least) by a one-dimensional subalgebra
$\mathbb{R} l$ of $\mh_{E_{10}}$ that is timelike.  One can take for
$l$ the Cartan element ${K^4}_4 + {K^5}_5 + {K^6}_6 + {K^7}_7 +
{K^8}_8 + {K^9}_9 + {K^{10}}_{10}$, which ensures isotropy in the
directions not supporting the electric field. Thus, the
appropriate regular subalgebra of $E_{10}$ in this case is $A_1
\oplus \mathbb{R}l$. This construction reproduces the ``SM2-brane"
solution given in section 3 of \cite{Demaret:1985js}, describing
two asymptotic Kasner regimes separated by a collision against an
electric wall.

The need to enlarge the algebra $A_1$ was discussed in the paper
\cite{Kleinschmidt:2005gz} where a group theoretical
interpretation of some cosmological solutions of eleven
dimensional supergravity was given.  In that paper, it was also
observed that $\mathbb{R}l$ can be viewed as the Cartan subalgebra
of the (non regularly embedded) subalgebra $A_1$ associated with
an imaginary root at level $21$, but since the corresponding field
is not excited, the relevant subalgebra is really $\mathbb{R}l$.

\subsection{Geometric Configuration $(6_{1},2_{3})$}

\begin{figure}[ht]
\begin{center}
\includegraphics[width=50mm]{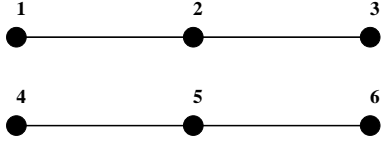}
\caption{$(6_{1},2_{3})$: The simplest configuration allowed for $n=6$.}
\label{figure:G61}
\end{center}
\end{figure}

For $n=6$ we start with the double-line configuration,
$(6_{1},2_{3})$, in Figure \ref{figure:G61}.

\begin{figure}[ht]
\begin{center}
\includegraphics[width=30mm]{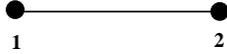}
\caption{The Dynkin diagram of $A_{2}$, dual to the configuration $(6_{1},2_{3})$.}
\label{figure:A2}
\end{center}
\end{figure}

\noindent This graph yields the generators
\beqa
& & e_{1}\equiv E^{123}, \;
\; f_{1}\equiv F_{123}, \;   \; h_{1}\equiv -\f{1}{3}\sum_{a\neq
1,2,3} {K^{a}}_{a}+\f{2}{3}({K^{1}}_{1}+{K^{2}}_{2}+{K^{3}}_{3})
\nn \\
& & e_{2}\equiv E^{456} , \;   \; f_{2}\equiv F_{456} ,  \;  \;
h_{2}\equiv -\f{1}{3}\sum_{a\neq 4,5,6}
{K^{a}}_{a}+\f{2}{3}({K^{4}}_{4}+{K^{5}}_{5}+{K^{6}}_{6})
\eqnlab{A2generators} \eqa with the following commutators, \beq
[e_{i},f_{i}]=h_{i} \qquad [h_{i},e_{i}]=2e_{i} \qquad
[h_{i},f_{i}]=-2f_{i}\qquad (i=1,2). \eqnlab{A2commutators} \eeq
Using the rules outlined above, the Cartan matrix is easily found
to be\beq A_{(6_{1},2_{3})}=\left( \begin{array}{cc}
2 & -1\\
-1 & 2\\
\end{array} \right),
\eqnlab{A2CartanMatrix} \eeq which is the Cartan matrix of $A_2$,
$A_{(6_{1},2_{3})} = A_2$.  Thus,  the configuration
$(6_{1},2_{3})$ in Figure \ref{figure:G61} is dual to  $A_{2}$,
whose Dynkin diagram is shown in Figure \ref{figure:A2}.

\noindent Note that the roots $\alpha_{123}$ and $\alpha_{456}$
are $\alpha_{10}$ and
$\al_{10}+\al_{1}+2\al_{2}+3\al_{3}+2\al_{4}+\al_{5}$ so that this
subalgebra is in fact already a (non maximal) regular subalgebra
of $E_6$. The generator corresponding to the highest root,
$\theta$, of $A_{2}$ arises naturally as the level $2$ generator
$E^{123456}$, i.e. \beq e_{\theta}\equiv
E^{123456}=[E^{123},E^{456}]. \eqnlab{A2highestroot} \eeq Although
they are guaranteed to hold from the general argument given above,
it is instructive and easy to verify explicitly the Serre
relations. These read, \beq
[e_{1},[e_{1},e_{2}]]=[e_{2},[e_{2},e_{1}]]=0 \qquad
[f_{1},[f_{1},f_{2}]]=[f_{2},[f_{2},f_{1}]]=0,
\eqnlab{A2SerreRelations} \eeq and are satisfied since the level
$3$-generators are killed because of antisymmetry, e.g. \beq
[E^{123},[E^{123},E^{456}]=[E^{123},E^{123456}]=
E^{1|23123456}+E^{2|31123456}+E^{3|12123456}=0,
\eqnlab{A2SerreRelations2} \eeq where each generator in the last
step vanished individually.

Since the Killing form on the Cartan subalgebra of $A_2$ has
Euclidean signature, one must extend $A_2$ by an appropriate
one-dimensional timelike  subalgebra $\mathbb{R}l$ of
$\mh_{E_{10}}$. We take $l={K^7}_7 + {K^8}_8 + {K^9}_9 +
{K^{10}}_{10}$.  In $A_2 \oplus \mathbb{R}l$, the Hamiltonian
constraint can be fulfilled. Furthermore, since $A_2 \oplus
\mathbb{R}l$ has generators only up to level two, the
$\sigma$-model equations of motion are equivalent to the dynamical
supergravity equations without need to implement an additional
level truncation.

The fact that there is a level 2 generator implies the generic
presence of a non-zero magnetic field.  The momentum constraint
and Gauss' law are automatically fulfilled because the only
non-vanishing components of the $4$-form $F_{\alpha \beta \gamma
\delta}$ are $F_{0 123}$, $F_{0 456}$ and $F_{7 8 9 (10)}$.

\subsection{Geometric Configuration $(6_{2},4_{3})$}

\begin{figure}[ht]
\begin{center}
\includegraphics[width=50mm]{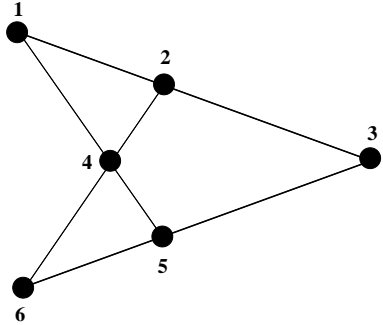}
\caption{$(6_{2},4_{3})$: The first configuration with intersecting lines.}
\label{figure:G62}
\end{center}
\end{figure}

We now treat the configuration $(6_{2},4_{3})$, shown in Figure
\ref{figure:G62}. Although the graph is more complicated, the corresponding algebra
is actually a lot simpler.

\noindent The generators associated to the simple
roots are \beq e_{1}=E^{123} \qquad e_{2}=E^{145} \qquad
e_{3}=E^{246} \qquad e_{4}=E^{356}. \eqnlab{G62generators} \eeq
The first thing to note is that in contrast to the previous case,
all generators now have one index in common since in the graph any
two lines share one node. This implies that the 4 lines in
$(6_{2},4_{3})$ define  $4$ \emph{commuting} $A_1$ subalgebras,
\beq (6_{2},4_{3}) \qquad \Longleftrightarrow \qquad
\mf{g}_{(6_{2},4_{3})}=A_{1}\oplus A_{1} \oplus A_{1}\oplus A_{1}.
\eqnlab{A1^4} \eeq

\noindent Again, although this is not necessary, one can make sure that the
Chevalley-Serre relations are indeed fulfilled. For instance, the
Cartan element $h = [E^{b_1 b_2 b_3}, F_{b_1 b_2 b_3}]$ (no
summation on the fixed, distinct indices $b_{1},b_{2},b_{3}$)
reads \beq h=-\f{1}{3}\sum_{a\neq
b_{1},b_{2},b_{3}}{K^{a}}_{a}+\f{2}{3}({K^{b_{1}}}_{b_{1}}+
{K^{b_{2}}}_{b_{2}}+{K^{b_{3}}}_{b_{3}}).
\eqnlab{GeneralCartanElement} \eeq Hence, the commutator
$[h,E^{b_{i} cd}]$ vanishes whenever $E^{b_{i} cd}$ has only one
$b$-index, \beqa {} [h,E^{b_{i}
cd}]&=&-\f{1}{3}[({K^{c}}_{c}+{K^{d}}_{d}),E^{b_{i}cd}]+\f{2}{3}
[({K^{b_{1}}}_{b_{1}}+{K^{b_{2}}}_{b_{2}}+{K^{b_{3}}}_{b_{3}}),E^{b_{i}cd}]
\nn \\
{}æ&=& (-\f{1}{3}-\f{1}{3}+\f{2}{3})E^{b_{i}cd}=0 \qquad
(i=1,2,3). \eqnlab{vanishingcommutator} \eqa Furthermore, multiple
commutators of the step operators are immediately killed at level
$2$ whenever they have one index or more in common, e.g. \beq
[E^{123},E^{145}]=E^{123145}=0. \eqnlab{vanishingcommutator2} \eeq

\noindent To fulfill the Hamiltonian constraint, one must extend the algebra
by taking a direct sum with $\mathbb{R}l$, $l={K^7}_7 + {K^8}_8 +
{K^9}_9 + {K^{10}}_{10}$. Accordingly, the final algebra is
$A_{1}\oplus A_{1} \oplus A_{1}\oplus A_{1} \oplus \mathbb{R}l$.
Because there is no magnetic field, the momentum constraint and
Gauss' law are identically satisfied.\\
\indent The gravitational solution associated to this configuration generalizes the one found in \cite{Demaret:1985js}. In fact, using the terminology of \cite{Strominger}, the solution describes a set of four intersecting $SM2$-branes, with a five-dimensional transverse spacetime in the directions $t, x^{7},x^{8},x^{9},x^{10}$. We postpone a more detailed discussion of this solution to Section $8$.

\subsection{Geometric Configuration $(7_{3},7_{3})$}

\begin{figure}[ht]
\begin{center}
\includegraphics[width=70mm]{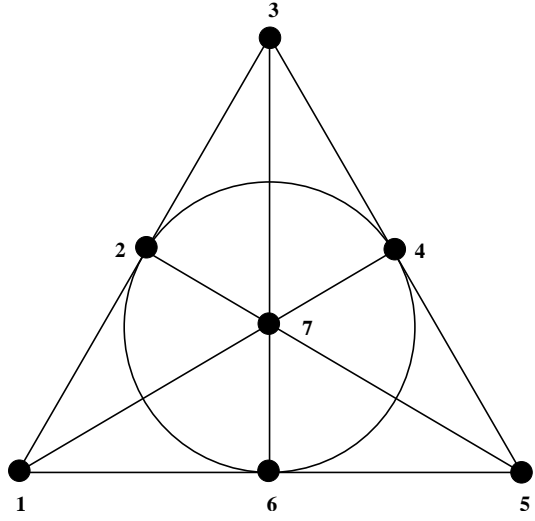}
\caption{$(7_{3},7_{3})$: The
\emph{Fano plane}, dual to the Lie algebra $A_{1}\oplus
A_{1}\oplus A_{1} \oplus A_{1}\oplus A_{1}\oplus A_{1}\oplus
A_{1}$.}
\label{figure:G7}
\end{center}
\end{figure}

We now turn to the only existing configuration for $n=7$, which
has seven lines and it accordingly denoted $(7_{3},7_{3})$. The
graph is shown in Figure \ref{figure:G7}. Readers familiar with
the octonions will recognize this as the so-called \emph{Fano
plane}, encoding the complete multiplication table of the
octonions (see e.g. \cite{Baez,Schafer} for an introduction).

\noindent We see from Figure \ref{figure:G7} that any two lines
have exactly one node in common and hence the corresponding
algebras will necessarily be commuting. Since the graph has $7$
lines we conclude that the Fano plane is dual to the direct sum of
seven $A_{1}$ algebras, \beq (7_{3},7_{3}) \qquad
\Longleftrightarrow \qquad \mf{g}_{\mathrm{Fano}}=A_{1}\oplus
A_{1}\oplus A_{1} \oplus A_{1}\oplus A_{1}\oplus A_{1}\oplus
A_{1}. \eqnlab{FanoAlgebra} \eeq Because there are seven points,
the algebra is in fact embedded in $E_7$.  Note that the $A_1$'s
are NOT the $\mf{sl}(2)$ subalgebras associated with the simple
roots of $E_7$ since the $A_1$'s in $\mf{g}_{\mathrm{Fano}}$ are
commuting.

Although of rank 7, the regular embedding of
$\mf{g}_{\mathrm{Fano}}$ in $E_7$ is not maximal, but is part of
the following chain of maximal regular embeddings: \beq
A_{1}\oplus A_{1}\oplus A_{1} \oplus A_{1}\oplus A_{1}\oplus
A_{1}\oplus A_{1} \subset  A_{1}\oplus A_{1}\oplus A_{1} \oplus
D_4  \subset  A_{1}\oplus D_6 \subset E_7 \label{chain1}\eeq as
can be verified by using the Dynkin argument based on the highest
root \cite{Dynkin}. The intermediate algebras occurring in
(\ref{chain1}) have as raising operators (with the choice of
$A_{1}\oplus A_{1}\oplus A_{1} \oplus A_{1}\oplus A_{1}\oplus
A_{1}\oplus A_{1}$ captured by Figure \ref{figure:G7}) $E^{123}$,
$E^{174}$, $E^{165}$, $K^{3}_{\; \;2}$, $K^{6}_{\; \;5}$,
$K^{7}_{\; \;4}$, $E^{245}$ (for $A_{1}\oplus A_{1}\oplus A_{1}
\oplus D_4$) and $E^{174}$, $K^{4}_{\; \;1}$, $K^{7}_{\; \;4}$,
$K^{5}_{\; \;3}$, $K^{3}_{\; \;2}$, $K^{6}_{\; \;5}$, $E^{123}$
(for $A_{1}\oplus D_6$).

The algebra being finite-dimensional, one needs to supplement it
by a one-dimensional timelike subalgebra $\mathbb{R} l$ of
$\mh_{E_{10}}$ in order to fulfill the Hamiltonian constraint.  One
can take $l={K^8}_8 + {K^9}_9 + {K^{10}}_{10}$, which is
orthogonal to it. Finally, because there is no level two element,
there is again no magnetic field and hence no momentum or Gauss
constraint to be concerned about. The solutions of the $\sigma$-model for
$A_{1}\oplus A_{1}\oplus A_{1} \oplus A_{1}\oplus A_{1}\oplus
A_{1}\oplus A_{1} \oplus \mathbb{R}l$ fulfilling the Hamiltonian
constraint all define solutions of eleven-dimensional
supergravity. Since the $A_1$-algebras commute the corresponding
solutions describe a set of seven intersecting $SM2$-branes.

\subsection{Geometric Configuration $(8_{3},8_{3})$}

\begin{figure}[ht]
\begin{center}
\includegraphics[width=100mm]{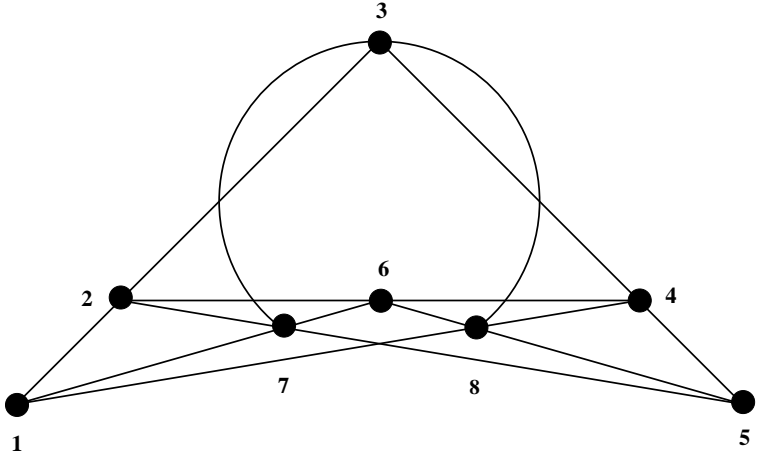}
\caption{The geometric configuration
$(8_{3},8_{3})$, dual to the Lie algebra $A_{2}\oplus A_{2}\oplus
A_{2} \oplus A_{2}$.}
\label{figure:G8}
\end{center}
\end{figure}

The last finite-dimensional case is provided by the geometric
configuration $(8_{3},8_{3})$.  Since there are eight points and
eight lines, the dual algebra is a rank-eight algebra regularly
embedded in $E_8$. Applying the rules derived above to the
geometric configuration $(8_{3},8_{3})$, depicted in Figure
\ref{figure:G8}, one easily finds \beq \mf{g}_{(8_{3},8_{3})}=
A_2 \oplus A_2 \oplus A_2 \oplus A_2. \eeq

\noindent Although this is a rank-eight subalgebra of $E_8$, it is not a
maximal regular subalgebra, but part of the chain of regular
embeddings \beq A_2 \oplus A_2 \oplus A_2 \oplus A_2 \subset A_2
\oplus E_6 \subset E_8 . \eeq  With the numbering of the lines of
Figure \ref{figure:G8}, the intermediate algebra $A_2 \oplus E_6$
may be taken to have as raising operators $E^{123}, E^{568}$ (for
$A_2$) and $K^{2}_{\; \;1}$, $K^{3}_{\; \;2}$, $K^{6}_{\; \;5}$,
$K^{8}_{\; \;6}$, $K^{7}_{\; \;4}$, $E^{145}$ for $E_6$.

In order to fulfill the Hamiltonian constraint, we add
$\mathbb{R}l$ with $l= {K^9}_9 + {K^{10}}_{10}$.  The final
algebra is thus $A_{2}\oplus A_{2}\oplus A_{2} \oplus A_{2} \oplus
\mathbb{R} l$.

There is no level-3 field in $A_2 \oplus A_2 \oplus A_2 \oplus
A_2$, so level-3 truncation is automatic in this model.  However,
because of the level-2 magnetic field generically present, the
momentum and Gauss constraints need to be analyzed.  The only
non-vanishing components of the magnetic field arising in the
model are $F_{479(10)}$, $F_{289(10)}$, $F_{369(10)}$ and $F_{159(10)}$.
Because these have always at least two indices ($9$ and $10$)
distinct from the indices on the electric field components, the
momentum constraint is satisfied.  Furthermore, because they share
the pair $(9,10)$, Gauss' law is also fulfilled. Accordingly, the
solutions of the $\sigma$-model for $A_2 \oplus A_2 \oplus A_2 \oplus
A_2 \oplus \mathbb{R}l$ fulfilling the Hamiltonian constraint all
define solutions of eleven-dimensional supergravity.

\begin{center}
\begin{table}
  \begin{tabular}{ |m{20mm}|m{45mm}|m{30mm}|m{35mm}|}
\hline
\multicolumn{2} {|c|} {Configuration}  & Dynkin diagram & Lie algebra \\
    \hline
     \hline
    $(3_1,1_3)$ & \includegraphics[width=30mm]{G31} &
      \includegraphics[width=1.5mm]{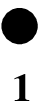} & $\mg_{(3_1,1_3)}=A_{1}$  \\
    \hline
      $(6_1,2_3)$ &  \includegraphics[width=30mm]{G61} &
      \includegraphics[width=15mm]{A2} & $\mg_{(6_1,2_3)}=A_{2}$ \\
      \hline
       $(6_2,4_3)$ &  \includegraphics[width=30mm]{G62} &
      \includegraphics[width=20mm]{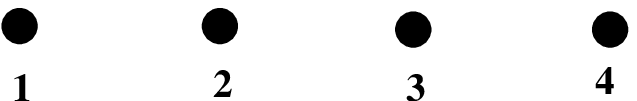} & $\mg_{(6_2,4_3)}=$\phantom{hhhhhhhh} $A_{1}\oplus A_1\oplus A_1 \oplus A_1$ \\
      \hline
      $(7_3,7_3)$ &  \includegraphics[width=40mm]{G7} &
      \includegraphics[width=30mm]{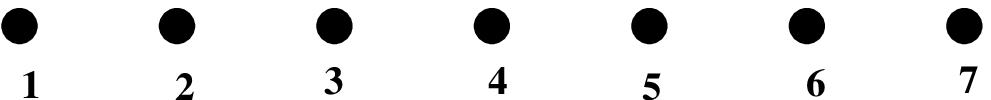} & $\mg_{(7_3,7_3)}=$\phantom{hhhhhhhh} $A_{1}\oplus A_1\oplus A_1 \oplus A_1\oplus A_1\oplus A_1 \oplus A_1$\\
      \hline
       $(8_3,8_3)$ &  \includegraphics[width=45mm]{G8} &
      \includegraphics[width=17mm]{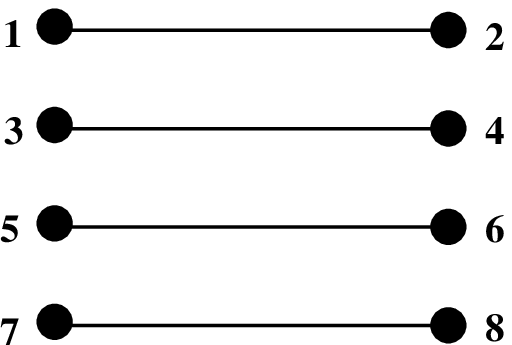} & $\mg_{(8_3,8_3)}=$\phantom{hhhhhhhh} $A_{2}\oplus A_2\oplus A_2\oplus A_2$\\
      \hline
 \end{tabular}
\caption{All configurations for $n\leq 8$ and their dual finite dimensional Lie algebras.}
\end{table}
\end{center}

\section{Geometric Configurations $(9_{m},g_{3})$}
\label{Affine} \setcounter{equation}{0}

All algebras that arise from $n=9$ configurations are naturally
embedded in $E_{9}$.  They turn out to be infinite dimensional
contrary to the cases with $n \leq 8$. Furthermore, because they
involve, as we shall see, affine algebras and degenerate Cartan
matrices, they turn out to be obtained from Kac-Moody algebras
through non trivial quotients. In total, there are $7$ different
configurations with nine nodes, which we consider in turn.

Because the algebras are infinite-dimensional, one must truncate
to level 2 in order to match the Bianchi I supergravity equations
with the $\sigma$-model equations.  Furthermore, if taken to be
non zero, the magnetic field must fulfill the relevant momentum
and Gauss constraints.

\subsection{Geometric Configurations $(9_{1},3_{3})$ and $(9_2,6_3)$}
\subsubsection{Geometric Configuration $(9_{1},3_{3})$}

\begin{figure}[ht]
\begin{center}
\includegraphics[width=60mm]{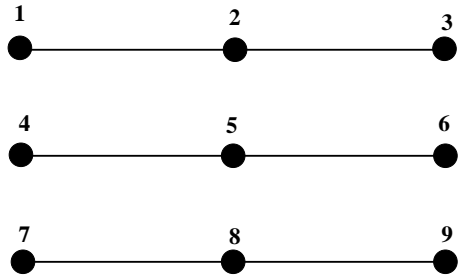}
\caption{The geometric configuration $(9_{1},3_{3})$.}
\label{figure:G91}
\end{center}
\end{figure}

The geometric configuration $(9_{1},3_{3})$ is somewhat trivial
and is given in Figure \ref{figure:G91}.

\begin{figure}[ht]
\begin{center}
\includegraphics[width=30mm]{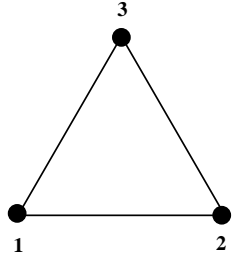}
\caption{The Dynkin diagram of $A_{2}^{+}$.}
\label{figure:A2p}
\end{center}
\end{figure}

\noindent By direct application of the rules, we deduce that the associated
Dynkin diagram consists of three nodes (corresponding to the three
lines $\{1,2,3\}$, $\{4,5,6\}$ and $\{7,8,9\}$) that must all be
connected since the corresponding lines in the configuration are
parallel. This gives the Cartan matrix of $A_{2}^{+}$, i.e. the
untwisted affine extension of $A_{2}$, \beq A(A_{2}^{+})=\left(
\begin{array}{ccc}
2 & -1 & -1\\
-1 & 2 & -1\\
-1 & -1 & 2\\
\end{array} \right)
\eqnlab{A2+Matrix} \eeq whose Dynkin diagram is shown in Figure
\ref{figure:A2p}.
\noindent The Cartan matrix of $A_{2}^{+}$ is degenerate and so, it is not
guaranteed that the algebra generated by the raising operators
\beq e_{1}=E^{123}\qquad e_{2}=E^{456}\qquad  e_{3}=E^{789},
\eqnlab{A2pGenerators} \eeq the lowering operators \beq
f_{1}=F_{123}\qquad f_{2}=F_{456}\qquad  f_{3}=F^{789},
\eqnlab{A2pGeneratorsf} \eeq and the corresponding Cartan elements
given by \Eqnref{GeneralCartanElement}, \beqa &&h_{1} = [e_1,f_1]
= -\f{1}{3}\sum_{a\neq 1,2,3}
{K^{a}}_{a}+\f{2}{3}({K^{1}}_{1}+{K^{2}}_{2}+{K^{3}}_{3}),\\
&&h_{2} = [e_2,f_2] = -\f{1}{3}\sum_{a\neq 4,5,6}
{K^{a}}_{a}+\f{2}{3}({K^{4}}_{4}+{K^{5}}_{5}+{K^{6}}_{6}), \\
&&h_{3} = [e_3,f_3] = -\f{1}{3}\sum_{a\neq 7,8,9}
{K^{a}}_{a}+\f{2}{3}({K^{7}}_{7}+{K^{8}}_{8}+{K^{9}}_{9}),\eqa is
the untwisted Kac-Moody algebra $A_2^+$.

To investigate this issue, we recall some properties of untwisted
affine Kac-Moody algebras.  Consider a finite-dimensional, simple
Lie algebra $\mf{g}$. One can associate with it three related
infinite-dimensional Lie algebras:
\begin{itemize} \item the (untwisted) affine Kac-Moody algebra $\mf{g}^+$,
\item the current algebra $\mf{g}^J$, \item the loop algebra
$\tilde{\mf{g}}$.
\end{itemize}
If $\{T_A\}$ is a basis of $\mf{g}$ with structure constants
${C^A}_{BC}$, the loop algebra $\tilde{\mf{g}}$ has basis
$\{T_A^n\}$ ($n \in \mathbb{Z}$) with commutation relations \beq
[T_B^n, T_C^m] = {C^A}_{BC} \, T_A^{m+n}, \label{loop}\eeq the
current algebra $\mf{g}^J$ has basis $\{T_A, c\}$ with commutation
relations \beq [T_B^n, T_C^m] = {C^A}_{BC} \, T_A^{m+n} + n \,
\delta_{n+m,0} \, k_{BC}\, c,  \; \; \; \; [T_A^n, c] = 0,
\label{current}\eeq while the Kac-Moody algebra $\mf{g}^+$ has
basis $\{T_A, c, d\}$ with commutation relations \beqa && [T_B^n,
T_C^m] = {C^A}_{BC} \, T_A^{m+n} + n \, \delta_{n+m,0} \, k_{BC}\,
c,  \; \; \; \; [T_A^n, c] = 0, \\ && [d ,T^n_A] = n T^n_A, \; \;
\; \; [d,c] = 0.\eqa  Here, $k_{AB}$ is an invariant form (Killing
form) on $\mf{g}$. As vector spaces, $\mf{g}^+ = \mf{g}^J \oplus
\mathbb{R}d = \tilde{\mf{g}} \oplus \mathbb{R}c \oplus
\mathbb{R}d$ and $\mf{g}^J = \tilde{\mf{g}} \oplus \mathbb{R}c$.
The algebra $\mf{g}^+$ is the Kac-Moody algebra associated to the
Cartan matrix $A^+_{ij}$ obtained from the Cartan matrix of
$\mf{g}$ by adding minus the affine root.  Because the Cartan
matrix $A^+_{ij}$ has vanishing determinant, the construction of
$\mf{g}^+$ involves a non trivial ``realization of $A^+_{ij}$"
\cite{Kac}, which is how the scaling operator $d$ enters.

The operator $d$ is in the Cartan subalgebra of $\mf{g}^+$ and has
the following scalar products with all the Cartan generators
\cite{Kac},
\beq \left\{
\begin{array}{cccc}
\left<d | h_{a} \right> & = & 0 & \hbox{(for $h_a$ in the Cartan subalgebra of $\mf{g}$})\\
\left<d | d\right> & = & 0 & \phantom{h} \\
\left<d | c\right>  & = & 1. & \phantom{h} \\
\end{array} \right.
\eqnlab{leveloperator}
\eeq

\noindent Note also that $\left<c | c\right> = 0$. The root
lattice, $\Lambda_{\mf{g}^{+}}$, of $\mf{g}^{+}$ is constructed by
adding to the root lattice of $\mf{g}$ a null vector $\de\in
{\Pi}_{1,1}$, where ${\Pi}_{1,1}$ is the $2$-dimensional self-dual
Lorentzian lattice \cite{Kac}. Thus the root lattice of
$\mf{g}^{+}$ is contained in the direct sum of $\Lambda_{\mf{g}}$
with ${\Pi}_{1,1}$, i.e. \beq \Lambda_{\mf{g}^{+}}\subset
\Lambda_{\mf{g}}\oplus {\Pi}_{1,1}. \eqnlab{affinerootlattice}
\eeq The affine root is given by \beq \al_{0}\equiv \de-\theta,
\eqnlab{affineroot} \eeq where $\theta$ is the highest root of
$\mf{g}$. The scaling generator $d$ counts the number of times the
raising operator corresponding to the affine root $\al_{0}$
appears in any multiple commutator in $\mf{g}^{+}$.

When $\mf{g}$ is simple, as is the algebra $A_2$ relevant for the
$(9_1,3_3)$ configuration, the current algebra $\mf{g}^J$ is the
derived algebra of $\mf{g}^+$, $\mf{g}^J = [\mf{g}^+, \mf{g}^+]
\equiv (\mf {g}^+)'$.   The current algebra coincides with the
algebra generated by the Chevalley-Serre relations associated with
the given Cartan matrix $A_{ij}^+$, and not with its realization.
Furthermore, the center of $\mf{g}^J$ and $\mf{g}^+$ is
one-dimensional and given by $\mathbb{R} c$.  The loop algebra
$\tilde{\mf{g}}$ is the quotient of the current algebra $\mf{g}^J$
by its one-dimensional center $\mathbb{R} c$.

In fact, according to Theorem 1.7 in \cite{Kac}, the only ideals
of a Kac-Moody algebra $\mf{a}$ with non-decomposable Cartan
matrix either contain its derived algebra $\mf{a}'$ or are
contained in its center.

In the case of the configuration $(9_1,3_3)$, we have all the
generators $\{h_i, e_i,f_i \}$ fulfilling the Chevalley-Serre
relations associated with the Cartan matrix of $A_2^+$, without
enlargement of the Cartan subalgebra to contain the scaling
operator $d$.  Hence, the algebra dual to $(9_1,3_3)$ must either
be the current algebra $A_2^J$ or its quotient by its center - the
loop algebra $\tilde{A_2}$.  This would be the case if the center
were represented trivially.  But the central charge is not trivial
and given by \beq c\equiv h_{1}+h_{2}+h_{3}=-{K^{10}}_{10},
\eqnlab{centralcharge} \eeq which does not vanish.  Hence, the
relevant algebra is the current algebra $A_2^J \equiv (A_2^+)'$.

It is straightforward to verify that $c$ commutes with all the
generators of $A_2^J$.  It is also possible to define, within the
Cartan subalgebra of $E_{10}$, an element $d$ that plays the role
of a scaling operator.  This enlargement of $(A_2^+)'$ leads to the
full Kac-Moody algebra $A_2^+$. It is necessary in order to have a
scalar product on the Cartan subalgebra which is of Lorentzian
signature, as required if one wants to solve the Hamiltonian
constraint within the algebra.

Choosing $E^{123}$ as the ``affine'' generator, there exists a six-parameter family of scaling operators,
\beqa
{}Êd &=& a_1 {K^1}_1 + a_2 {K^2}_2 + a_3 {K^3}_3 + b_1 {K^4}_4+b_2 {K^5}_5+b_3 {K^6}_6
\nn \\
{}Ê& &  + c_1  {K^7}_7 +c_2 {K^8}_8 + c_3 {K^9}_9+p {K^{10}}_{10}
\eqnlab{A2familyscalingoperators}
\eqa
with
\beqa
{}Ê& & a_1 +a_2 +a_3 =1
\nn \\
{}Ê& & b_1 + b_2 + b_3 =c_1+c_2 +c_3=0
\nn \\
{}Ê& & a_1^2 + a_2^2 + a_3^2+b_1^2 + b_2^2 + b_3^2+c_1^2 + c_2^2 +
c_3^2=1-2p. \eqnlab{A2familyscalingoperators2} \eqa The simplest
and most convenient choice is to take for $d$, \beq d={K^{1}}_{1}.
\eqnlab{fixedleveloperator1} \eeq Let us check that the null
generator $e_\delta$, \beq e_{\delta}\equiv
[E^{123},[E^{456},E^{789}]], \eqnlab{nullgenerator} \eeq
associated with the null root \beq \de =
\al_{0}+\theta=\al_{1}+\al_{2}+\al_{3}, \eqnlab{A2pNullroot} \eeq
where $\theta=\al_{2}+\al_{3}$ is the highest root of $A_{2}$, is
indeed an eigenvector with eigenvalue 1 for the adjoint action of
$d$. To this end we observe that \beq
[d,E^{123}]=[{K^{1}}_{1},E^{123}]=E^{123}, \; \; \; [d,E^{456}] =
[d,E^{789}]= 0. \eqnlab{levelcomputation1} \eeq Hence, $d$ counts
indeed the number of times $E^{123}$ appears in any commutator so
that one gets \beq æ[d,e_{\de}] = e_{\delta},
\eqnlab{nullgeneratorcheck} \eeq as desired.

Note that among the momentum constraints and Gauss' law, the only
non identically vanishing condition on the magnetic field is the
$10$-th component of the momentum constraint.

\subsubsection{Geometric Configurations $(9_2,6_3)$}

\begin{figure}[ht]
\begin{center}
\includegraphics[width=60mm]{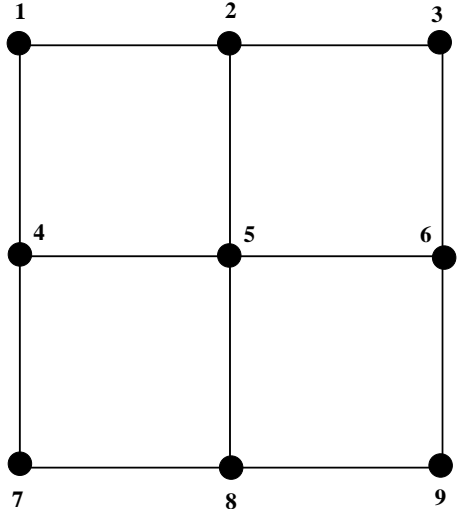}
\caption{The geometric configuration $(9_{2},6_{3})_1$.}
\label{figure:G92}
\end{center}
\end{figure}

There are two geometric configurations $(9_2,6_3)$.  We start
with $(9_2,6_3)_1$, shown in Figure \ref{figure:G92}.  The
configuration consists of two sets with three distinct triples in
each set: $S_{(1)}=\{(123),(456),(789)\}$,
$S_{(2)}=\{(147),(258),(369)\}$. By direct application of the
rules from above we can state that all generators associated with
each set will commute with the generators from the other set. Thus
the corresponding Cartan matrix is decomposable and equal to the
direct sum of two $A_2^+$'s.  This $6 \times 6$ matrix has rank
$4$.\\
\indent Since the generators $\{h_i, e_i, f_i\}$ ($i=1, \cdots, 6$) of
$\mf{g}_{(9_2,6_3)_1}$ fulfill the Chevalley-Serre relations
associated with the matrix $A_2^+ \oplus A_2^+$ (un-enlarged,
i.e., not its realization), the algebra $\mf{g}_{(9_2,6_3)_1}$
dual to $(9_2,6_3)_1$ is either the derived algebra $(A_2^+ \oplus
A_2^+)' = (A_2^+)' \oplus (A_2^+)'$ or a quotient of this algebra
by a subspace of its center.  The center of $(A_2^+)' \oplus
(A_2^+)'$ is two-dimensional and generated by the two central
charges $c_{(1)}= h_1 +h_2 + h_3$ and $c_{(2)} = h_4 + h_5 +h_6$.
It is clear that these central charges are not independent in the
algebra $\mf{g}_{(9_2,6_3)_1}$ since \beq c_{(1)} = c_{(2)}
=-{K^{10}}_{10}. \eeq  The two $(A_2^+)'$'s share therefore the
same central charge.  Hence $\mf{g}_{(9_2,6_3)_1}$ is the quotient
of $(A_2^+)' \oplus (A_2^+)'$ by the ideal $\mathbb{R}(c_{(1)} -
c_{(2)})$, \beq \mf{g}_{(9_2,6_3)_1} = \frac{(A_2^+)' \oplus
(A_2^+)'}{\mathbb{R}(c_{(1)} - c_{(2)})}.\eeq  This is the current
algebra $(A_2 \oplus A_2)^J$ of $A_2 \oplus A_2$ with a single
central charge ((\ref{current}) with a single $c$).

We can again introduce a (single) scaling element within the
Cartan subalgebra of $E_{10}$. Taking the affine roots to be
$\alpha_{123}$ and $\alpha_{147}$ (with generators $E^{123}$ and
$E^{147}$), one may choose \beq d = {K^1}_1,  \eeq as before. In
the algebra enlarged with the scaling operator, the Hamiltonian
constraint can be satisfied since the metric in the Cartan
subalgebra has Lorentzian signature.

An interesting new phenomenon occurs also for this configuration,
namely that the null roots of both algebras are equal (and equal to
$\beta^1 + \beta^2 + \beta^3 + \beta^4 + \beta^5 + \beta^6 +
\beta^7 + \beta^8 + \beta^9$ in the billiard parametrization).
Hence the vector space spanned by the roots of $\frac{(A_2^+)'
\oplus (A_2^+)'}{\mathbb{R}(c_{(1)} - c_{(2)})}$ in the space of
the roots of $E_{9}$ is 5-dimensional. This ``disappearance of one
dimension" is compatible with the fact that both null roots have
the same scaling behaviour under $d$ and is possible because we do
not have an embedding of the full Kac-Moody algebra $A_2^+ \oplus
A_2^+$ with two independent scaling operators under which the two
null roots behave distinctly. Note that, of course, the
corresponding generators $[E^{123}, [E^{456}, E^{789}]]$ and
$[E^{147}, [E^{258}, E^{369}]]$ are linearly independent.

\begin{figure}[ht]
\begin{center}
\includegraphics[width=80mm]{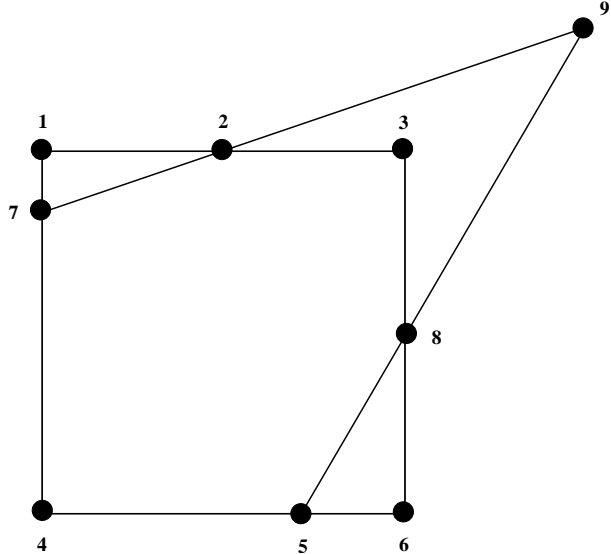}
\caption{The geometric configuration $(9_{2},6_{3})_2$.}
\label{figure:G93}
\end{center}
\end{figure}

\begin{figure}[ht]
\begin{center}
\includegraphics[width=40mm]{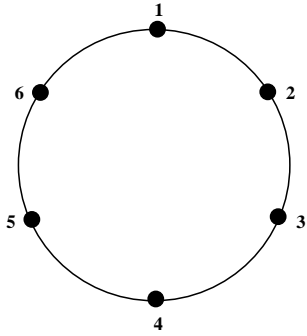}
\caption{The Dynkin diagram of
$A^+_5$, dual to the geometric configuration $(9_{2},6_{3})_{2}$.}
\label{figure:A5p}
\end{center}
\end{figure}

\vspace{.5cm} The other configuration $(9_2,6_3)$ is the
configuration $(9_{2},6_{3})_2$, depicted in Figure
\ref{figure:G93}. The analysis proceeds as for the configuration $(9_1,3_3)$. The
computation of the Cartan matrix is direct and yields the Dynkin
diagram shown in Figure \ref{figure:A5p}, which is recognized as being the
diagram of the untwisted affine extension $A^{+}_5$ of $A_5$.

The dual algebra is now the current algebra $(A^{+}_5)'$, with
central charge $c = -{K^{10}}_{10}$. If one regards $\alpha_{123}$
as the affine root, one can add the scaling operator $d= {K^1}_1$
to get the complete Kac-Moody algebra $A^{+}_5$.

\subsection{Geometric Configurations $(9_{3},9_{3})$}

\begin{figure}[ht]
\begin{center}
\includegraphics[width=60mm]{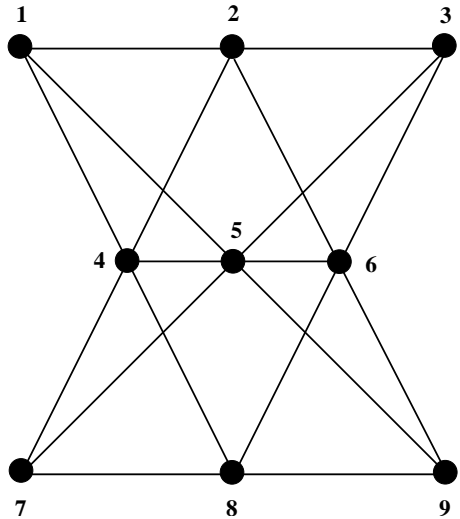}
\caption{$(9_{3},9_{3})_{1}$: This
is the so-called \emph{Pappus configuration}.}
\label{figure:G94}
\end{center}
\end{figure}

There are three geometric configurations $(9_{3},9_{3})$.  Their
treatment is a direct generalization of what we have discussed before.  Let us
consider first the configuration $(9_{3},9_{3})_{1}$, displayed in
Figure \ref{figure:G94}.
\noindent This configuration was constructed by \emph{Pappus of
Alexandria} during the $3^{\mathrm{rd}}$ century A.D. with the
purpose of illustrating the following theorem (adapted from
\cite{Page}):

\begin{quote}
\emph{Let three points $\{1,2,3\}$ lie in consecutive order on a
single straight line and let three other points $\{7,8,9\}$ lie in
consecutive order on another straight line. Then the three
pairwise intersections $4=\{1,2\}\cap \{7,8\}, 5=\{1,3\}\cap
\{7,9\}$ and $6=\{2,3\}\cap \{8,9\}$ are collinear. }
\end{quote}

\noindent The configuration is also called the Brianchon-Pascal
configuration \cite{Hilbert}.

By inspecting the Pappus configuration we note that it consists of
three sets with three distinct triples in each set:
$S_{1}=\{(123),(456),(789)\}, S_{2}=\{(159),(368),(247)\}$,
$S_{3}=\{(269),(357),(148)\}$. Hence its Cartan matrix is
decomposable and the direct sum of three times the Cartan matrix
$A^+_2$ of the untwisted affine extension of $A_2$. It has rank 6.
As in the previous examples, this does not imply, however, that
the complete algebra associated to the Pappus configuration is a
direct sum of $A_{2}^{+}$ algebras, or the derived algebra. One
has non trivial quotients because the three $(A_{2}^{+})'$ share
the same central charge.  Indeed, one finds again, just as above,
the relation \beq c_{(1)} = c_{(2)} = c_{(3)} =-{K^{10}}_{10}.
\eeq Hence $\mf{g}_{\mathrm{Pappus}} \equiv \mf{g}_{(9_3,9_3)_1}$
is the quotient of $(A_{2}^{+})' \oplus (A_{2}^{+})' \oplus
(A_{2}^{+})'$ by the ideal $\mathbb{R} (c_{(1)}- c_{(2)}) \oplus
\mathbb{R}(c_{(1)}- c_{(3)})$, i.e., the current algebra $(A_2
\oplus A_2 \oplus A_2)^J$ with only one central charge, \beq
\mf{g}_{\mathrm{Pappus}} = \frac{(A_2^+)' \oplus (A_2^+)' \oplus
(A_2^+)'}{\mathbb{R}(c_{(1)} - c_{(2)}) \oplus \mathbb{R}(c_{(1)}
- c_{(3)})} = (A_2 \oplus A_2 \oplus A_2)^J .\eeq

Regarding the affine roots as being $\alpha_{123}$, $\alpha_{159}$
and $\alpha_{148}$, one can add to the algebra the scaling
operator $d = {K^1}_1$, a task necessary to be able to get non
trivial solutions of the Hamiltonian constraint within the
algebra.

\begin{figure}[ht]
\begin{center}
\includegraphics[width=60mm]{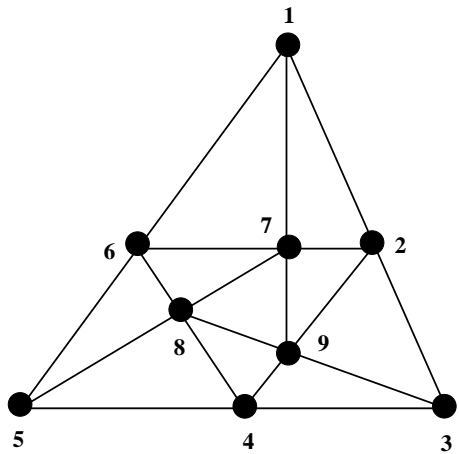}
\caption{The geometric
configuration $(9_{3},9_{3})_{2}$.}
\label{figure:G95}
\end{center}
\end{figure}

\begin{figure}[ht]
\begin{center}
\includegraphics[width=60mm]{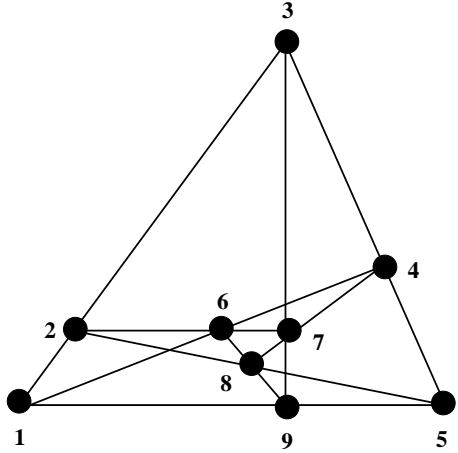}
\caption{The geometric
configuration $(9_{3},9_{3})_{3}$.}
\label{figure:G96}
\end{center}
\end{figure}

The two remaining $n=9$ configurations with $9$ lines are shown in
Figures \ref{figure:G95} and \ref{figure:G96}, respectively.

The configuration $(9_{3},9_{3})_{2}$ leads to the Dynkin diagram
of $A_8^+$ and to the derived algebra $(A_8^+)' \equiv A_8^J$.
Taking $\alpha_{123}$ as the affine root, one can add the scaling
element
\beq
d=\f{1}{6}(2 {K^{1}}_{1}+2 {K^{2}}_2 +2 {K^3}_3 -{K^4}_4 -{K^5}_5-{K^6}_6 -{K^7}_7+2 {K^8}_8 -{K^9}_9 -2 {K^{10}}_{10})
\eqnlab{scalingelement(9_3,9_3)2}
\eeq
to get the complete Kac-Moody algebra $A_8^+$.

\begin{figure}[ht]
\begin{center}
\includegraphics[width=50mm]{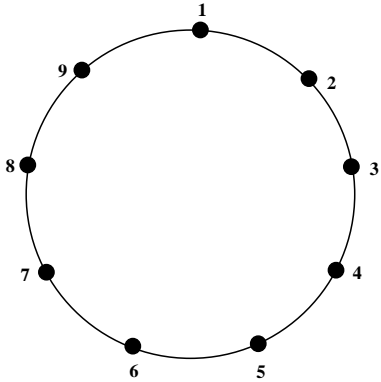}
\caption{The Dynkin diagram of
$A_8^+$ associated with the configuration $(9_{3},9_{3})_{2}$.}
\label{figure:A8p}
\end{center}
\end{figure}

\begin{figure}[ht]
\begin{center}
\includegraphics[width=80mm]{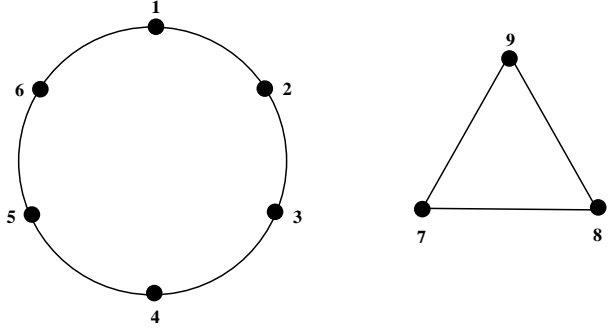}
\caption{The Dynkin diagram
of $A_2^+ \oplus A_5^+$ associated with the configuration
$(9_{3},9_{3})_{3}$.}
\label{figure:(A5p)(A2p)}
\end{center}
\end{figure}

The configuration $(9_{3},9_{3})_{3}$ leads to the Dynkin diagram
of $A_2^+ \oplus A_5^+$ and to the algebra $\frac{(A_2^+)' \oplus
(A_5^+)'}{\mathbb{R}(2 c_{(1)} - c_{(2)})}$ with only one central
charge. Taking $\alpha_{123}$ and $\alpha_{146}$ as the affine
roots, there exists a one-parameter family of scaling operators of the form
\beqa
{}Êd &=& \f{1}{6}(5 {K^{1}}_1 +4 {K^2}_2 +3 {K^3}_3 +2 {K^4}_4 -5 {K^5}_5 -{K^6}_6 -3 {K^7}_7 +{K^8}_8)
\nn \\
{}Ê& & +\f{p}{2} (-{K^1}_1+2 {K^2}_2 -{K^3}_3+2{K^4}_4 -{K^5}_5 -{K^6}_6 -{K^7}_7 -{K^8}_8 + 2 {K^9}_9)
\nn \\
{}Ê& & + \f{1}{4}(3 + 4 p + 9 p^2) {K^{10}}_{10}.
\eqnlab{scalingelement(9_3,9_3)3} \eqa Note that this operator
actually counts the number of roots $\al_{146}$ but twice the number
of roots $\al_{123}$. This is in accordance with the fact that the
ideal is of the form $\mathbb{R}(2 c_{(1)} - c_{(2)})$. The
corresponding Dynkin diagrams are shown in Figures \ref{figure:A8p}
and \ref{figure:(A5p)(A2p)}, respectively.

\subsection{Geometric Configuration $(9_{4},12_{3})$}

The geometric configuration is shown in Figure
\ref{figure:(9_4,12_3)}. We find the Dynkin diagram of $A_2^+ \oplus A_2^+ \oplus
A_2^+ \oplus A_2^+$.  The relevant algebra is then the direct sum
of the corresponding derived algebras with same central charge,
i.e. \beq \mf{g}_{(9_4,12_3)} = \frac{(A_2^+)' \oplus (A_2^+)'
\oplus (A_2^+)' \oplus (A_2^+)'}{\mathbb{R}(c_{(1)} - c_{(2)})
\oplus \mathbb{R}(c_{(1)} - c_{(3)}) \oplus \mathbb{R}(c_{(1)} -
c_{(4)})}.\eeq  The scaling operator $d = {K^1}_1$ can be added to
the algebra.

This result on $(9_4,12_3)$ is intimately connected with the
analysis of the geometric configuration $(8_3,8_3)$, for which the
algebra is $A_2 \oplus A_2 \oplus A_2 \oplus A_2$.  This algebra
can be embedded in $E_8$ and, accordingly, the corresponding
current algebra with a single central charge can be embedded in
the current algebra $E_9'\equiv (E_8^+)'$ of $E_8$.  On the side
of the geometric configurations, the affinization of $A_2 \oplus
A_2 \oplus A_2 \oplus A_2$ corresponds to adding one point, say 9,
to $(8_3,8_3)$ and drawing the four lines connecting this new
point to the four pairs of unconnected points of $(8_3,8_3)$.
This yields $(9_{4},12_{3})$.  Note that this is the only case for
which it is possible to extend an $n=p$ configuration to an
$n=p+1$ configuration
through the inclusion of an additional point directly in the configuration.  \\
\indent The configuration $(9_4,12_3)$ also has an interesting
interpretation in terms of points of inflection of third-order
plane curves \cite{Hilbert}. \begin{figure}
\begin{center}
\includegraphics[width=60mm]{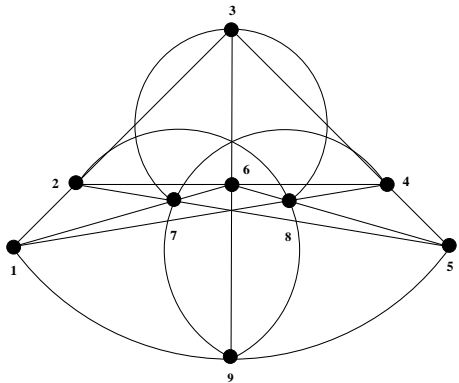}
\caption{The geometric
configuration $(9_{4},12_{3})$.}
\label{figure:(9_4,12_3)}
\end{center}
\end{figure}

\begin{center}
\begin{table}
  \begin{tabular}{ |m{20mm}|m{45mm}|m{30mm}|m{40mm}|}
\hline
\multicolumn{2} {|c|} {Configuration}  & Dynkin diagram & Lie algebra \\
    \hline
     \hline
    $(9_1,3_3)$ & \includegraphics[width=30mm]{G91} &
      \includegraphics[width=15mm]{A2p} & $\mg_{(9_1,3_3)}=A_{2}^{J}$  \\
    \hline
      $(9_2,6_3)_1$ &  \includegraphics[width=30mm]{G92} &
      \includegraphics[width=30mm]{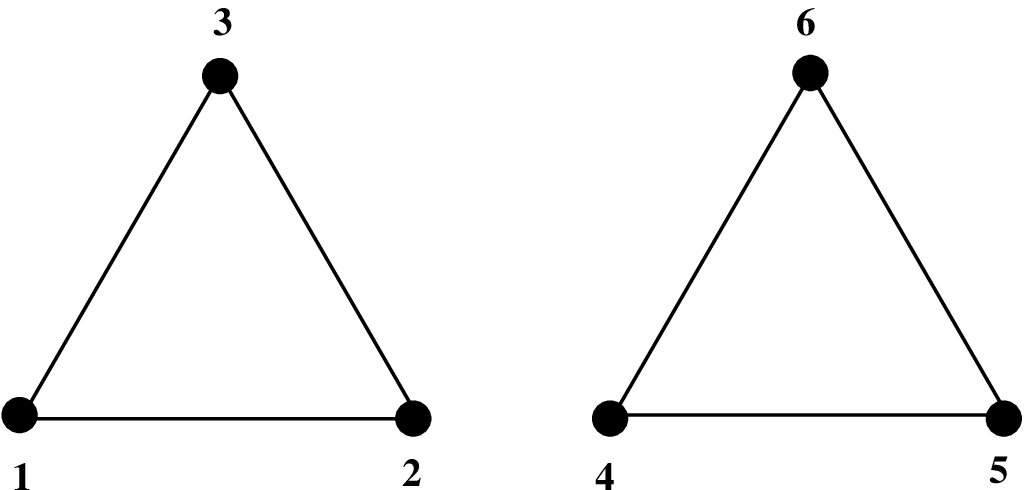} & $\mg_{(9_2,6_3)_1}=(A_{2}\oplus A_{2})^{J}$ \\
      \hline
       $(9_2,6_3)_2$ &  \includegraphics[width=40mm]{G93} &
      \includegraphics[width=25mm]{A5p} & $\mg_{(9_2,6_3)_2}=A_{5}^{J}$ \\
      \hline
       $(9_3,9_3)_1$ &  \includegraphics[width=30mm]{G94} &
      \includegraphics[width=30mm]{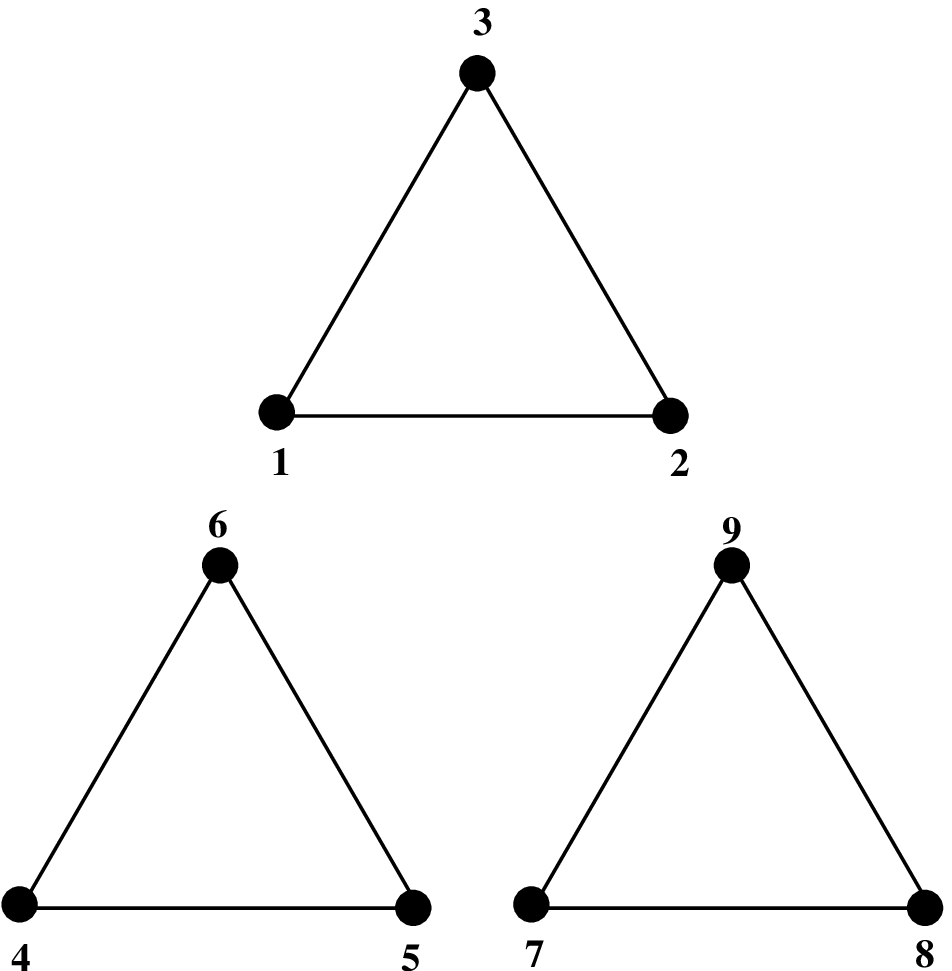} & $\mg_{(9_3,9_3)_1}=$\phantom{hhhhhhhhhiii} $(A_2\oplus A_{2}\oplus A_2)^{J}$\\
      \hline
      $(9_3,9_3)_2$ &  \includegraphics[width=30mm]{G95} &
      \includegraphics[width=25mm]{A8p} & $\mg_{(9_3,9_3)_{2}}=A_{8}^{J}$\\
      \hline
  \end{tabular}
         \caption{$n=9$ configurations and their dual affine Kac-Moody algebras.}
\end{table}
\end{center}
\begin{center}
\begin{table}
  \begin{tabular}{ |m{20mm}|m{45mm}|m{30mm}|m{40mm}|}
\hline
\multicolumn{2} {|c|} {Configuration}  & Dynkin diagram & Lie algebra \\
    \hline
\hline
         $(9_3,9_3)_3$ &  \includegraphics[width=30mm]{G96} &
      \includegraphics[width=30mm]{A5pA2p} & $\mg_{(9_3,9_3)_{3}}=(A_{5}\oplus A_{2})^{J}$\\
      \hline
       $(9_4,12_3)$ &  \includegraphics[width=45mm]{9_4,12_3} &
      \includegraphics[width=30mm]{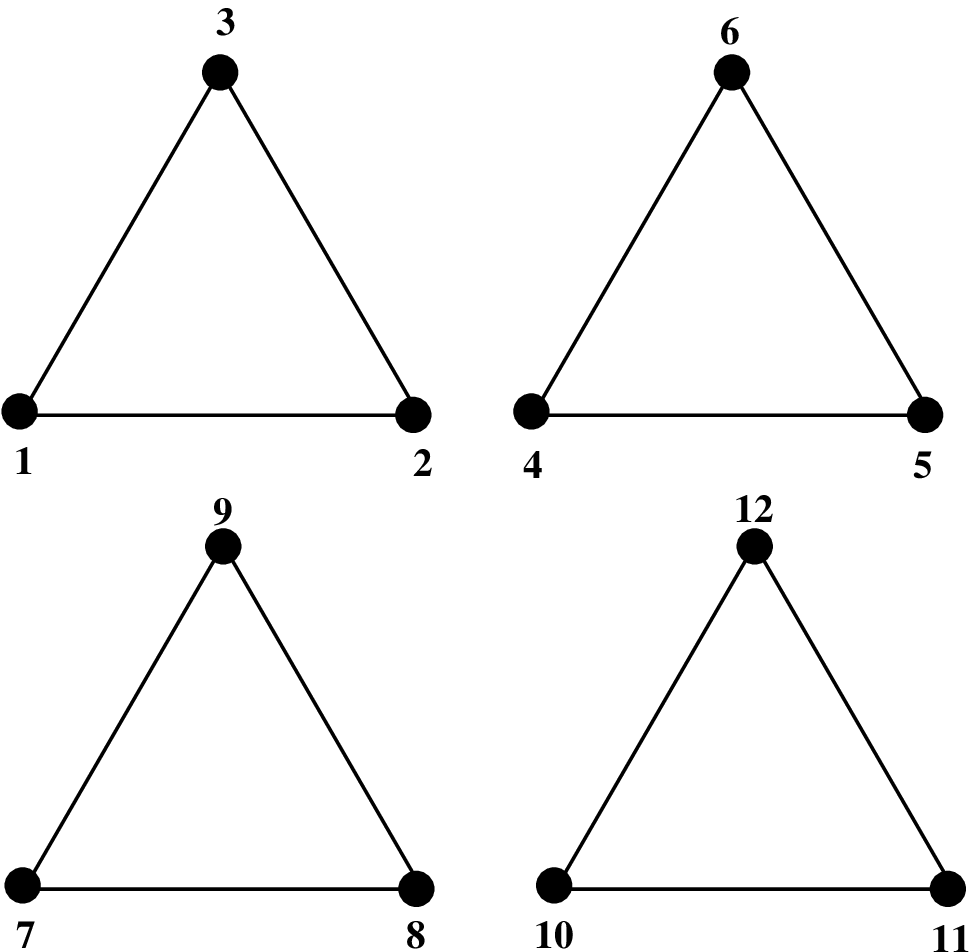} & $\mg_{(9_4,12_3)} =$\phantom{hhhhhhiiii =} $(A_{2}\oplus A_{2}\oplus A_2\oplus A_2)^{J}$\\
      \hline
 \end{tabular}
         \caption{$n=9$ configurations and their dual affine Kac-Moody algebras.}
\end{table}
\end{center}

\newpage

\section{Geometric Configurations $(10_{m},g_{3})$}
\label{Rank10} \setcounter{equation}{0}

As we shall see, subalgebras constructed from configurations with
ten nodes give rise to Lorentzian subalgebras of $E_{10}$, except
in two cases, denoted $(10_3,10_3)_4$ and $(10_3,10_3)_7$ in Tables $4$
and $5$ and in \cite{Demaret:1985js}, for
which the Cartan matrix has zero determinant.

Because the rank-10 algebras are infinite-dimensional, one must
again truncate to level 2 in order to match the Bianchi I
supergravity equations with the $\sigma$-model equations.
Furthermore, if taken to be non zero, the magnetic field must
fulfill the relevant momentum and Gauss constraints.

\subsection{The Petersen Algebra}

\begin{figure}[ht]
\begin{center}
\includegraphics[width=80mm]{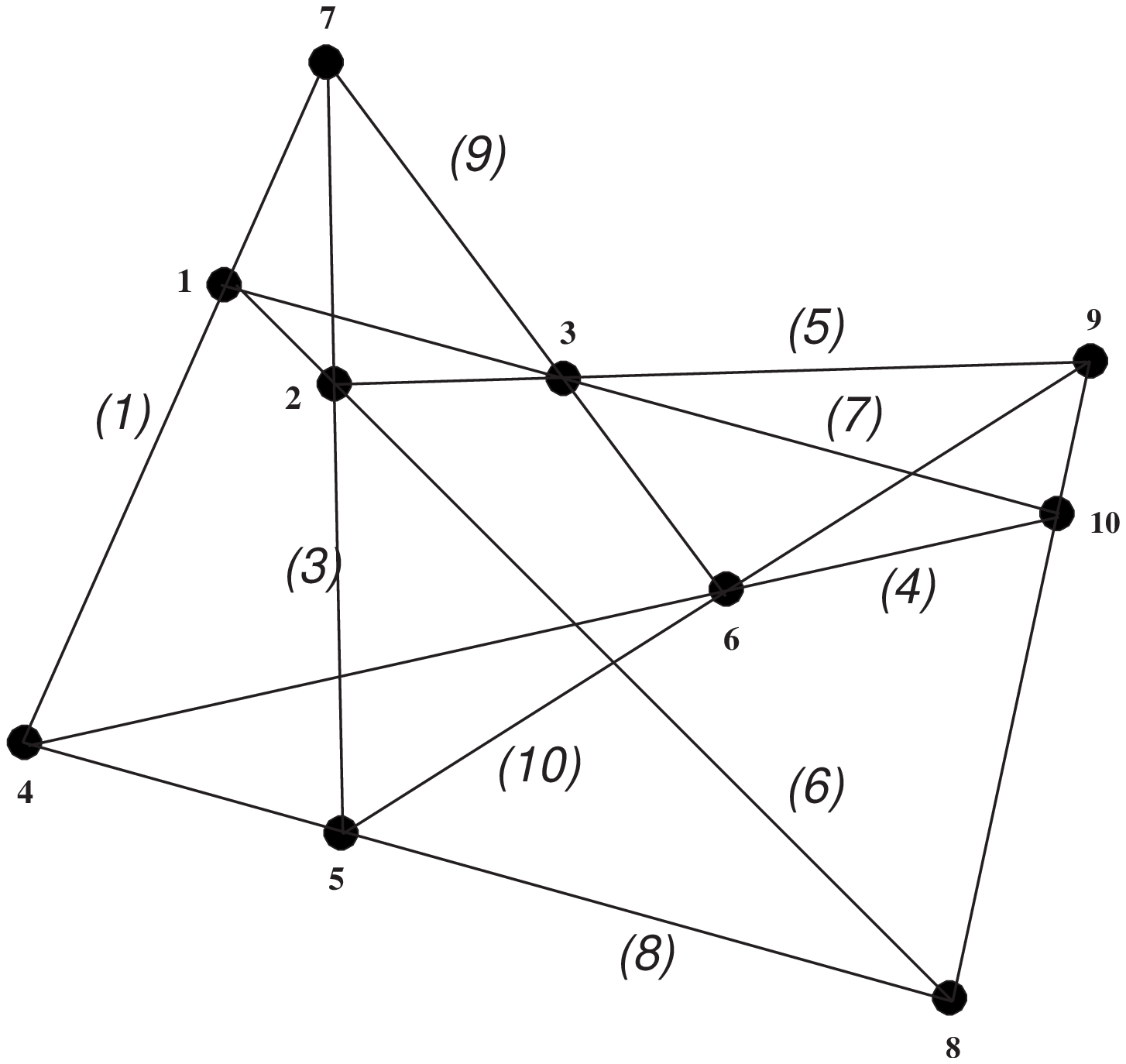}
\caption{$(10_{3},10_{3})_{3}$: The Desargues configuration, dual
to the Petersen graph.} \label{figure:GDesargues}
\end{center}
\end{figure}

\begin{figure}[ht]
\begin{center}
\includegraphics[width=70mm]{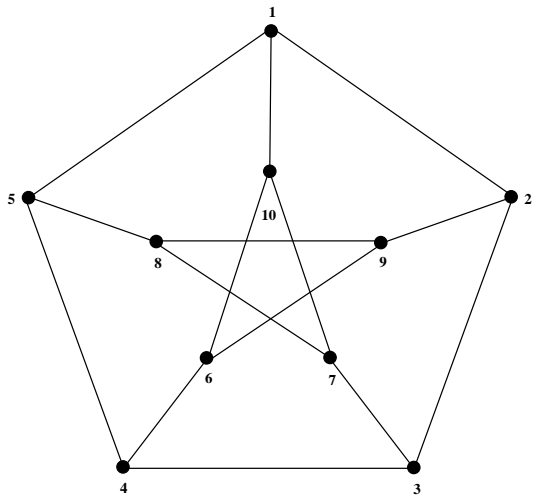}
\caption{This is the so-called
\emph{Petersen graph}. It is the Dynkin diagram dual to the
Desargues configuration, and is in fact a geometric configuration
itself, denoted $(10_{3},15_{2})$.}
\label{figure:D2(Petersen)}
\end{center}
\end{figure}

We first illustrate the simple situation with a non degenerate
Cartan matrix, for which the above theorem of Section
\ref{Regular} applies directly. We consider explicitly the
well-known \emph{Desargues configuration}, denoted
$(10_{3},10_{3})_{3}$, for which a new fascinating feature
emerges, namely that the Dynkin diagram dual to it \emph{also}
corresponds in itself  to a geometric configuration. In fact, the
dual Dynkin diagram turns out to be the famous \emph{Petersen
graph}, denoted $(10_{3},15_{2})$. These are displayed in Figures
\ref{figure:GDesargues} and \ref{figure:D2(Petersen)},
respectively. \\
\indent The configuration $(10_{3},10_{3})_{3}$ is associated with
the $17^{\mathrm{th}}$ century French mathematician
\emph{G\'{e}rard Desargues} to illustrate the following
``Desargues theorem" (adapted from \cite{Page}):

\begin{quote}
\emph{Let the three lines defined by $\{4,1\},\{5,2\}$ and
$\{6,3\}$ be concurrent, i.e. be intersecting at one point, say
$\{7\}$. Then the three intersection points $8\equiv
\{1,2\}\cap\{4,5\}, 9\equiv \{2,3\}\cap\{5,6\}$ and $10\equiv
\{1,3\}\cap\{4,6\}$ are collinear. }
\end{quote}

\noindent Another way to say this is that the two triangles
$\{1,2,3\}$ and $\{4,5,6\}$ in Figure \ref{figure:GDesargues} are
in perspective from the point $\{7\}$ and in perspective from the
line $\{8,10,9\}$.

To construct the Dynkin diagram we first observe that each line in
the configuration is disconnected from three other lines, e.g.
$\{4,1,7\}$ have no nodes in common with the lines $\{2,3,9\}$,
$\{5,6,9\}$, $\{8,10,9\}$. This implies that all nodes in the
Dynkin diagram will be connected to three other nodes.  Proceeding
as in the previous section leads to the Dynkin diagram in Figure
\ref{figure:D2(Petersen)}, which we identify as the Petersen
graph. The corresponding Cartan matrix is \beq
A(\mf{g}_{\mathrm{Petersen}})=\left(
\begin{array}{cccccccccc}
2 & -1 & 0 & 0 & 0 & 0 & 0 & 0 & -1 & -1 \\
-1 & 2 & -1 & 0 & 0 & -1 & 0 & 0 & 0 & 0 \\
0 & -1 & 2 & -1 & 0 & 0 & 0 & -1 & 0 & 0 \\
0 & 0 & -1 & 2 & -1 & 0 & 0 & 0 & 0 & -1 \\
0 & 0 & 0 & -1 & 2 & -1 & 0 & 0 & -1 & 0 \\
0 & -1 & 0 & 0 & -1 & 2 & -1 & 0 & 0 & 0 \\
0 & 0 & 0 & 0 & 0 & -1 & 2 & -1 & 0 & -1 \\
0 & 0 & -1 & 0 & 0 & 0 & -1 & 2 & -1 & 0 \\
-1 & 0 & 0 & 0 & -1 & 0 & 0 & -1 & 2 & 0 \\
-1 & 0 & 0 & -1 & 0 & 0 & -1 & 0 & 0 & 2 \\
\end{array} \right),
\eqnlab{PetersenCartanMatrix} \eeq which is of Lorentzian
signature with \beq \det A(\mf{g}_{\mathrm{Petersen}})=-256.
\eqnlab{PetersenDeterminant} \eeq The Petersen graph was invented
by the Danish mathematician \emph{Julius Petersen} in the end of
the $19^{\mathrm{th}}$ century. It has several embeddings on the
plane, but perhaps the most famous one is as a star inside a
pentagon as depicted in Figure \ref{figure:D2(Petersen)}. One of its
distinguishing features from the point of view of graph theory is
that it contains a \emph{Hamiltonian path} but no
\emph{Hamiltonian cycle}.\footnote{We recall that a
Hamiltonian path is defined as a path in an undirected graph which
intersects each node once and only once. A Hamiltonian cycle is
then a Hamiltonian path which also returns to its initial node.}
\begin{figure}[ht]
\begin{center}
\includegraphics[width=50mm]{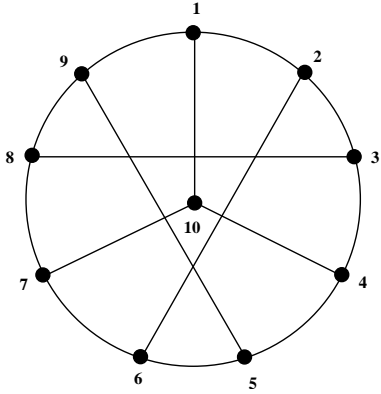}
\caption{An alternative drawing of the Petersen graph in the
plane. This embedding reveals an $S_3$ permutation symmetry about
the central point.  Yet another drawing is given in  Tables {\bf
4} and {\bf 5} summarizing the results for the $(10_3,10_3)$
configurations.} \label{figure:D2(Petersen)2}
\end{center}
\end{figure}
\noindent Because the algebra is Lorentzian  (with a metric that coincides
with the metric induced from the embedding in $E_{10}$), it does
not need to be enlarged by any further generator to be compatible
with the Hamiltonian constraint.\\
\indent It is interesting to examine the symmetries of the various
embeddings of the Petersen graph in the plane and the connection
to the Desargues configurations. The embedding in Figure
\ref{figure:D2(Petersen)} clearly exhibits a $\mathbb{Z}_{5}\times
\mathbb{Z}_2$-symmetry, while the Desargues configuration in
Figure \ref{figure:GDesargues} has only a $\mathbb{Z}_2$-symmetry.
Moreover, the embedding of the Petersen graph shown in Figure
\ref{figure:D2(Petersen)2} reveals yet another symmetry, namely an
$S_3$ permutation symmetry about the central point, labeled
``$10$''. In fact, the external automorphism group of the Petersen
graph is $S_5$ so what we see in the various embeddings are simply
subgroups of $S_5$ made manifest. It is not clear how these
symmetries are realized in the Desargues configuration that seems
to exhibit much less symmetry.

\subsection{A Degenerate Case}

\begin{figure}[ht]
\begin{center}
\includegraphics[width=90mm]{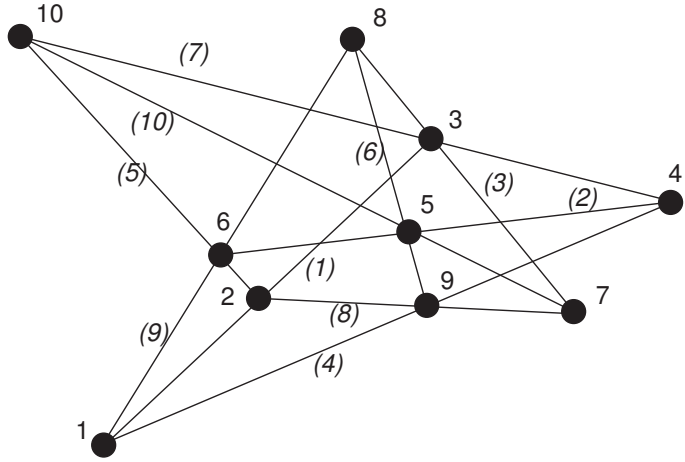}
\caption{$(10_{3},10_{3})_{4}$: This configuration gives rise to a
Dynkin diagram whose Cartan matrix has vanishing determinant, and
hence contains a non-trivial center.} \label{figure:G104}
\end{center}
\end{figure}

\begin{figure}[ht]
\begin{center}
\includegraphics[width=110mm]{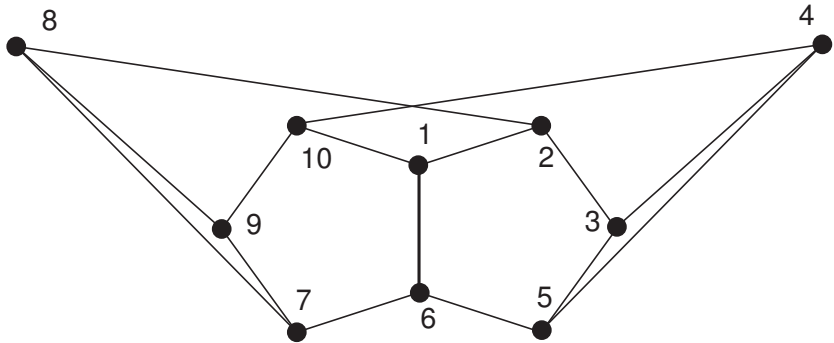}
\caption{The Dynkin diagram of $(10_{3},10_{3})_{4}$ corresponds
to a Cartan matrix with vanishing determinant and hence to an
algebra that contains a non-trivial ideal. } \label{figure:D3}
\end{center}
\end{figure}

We now discuss another interesting case, which requires a special
treatment because the corresponding Cartan matrix is degenerate.
It is the configuration $(10_{3},10_{3})_{4}$, shown in Figure
\ref{figure:G104}. By application of the rules, with the
generators chosen according to the numbering of the lines in
Figure \ref{figure:G104}, i.e. $(1)=123, (2)=456\dots$etc, we find
that this configuration gives rise to the Dynkin diagram shown in
Figure \ref{figure:D3}. The Cartan matrix takes the form \beq
A(\mf{g}_{(10_{3},10_{3})_{4}})=\left(
\begin{array}{cccccccccc}
2 & -1 & 0 & 0 & 0 & 0 & -1 & 0 & 0 & -1 \\
-1 & 2 & -1 & -1 & 0 & 0 & 0 & 0 & 0 & 0 \\
0 & -1 & 2 & -1 & 0 & 0 & 0 & 0 & -1 & 0 \\
0 & -1 & -1 & 2 & -1 & 0 & 0 & 0 & 0 & 0 \\
0 & 0 & 0 & -1 & 2 & -1 & 0 & 0 & 0 & -1 \\
0 & 0 & 0 & 0 & -1 & 2 & -1 & -1 & 0 & 0 \\
-1 & 0 & 0 & 0 & 0 & -1 & 2 & -1 & 0 & 0 \\
0 & 0 & 0 & 0 & 0 & -1 & -1 & 2 & -1 & 0 \\
0 & 0 & -1 & 0 & 0 & 0 & 0 & -1 & 2 & -1 \\
-1 & 0 & 0 & 0 & -1 & 0 & 0 & 0 & -1 & 2 \\
\end{array} \right),
\eqnlab{ReducibleCartanMatrix} \eeq which has vanishing
determinant \beq \det A(\mf{g}_{(10_{3},10_{3})_{4}})=0.
\eqnlab{VanishingDeterminant} \eeq This Cartan matrix has one
negative and one null eigenvalue while the rest of the eigenvalues
are positive. Hence, the algebra is indefinite type. The
eigenvector associated to the null eigenvalue is given explicitly
by \beq u= (0,1,1,1,0,-1,-1,-1,0,0). \eqnlab{nulleigenvector} \eeq
We then deduce that the corresponding algebra has a one
dimensional non-trivial center, $\mf{r}=\{k\}$, with \beq
k=\sum_{i=1}^{10} u_{i}h_{i}=
-h_{2}-h_{3}-h_{4}+h_{6}+h_{7}+h_{8}, \eqnlab{ideal} \eeq where
$u_{i}$ are the components of the null eigenvector and $h_{i}$ are
the generators of the Cartan subalgebra.  Making use of the
explicit form of $h_{i}$, \Eqnref{GeneralCartanElement}, we find
that $k$ vanishes identically in $E_{10}$\beq
k=-h_{2}-h_{3}-h_{4}+h_{6}+h_{7}+h_{8}=0. \eqnlab{zeroideal} \eeq
This shows that the embedding introduces a relation among the
generators of the Cartan subalgebra. Constructing the quotient
algebra \beq
\mf{g}_{(10_{3},10_{3})_{4}}=\frac{(\mathbb{KM}(A(\mf{g}_{(10_{3},10_{3})_{4}})))'}
{\mf{r}}, \eqnlab{quotientalgebra} \eeq corresponds to eliminating
one of the generators $h_{a}$ through \Eqnref{zeroideal}.  It is
that quotient algebra that is dual to the geometric configuration
$(10_{3},10_{3})_{4}$. Note that the roots are not linearly
independent but obeys the same linear relation as the Cartan
elements in \Eqnref{zeroideal}.

Note that that the Kac-Moody algebra associated with the
degenerate Cartan matrix \Eqnref{ReducibleCartanMatrix} along the
lines of \cite{Kac} involves augmenting the Cartan matrix to get a
``realization" and adding one more Cartan generator.  The metric
in the complete Cartan subalgebra of this Kac-Moody algebra has
signature $(-,-,+, +, \cdots, +)$.

\subsection{Dynkin Diagrams Dual to Configurations $(10_3,10_3)$}

We now give, in the form of a table, the list of all
configurations $(10_3,10_3)$ and the corresponding Dynkin
diagrams.  These are all connected. Note that some of the
configurations give rise to equivalent Dynkin diagrams. For
instance, the configuration $(10_3,10_3)_2$ and the Desargues
configuration $(10_3,10_3)_3$ (which is projectively self-dual)
both lead to the Petersen Dynkin diagram. Thus, although we have
ten configurations, we only find seven distinct rank 10
subalgebras of $E_{10}$ : six Lorentzian subalgebras and one
subalgebra with a Cartan matrix having zero determinant. The
degenerate case discussed above appears for two of the
configurations, $(10_3,10_3)_4$ and $(10_3,10_3)_7$. All other
cases give Cartan matrices with one negative and nine positive
eigenvalues. Only the first configuration, $(10_3,10_3)_1$, is
non-planar, i.e. cannot be realized with straight lines in the
plane \cite{Bokowski}. This
fact does not seem to manifest itself on the algebraic side.\\
\indent Since some of the configurations give rise to equivalent
Dynkin diagrams one might wonder if this means that two cosmological
solutions may seem different from the supergravity point of view but
are in fact equivalent in the coset construction. This is not true
because even though the Dynkin diagrams are the same, the embedding
in $E_{10}$ is not. Hence, when constructing a coset Lagrangian
based on the algebra associated to a given configuration, one must
choose the generators according to the numbering of the lines in the
configurations, and this uniquely determines the solution. This also
motivates the use of the word ``dual'' for the correspondences we
find.
\begin{center}
\begin{table}
  \begin{tabular}{ |m{20mm}|m{50mm}|m{30mm}|m{45mm}|}
\hline
\multicolumn{2} {|c|} {Configuration}  & Dynkin diagram  & Determinant of $A$\\
    \hline
     \hline
    $(10_3,10_3)_1$ & \includegraphics[width=45mm]{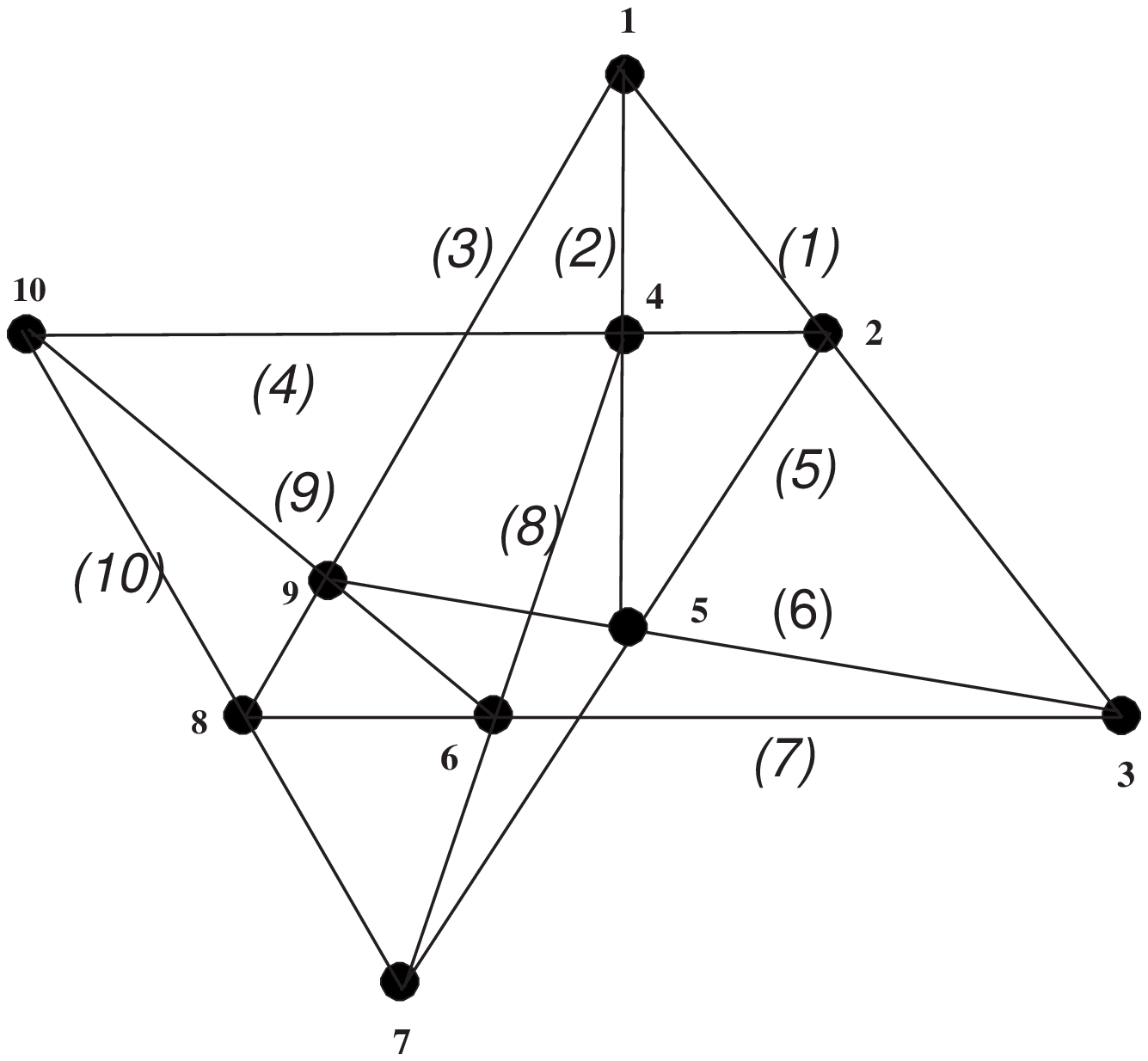} &
      \includegraphics[width=30mm]{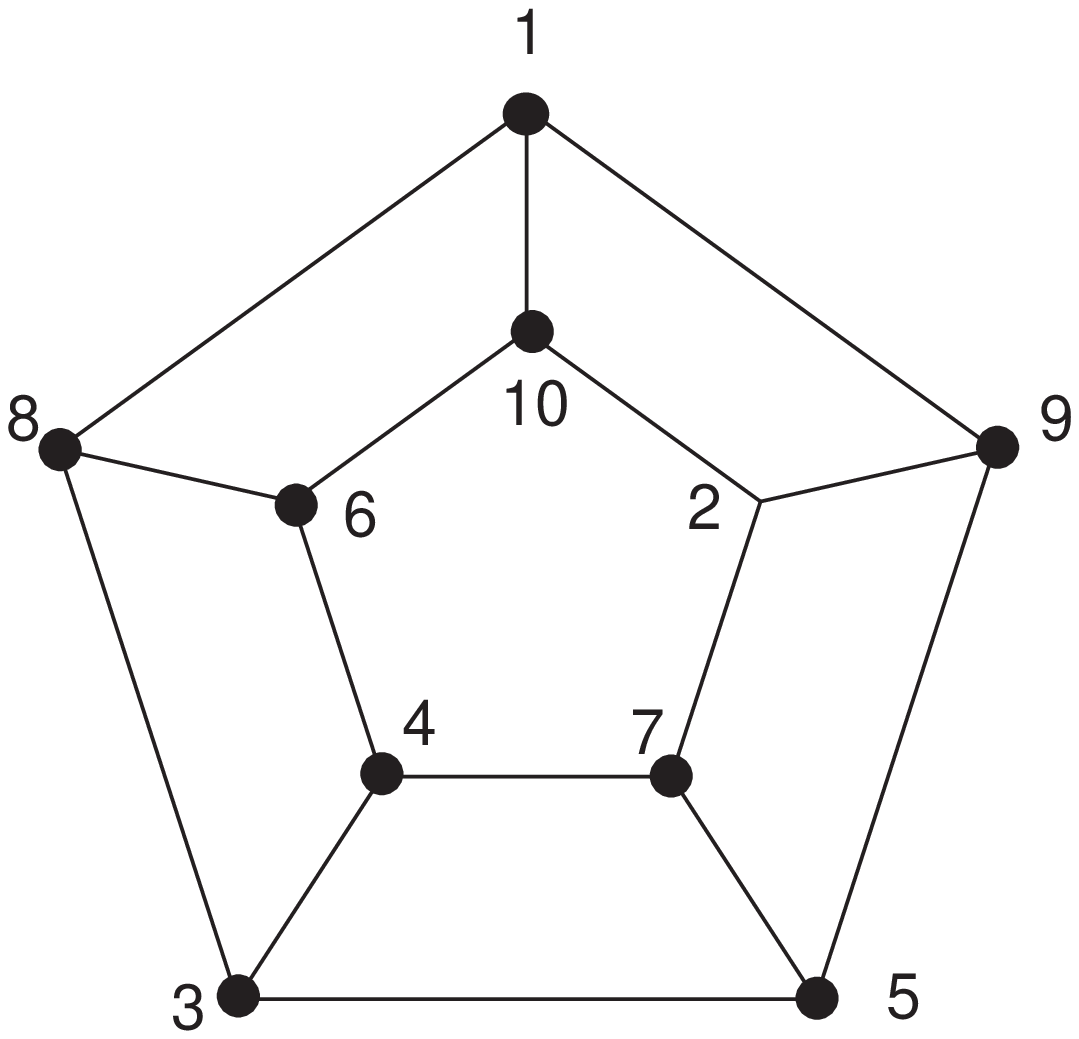} & $\det A(\mg_{(10_3,10_3)_1})=-121$  \\
    \hline
      $(10_3,10_3)_2$ &  \includegraphics[width=50mm]{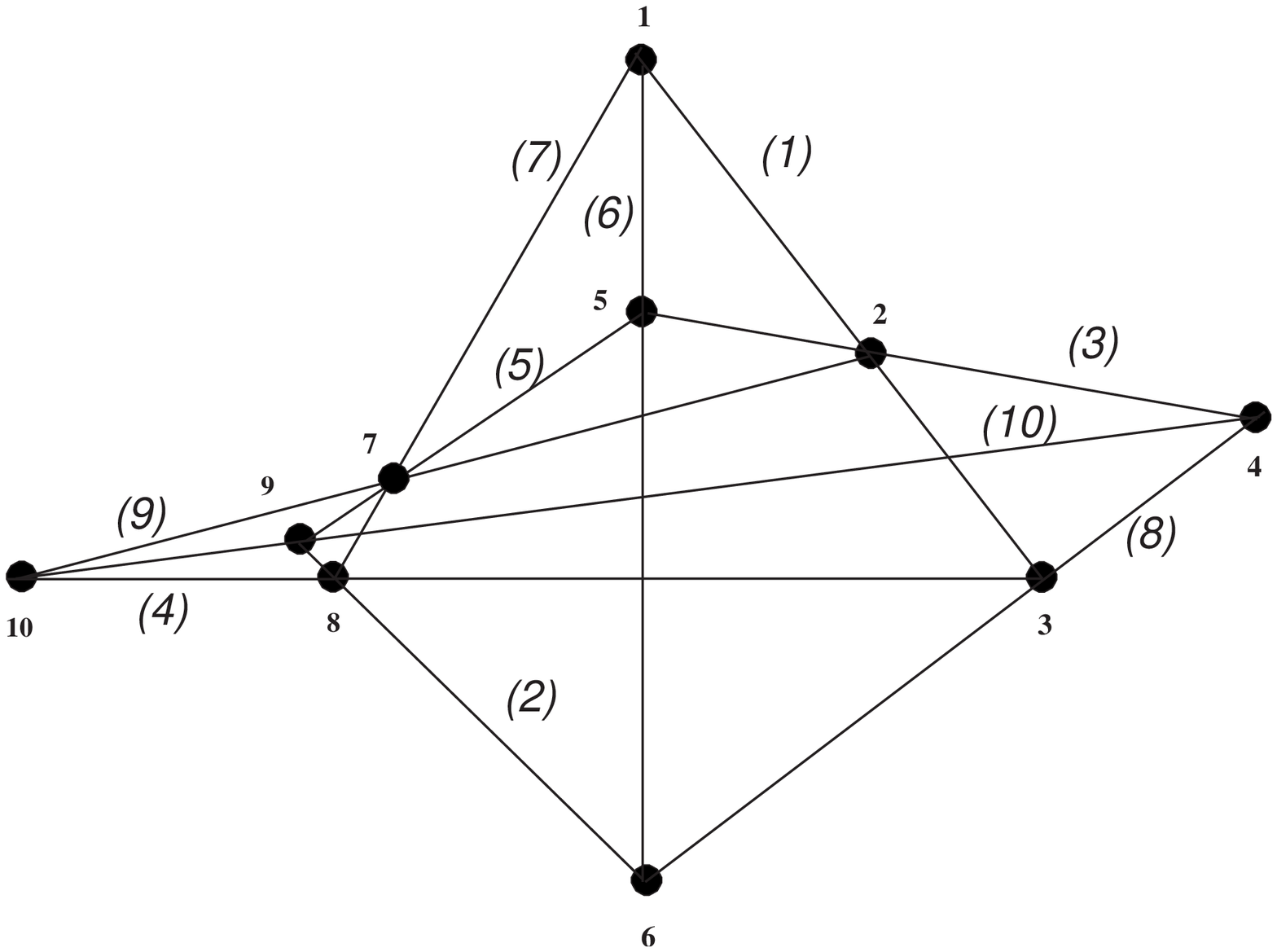} &
      \includegraphics[width=30mm]{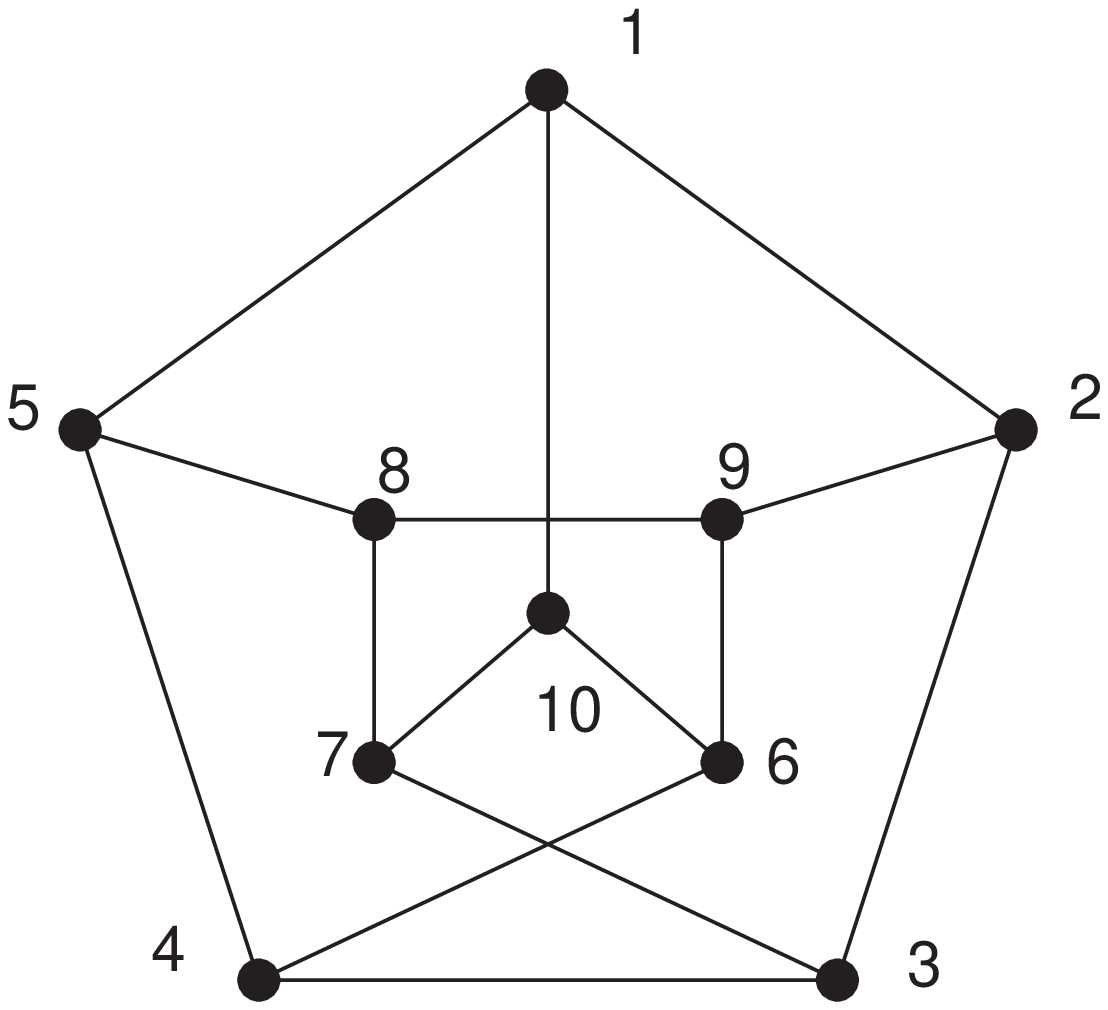} & $\det A(\mg_{(10_3,10_3)_2})=-256$ \\
      \hline
       $(10_3,10_3)_3$ &  \includegraphics[width=45mm]{C3} &
      \includegraphics[width=30mm]{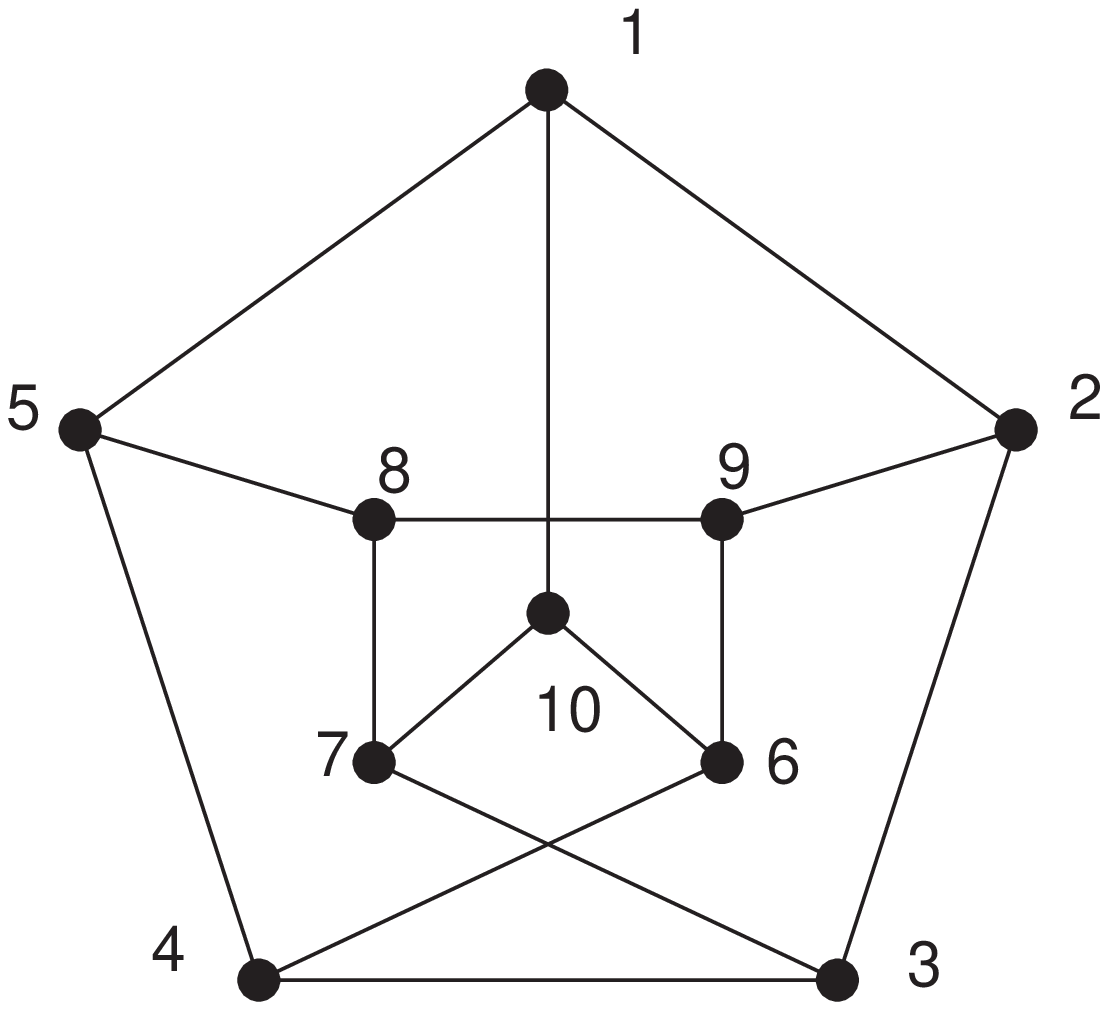} & $\det A(\mg_{(10_3,10_3)_3})=-256$ \\
      \hline
       $(10_3,10_3)_4$ &  \includegraphics[width=45mm]{C4} &
      \includegraphics[width=30mm]{D4} & $\det A(\mg_{(10_3,10_3)_4})=0$\\
      \hline
$(10_3,10_3)_5$ &  \includegraphics[width=45mm]{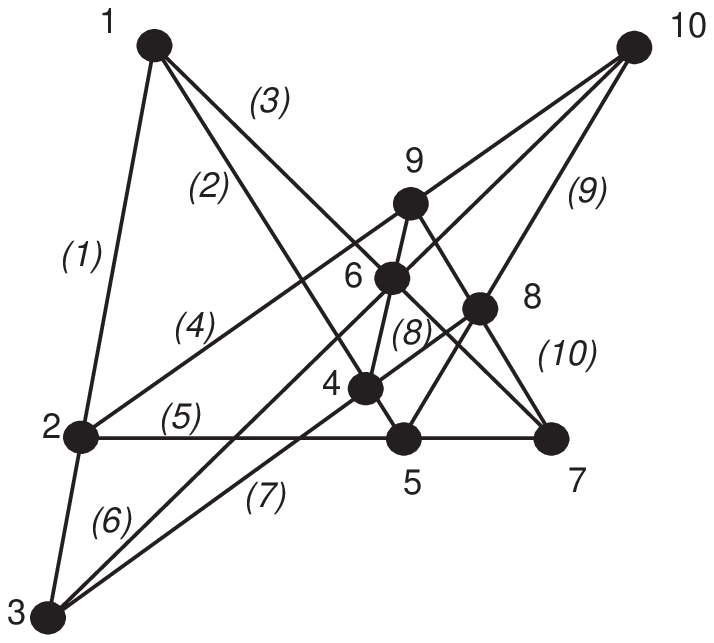} &
      \includegraphics[width=30mm]{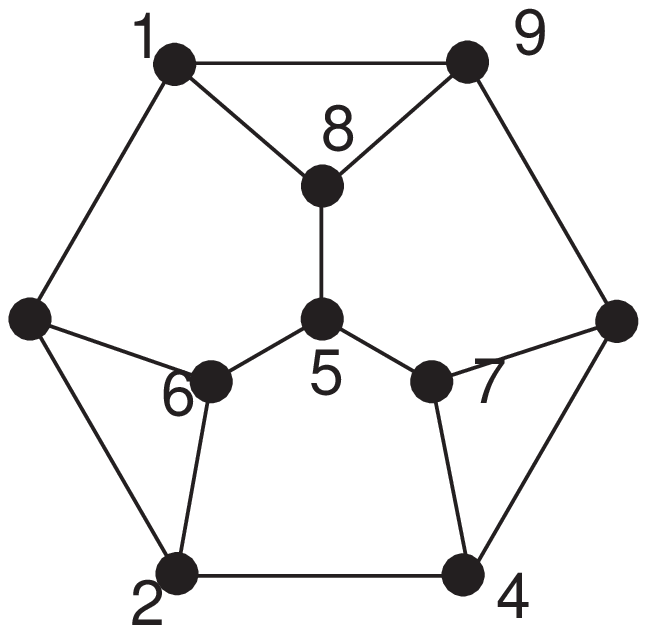} & $\det A(\mg_{(10_3,10_3)_{5}})=-16$\\
      \hline
  \end{tabular}
         \caption{$n=10$ configurations and their dual Lorentzian Kac-Moody algebras.
Note that some of the configurations give rise to equivalent
Dynkin diagrams.  In this table and in the next one, we have drawn
the Dynkin diagrams in a way that minimizes the crossing number
(i.e., the unwanted crossings of edges at points that do not
belong to the Dynkin diagram).}
\end{table}
\end{center}

\begin{center}
\begin{table}
  \begin{tabular}{ |m{20mm}|m{50mm}|m{30mm}|m{45mm}|}
\hline
\multicolumn{2} {|c|} {Configuration}  & Dynkin diagram  & Determinant of $A$\\
    \hline
     \hline
    $(10_3,10_3)_6$ & \includegraphics[width=45mm]{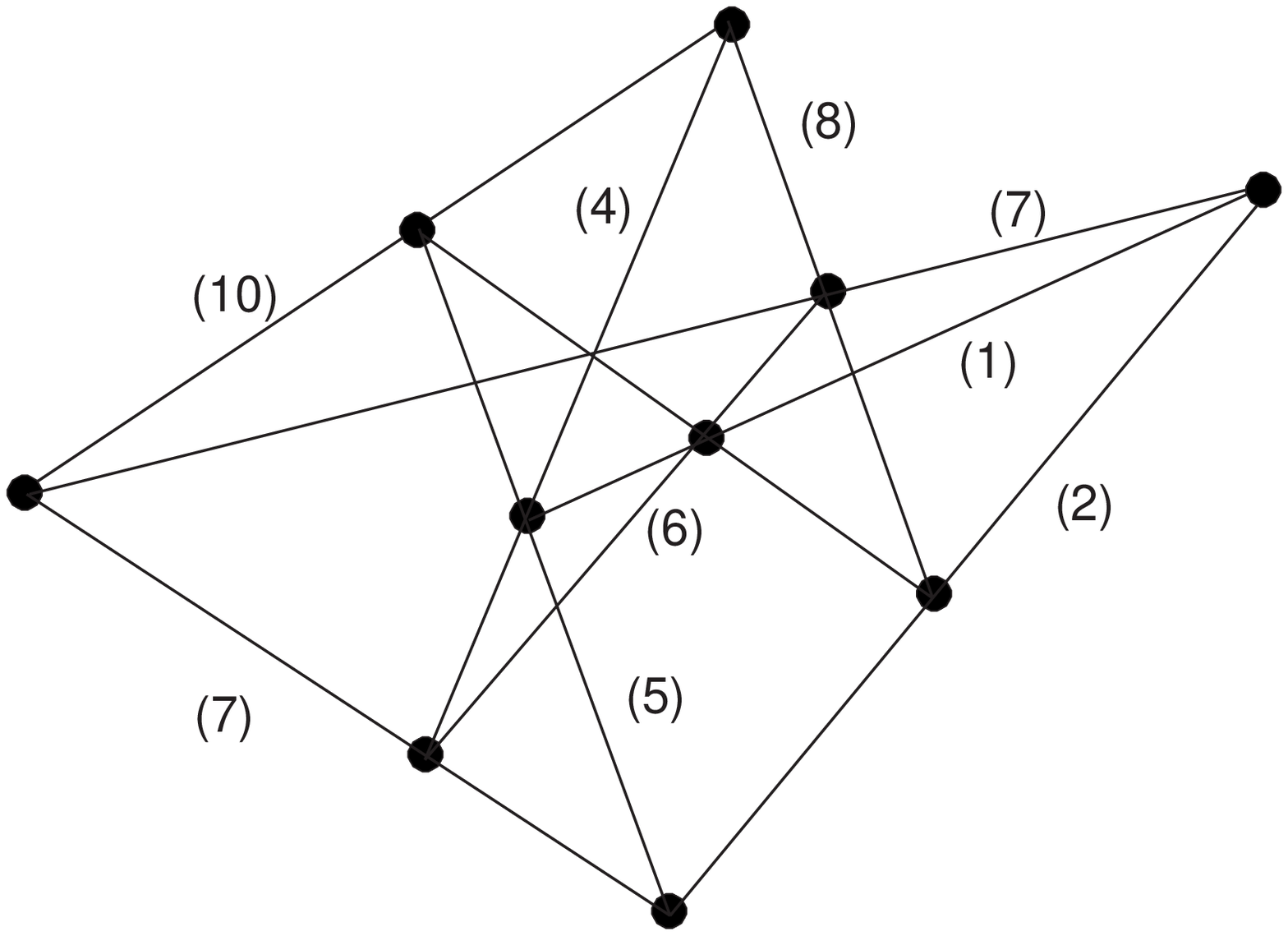} &
      \includegraphics[width=30mm]{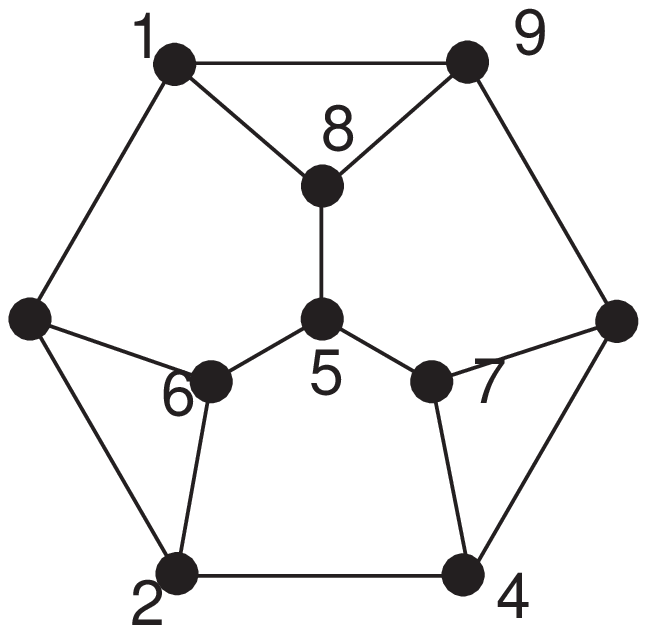} & $\det A(\mg_{(10_3,10_3)_6})=-16$  \\
    \hline
      $(10_3,10_3)_7$ &  \includegraphics[width=25mm]{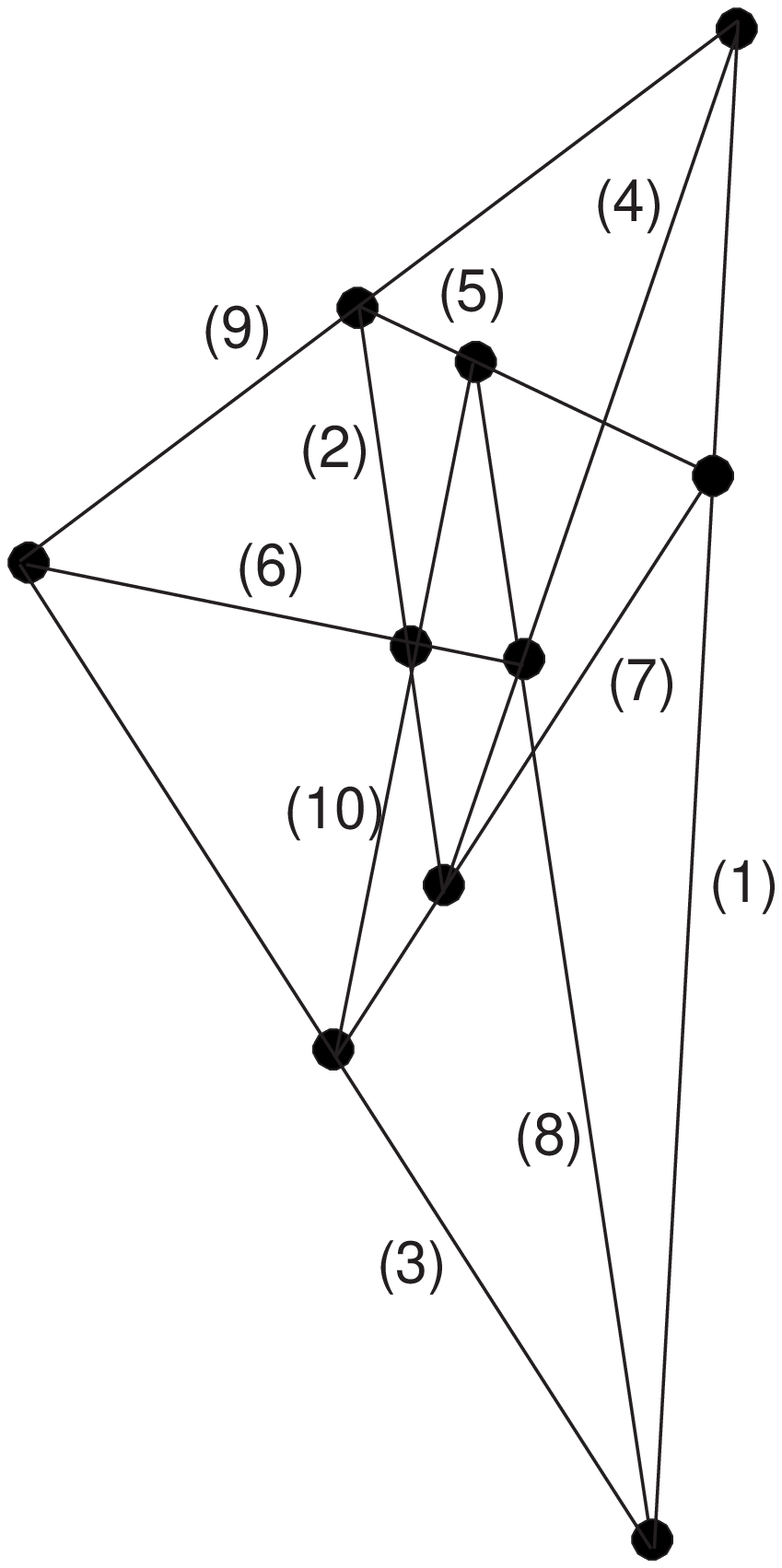} &
      \includegraphics[width=30mm]{D4} & $\det A(\mg_{(10_3,10_3)_7})=0$ \\
      \hline
       $(10_3,10_3)_8$ & \includegraphics[width=45mm]{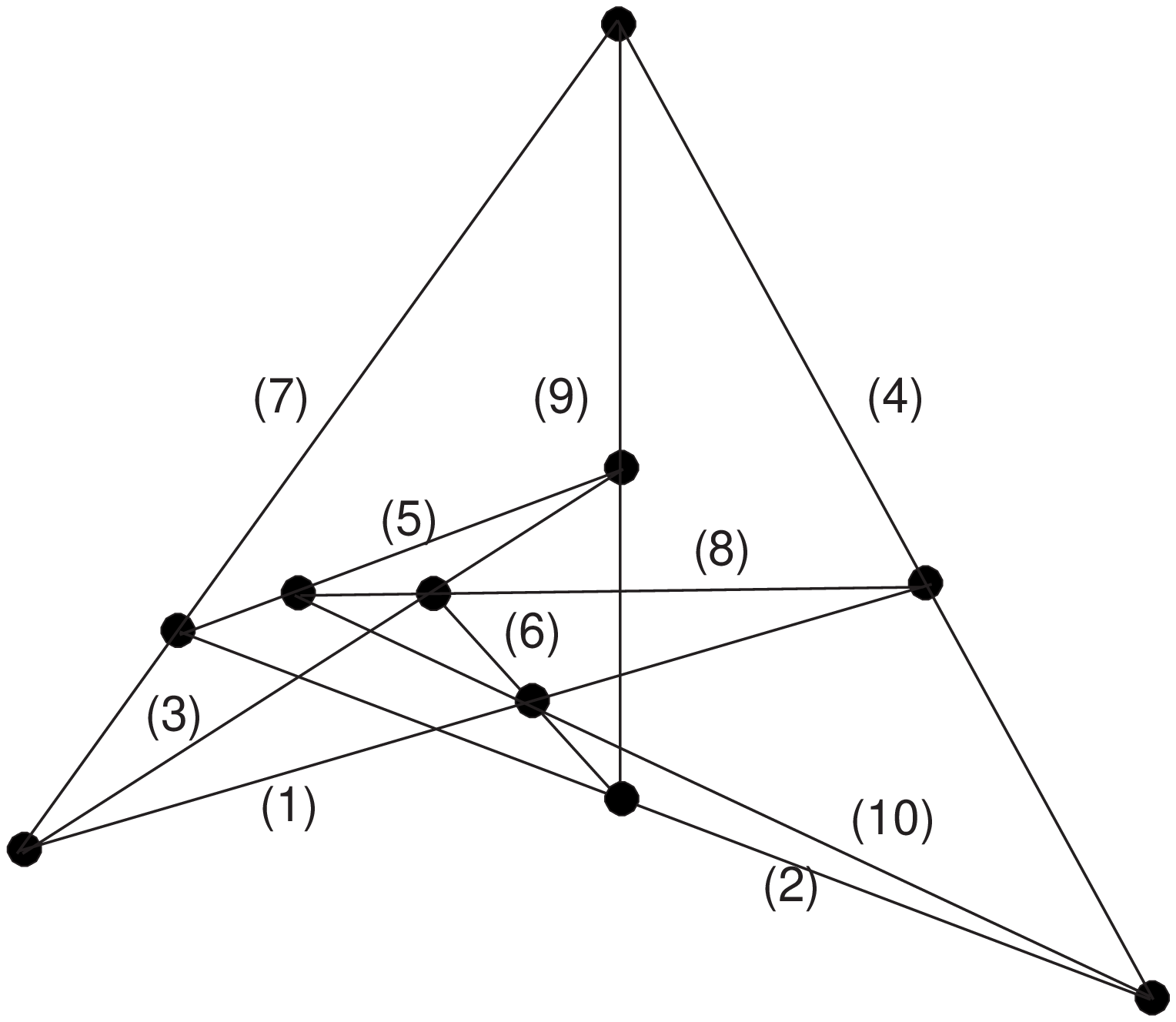} &
      \includegraphics[width=30mm]{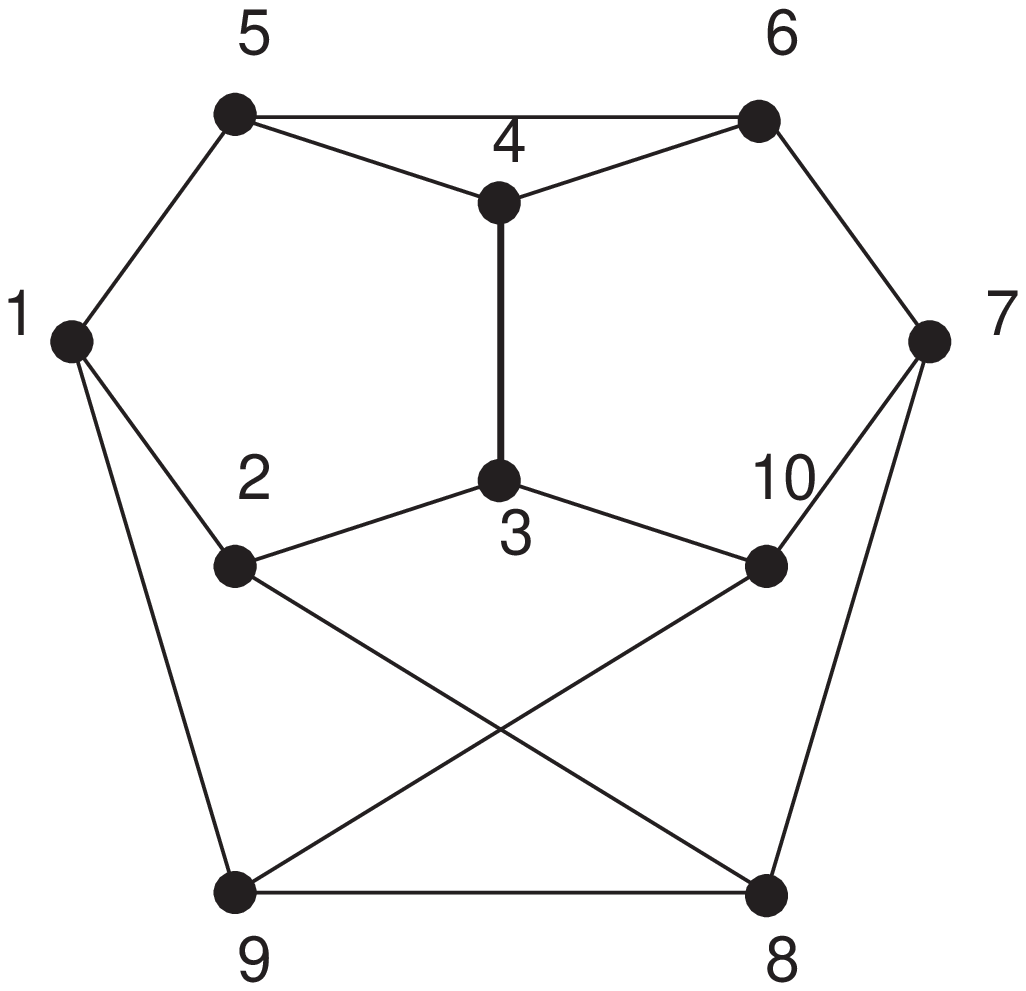} & $\det A(\mg_{(10_3,10_3)_8})=-64$  \\
    \hline
      $(10_3,10_3)_9$ &  \includegraphics[width=45mm]{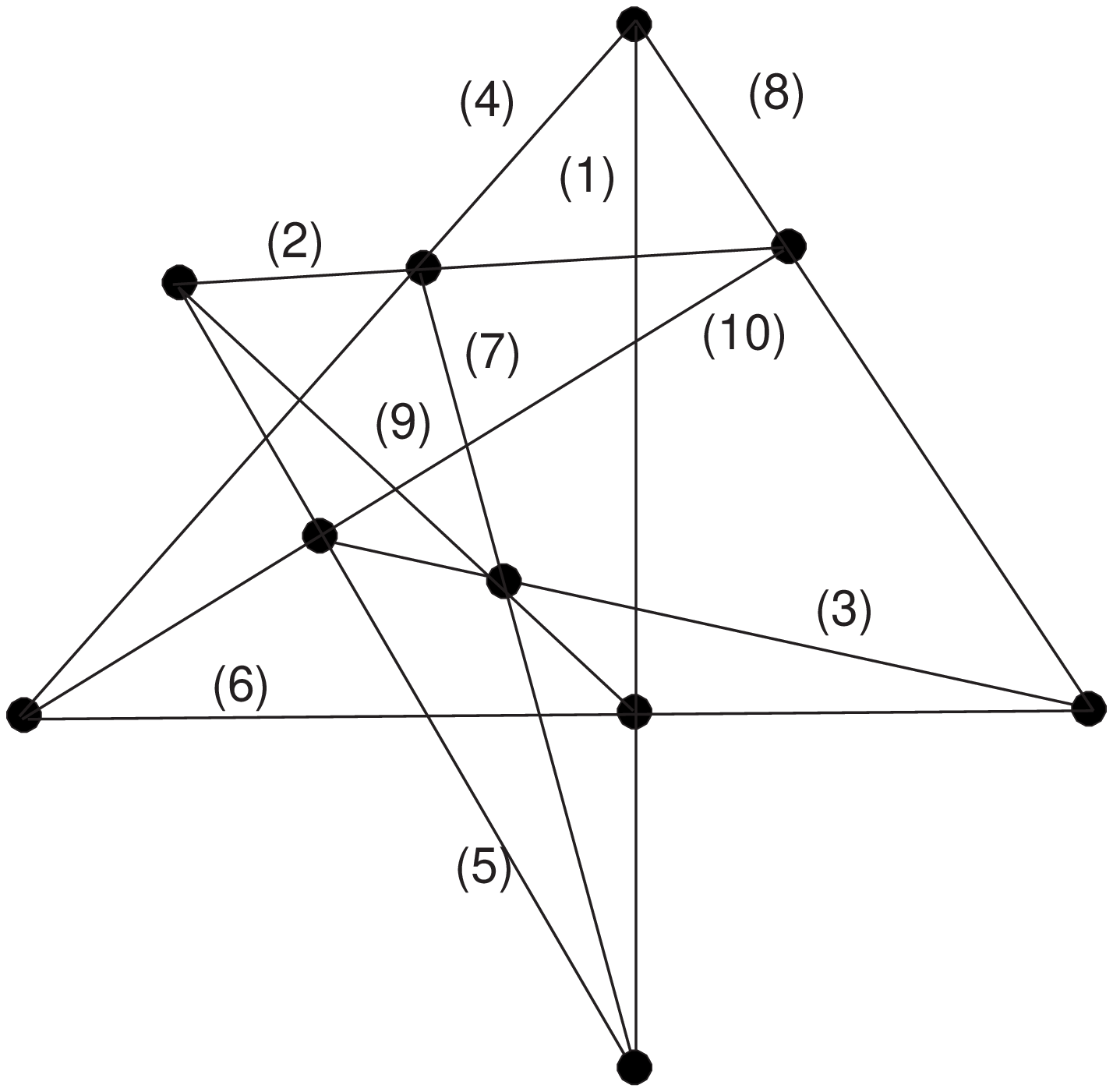} &
      \includegraphics[width=30mm]{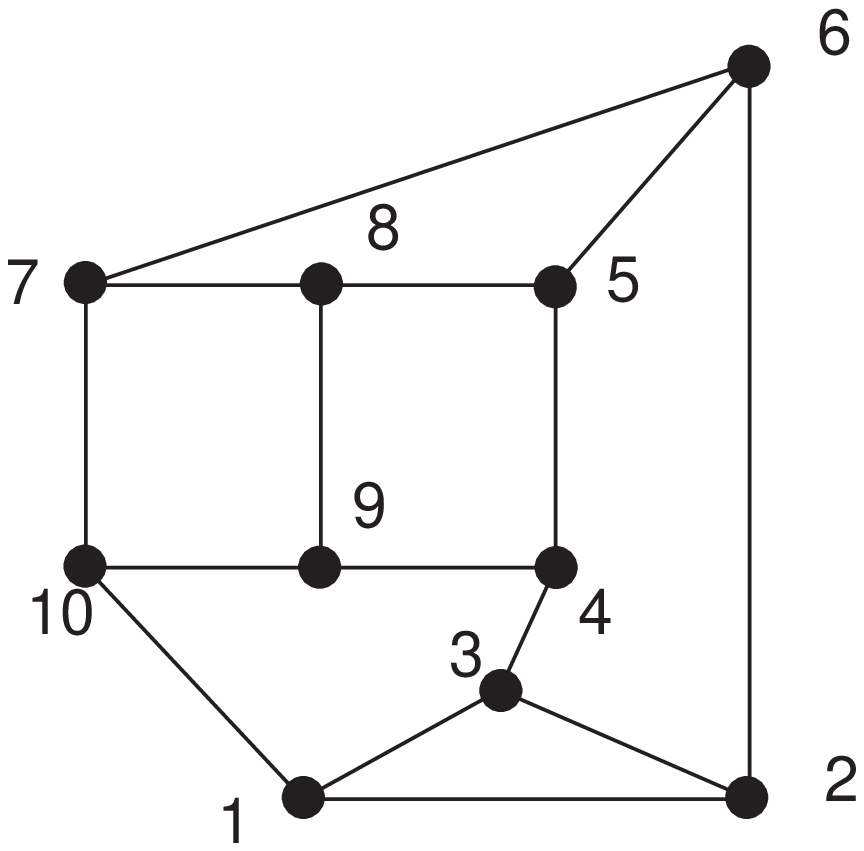} & $\det A(\mg_{(10_3,10_3)_9})=-49$ \\
      \hline
       $(10_3,10_3)_{10}$ &  \includegraphics[width=50mm]{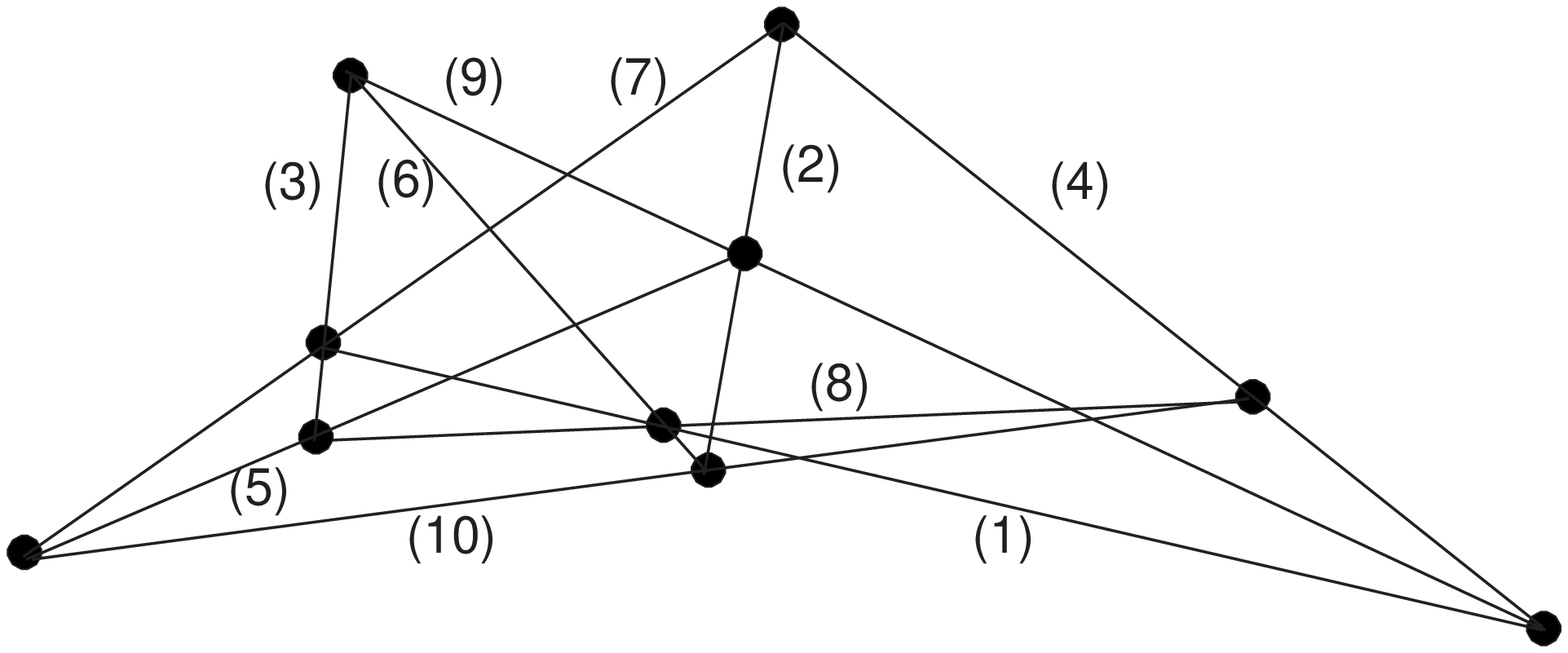} &
      \includegraphics[width=30mm]{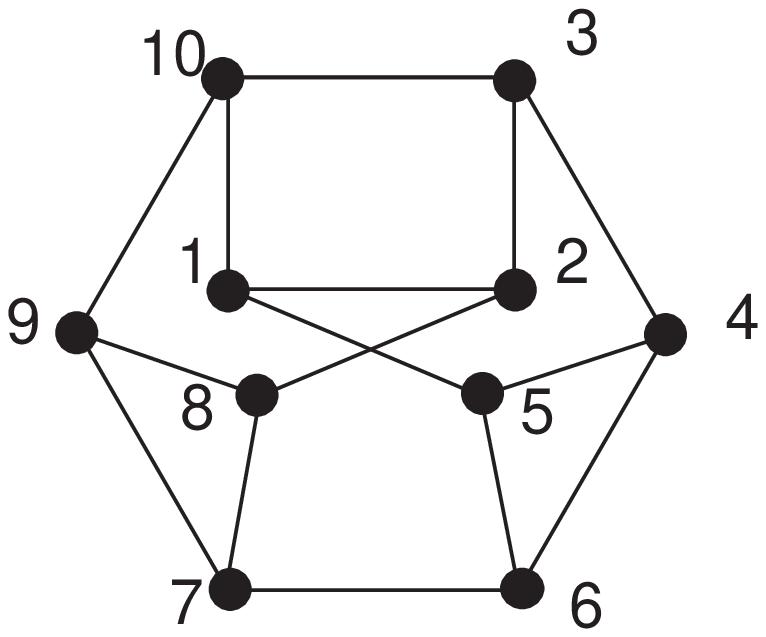} & $\det A(\mg_{(10_3,10_3)_{10}})=-25$ \\
      \hline
 \end{tabular}
         \caption{$n=10$ configurations and their dual Lorentzian Kac-Moody algebras.
         Note that some of the configurations give rise to equivalent Dynkin diagrams.  Here, we have
         ceased to number the points of the geometrical configurations as this information is not needed
         in order to draw the Dynkin diagram.}
\end{table}
\end{center}


\section{Conclusions}
\label{Conclusions} \setcounter{equation}{0}

In this paper, we have shed a new algebraic light on previous work
on M-theory cosmology \cite{Demaret:1985js}. This has been done by
associating to each geometric configuration $(n_m,g_3)$ a regular,
electric subalgebra of $E_{n}$, through the following rule: the
line incidence diagram of the geometric configuration $(n_m, g_3)$
is the Dynkin diagram of the corresponding regular subalgebra of
$E_n$. We have also derived explicitly which subalgebras arise for
all known geometric configurations with $n \leq 10$. In this
context, a particularly intriguing case was the realization of the
Petersen graph as the Dynkin diagram of a rank-$10$ Lorentzian
subalgebra of $E_{10}$.

These somewhat unexpected mathematical results encompass other
cosmological solutions besides those given in \cite{Demaret:1985js},
since the algebras that we have exhibited underlie many interesting,
time-dependent M-theory solutions. In particular, we found that
$\sigma$-models for commuting $A_1$-subalgebras of $E_{10}$ give
rise to intersecting $SM2$-brane solutions \cite{Strominger}. This
result is similar in spirit to that of
\cite{IntersectingEnglert,BPSEnglert} where it was discovered, in
the context of $\mf{g}^{+++}$-algebras, that the intersection rules
for $Mq$-branes are encoded in orthogonality conditions between the
various roots of $\mg^{+++}$. These intersection rules apply also to
spacelike branes \cite{IntersectingOhta} so they are of interest for
some of the solutions discussed in this paper. For two $Sq$-branes,
$A$ and $B$, in $M$-theory the rules are
\cite{IntersectingArgurio,Ohta:1997gw} \beq SMq_{A}\cap
SMq_{B}=\f{(q_{A}+1)(q_{B}+1)}{9}-1. \eqnlab{intersectionrules} \eeq
So, for example, if we have two $SM2$-branes the result is \beq
SM2\cap SM2=0, \eqnlab{intersectingSM2} \eeq which means that they
are allowed to intersect on a $0$-brane. Note that since we are
dealing with spacelike branes, a $0$-brane is extended in one
spatial direction so the two $SM2$-branes may therefore intersect in
one spatial direction only. Hence, the intersection rules are
fulfilled for the relevant configurations, namely $(3_1,1_3),
(6_{2},4_3)$ and $(7_3,7_3)$. So, in our treatment the orthogonality
conditions of \cite{IntersectingEnglert} are equivalent to only
exciting commuting $A_1$-algebras\footnote{This was also pointed out
in \cite{Kleinschmidt:2005gz}.}, regularly embedded in $E_{10}$.
This implies that the intersection rules are automatically fulfilled
for configurations with no parallel lines.

Our paper can be developed in various directions. One can look for
explicit new solutions using the sigma-model insight, for which
techniques have been developed. Systems with $n \leq 8$ are in
principle integrable so here we know that solutions can be found
in closed form.  However, also for $n = 9, 10$, simplifications
should arise since the algebras are affine or Lorentzian. In the
Lorentzian case the algebras are not hyperbolic so the associated
cosmological solutions must be non chaotic, and hence explicit
solutions should exist. Work along these lines is in progress. One
might also perhaps get new information on the meaning of the
higher level fields and the dictionary between supergravity and
the sigma-model in the context of these simpler algebras.

Another interesting possibility is to extend the approach taken in
this paper and consider ``magnetic algebras'' (for which the
simple roots are all magnetic) and their associated
configurations. This corresponds to exciting a set of fields at
level $2$ in the $E_{10}$-decomposition. The simplest case would
be to consider a configuration with one line through six points. A
possible choice of generators is \beqa
 {}Ê& & e=E^{123456}\quad f=F_{123456}
 \nn \\
 {}Ê& & h=[E^{123456},F_{123456}]=-\f{1}{6}\sum_{a\neq 1,\dots,6}
 {K^{a}}_{a}+ \f{1}{3}({K^{1}}_{1}+\cdots {K^{6}}_{6}).
 \eqnlab{SM5generators}
 \eqa
\noindent These generators constitute an $A_1$-subalgebra of
$E_{10}$ and the gravitational solution is precisely the
$SM5$-brane solution of \cite{Strominger}, i.e., in the billiard
language, a bounce against a magnetic wall. In a sense this gives
the simplest case of a duality between configurations with $3$
points and $6$ points. This can be seen as a manifestation of
electric-magnetic duality from an algebraic point of view. An
alternative approach could be to realize the magnetic algebras as
configurations with four points on each line, corresponding to the
spatial indices of the dual field strength, i.e. in the example
above we would then associate a configuration to $F_{789(10)}$
instead of $A_{123456}$.

A natural line of development is to further consider
configurations with $n>10$ since the association between geometric
configurations and regular subalgebras holds for the whole $E_n$
family. For instance, there exist $31$ configurations of type
$(11_3,11_3)$ \cite{Page} and these lead by our rules to rank-$11$
regular subalgebras of $E_{11}$ that are either of Lorentzian type
or, if their Cartan matrix is degenerate, of indefinite type with
Cartan subalgebra embeddable in a space of Lorentzian signature.
It would be of interest to investigate the solutions of
eleven-dimensional supergravity to which these algebras give rise.

One might wonder if new features arise if we relax some of the
rules defined in section $2.2$. For example, the condition of $m$
lines through each point was imposed mainly for aesthetical
reasons since this gives interesting configurations. Relaxing it
increases the number of different configurations for each $n$.
These also lead to regular subalgebras of $E_{10}$. For instance,
the set of 10 points with lines $(123)$, $(145)$, $(167)$, $(189)$
and $(79(10))$ yields the algebra $A_3 \oplus A_1 \oplus A_1$,
which is decomposable. However, these configurations with an
unequal number of lines through each point seem to give rise to
Dynkin diagrams with less structure.

\begin{figure}[ht]
\begin{center}
\includegraphics[width=60mm]{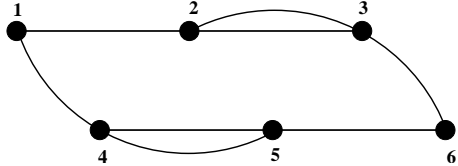}
\caption{This set of six points, four lines containing three
points each, with two lines through each point, is not a geometric
configuration because it violates Rule {\bf 3}:  two points may
determine more than one line.} \label{figure:G63}
\end{center}
\end{figure}

\begin{figure}[ht]
\begin{center}
\includegraphics[width=50mm]{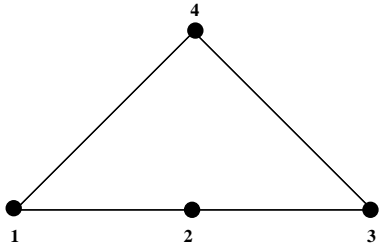}
\caption{The Dynkin diagram of $A_3^{+}$, associated with the
above set of points and lines through the rules outlined in the
text.} \label{figure:A3p}
\end{center}
\end{figure}

We could also consider going beyond positive regular subalgebras of
$E_{10}$.  A way to achieve this is to relax Rule {\bf \ref{rule3}},
i.e. that two points in the configuration determine at most one
line. Let us check this for a simple example. For $n=6$ a possible
configuration that violates Rule {\bf \ref{rule3}} is the set of six
points and four lines discussed in \cite{Demaret:1985js} and shown
in Figure \ref{figure:G63}.
\noindent In this case it is not interesting to define the
generators as we have previously done since this would give a
non-sensical Cartan matrix. For example, defining $e_1=E^{123}$
and $e_{2}=E^{236}$ gives commutators of the form $[h_1,e_2]=e_2$
and $[h_2,e_1]=e_1$ which give rise to a Cartan matrix with
positive entries. Instead, a reasonable choice of generators is
\beq e_{1}=E^{456}\qquad e_{2}=E^{123}\qquad e_{3}=F_{236} \qquad
e_{4}=F_{145}. \eqnlab{nonregularconfiguration} \eeq These yield
the Cartan matrix of the affine extension $A_3^{+}$ of $A_3$ (see
Figure \ref{figure:A3p}).
However, a new feature arises here because for example the
commutator $[e_{1},e_{4}]$ does not give a level $2$ generator but
instead we find an off-diagonal level $0$ generator \beq
[e_{1},e_{4}]=[E^{456},F_{145}]={K^{6}}_{1}. \eqnlab{off-diagonal}
\eeq In the gravitational solution this generator turns on an
off-diagonal metric component and so it corresponds to going
outside of the diagonal regime investigated in this paper (unless
one imposes further conditions as in \cite{Demaret:1985js}). We
further see that the embedding into $E_{10}$ is not regular since
positive root generators of the subalgebra are in fact negative
root generators of $E_{10}$, and vice versa.

{}Finally, on the mathematical side, we have uncovered seven
rank-10 Coxeter groups that are subgroups of the Weyl group of
$E_{10}$ and have the following properties: \begin{itemize} \item
Their Coxeter graphs (which coincide with the Dynkin diagrams of
the corresponding $E_{10}$-subalgebras in this ``simply-laced"
case) are connected. \item The only Coxeter exponents are 2 and 3
(i.e., the off-diagonal elements in the Cartan matrix are either 0
or -1). \item Each node is connected to exactly three other nodes.
\end{itemize}
Note that the determinants of their Cartan matrices are all minus
squared integers\footnote{Let us mention that among the conceivable
connected 10-points  Coxeter graphs (with each node linked to three
others), there are only four other cases with (non-positive)
determinants: 0,-125,-165 and  -192. These have only one negative
eigenvalue. One might wonder if they define also Coxeter subgroups
of the Weyl group of $E_{10}$.}.  It would be of interest to
determine the automorphisms of the corresponding Lie algebras and
investigate whether their embedding in $E_{10}$ is maximal. Also, we
have found two explicitly different embeddings for some of them and
one might inquire whether they are equivalent. Investigations of
these questions are currently in progress.

We also observe that the association of subalgebras of the relevant
Kac-Moody algebras $A_{D-3}^{++}$ to homogeneous cosmological models
has been done in \cite{deBuyl:2003za} in the context of pure gravity
in spacetime dimensions $D \leq 5$.

 \vspace{.5cm}
\noindent \textbf{Acknowledgments:} We thank A. Keurentjes for
discussions and for providing us with a copy of \cite{Dynkin}, and
J. Brown and A. Kleinschmidt for useful comments. D.P. would also
like to thank the following people for helpful discussions: Riccardo
Argurio, Sophie de Buyl, Jarah Evslin, Laurent Houart, Stanislav
Kuperstein, Carlo Maccaferri,
Jakob Palmkvist and Christoffer Petersson. \\
\indent Work supported in part by IISN-Belgium (convention
4.4511.06 (M.H. and P.S) and convention 4.4505.86 (M.H. and D.P)),
by the Belgian National Lottery, by the European Commission FP6
RTN programme MRTN-CT-2004-005104 (M.H., M.L. and D.P.), and by
the Belgian Federal Science Policy Office through the
Interuniversity Attraction Pole P5/27.
 Mauricio Leston is also supported in part by the
``FWO-Vlaanderen'' through project G.0428.06 and by a CONICET
graduate scholarship.


\end{document}